\documentclass[a4paper,usenatbib]{mnras}

\usepackage{times}

\usepackage[T1]{fontenc}
\usepackage{ae,aecompl}
\usepackage{multirow}

\usepackage{graphicx}	
\usepackage{amsmath}	
\usepackage{amssymb}	
\usepackage{amsfonts}

\usepackage{url}
\usepackage[flushleft]{threeparttable}

\usepackage{array}
\newcolumntype{C}[1]{>{\centering\let\newline\\\arraybackslash\hspace{0pt}}m{#1}}

\usepackage{etoolbox}

\hypersetup{colorlinks=true,linkcolor=blue,citecolor=blue,filecolor=blue,urlcolor=blue}
\makeatletter
\patchcmd{\NAT@citex}
  {\@citea\NAT@hyper@{%
     \NAT@nmfmt{\NAT@nm}%
     \hyper@natlinkbreak{\NAT@aysep\NAT@spacechar}{\@citeb\@extra@b@citeb}%
     \NAT@date}}
  {\@citea\NAT@nmfmt{\NAT@nm}%
   \NAT@aysep\NAT@spacechar\NAT@hyper@{\NAT@date}}{}{}

\patchcmd{\NAT@citex}
  {\@citea\NAT@hyper@{%
     \NAT@nmfmt{\NAT@nm}%
     \hyper@natlinkbreak{\NAT@spacechar\NAT@@open\if*#1*\else#1\NAT@spacechar\fi}%
       {\@citeb\@extra@b@citeb}%
     \NAT@date}}
  {\@citea\NAT@nmfmt{\NAT@nm}%
   \NAT@spacechar\NAT@@open\if*#1*\else#1\NAT@spacechar\fi\NAT@hyper@{\NAT@date}}
  {}{}
  
\usepackage{xspace}
  
\def\aj{AJ}

\def\apj{ApJ}
\def\apjl{ApJ}
\def\apjs{ApJS}
\def\ao{Appl.~Opt.}

\def\aap{A\&A}

\def\aaps{A\&AS}

\def\mnras{MNRAS}

\def\pasp{PASP}
\def\pasj{PASJ}

\def\nat{Nature}

\def\gca{Geochim.~Cosmochim.~Acta}

\def\jcp{J.~Chem.~Phys.}

\newcommand{\hi}{H\,{\sc i}}

\newcommand{\fexxvi}{Fe\,{\sc xxvi}}
\newcommand{\fexxv}{Fe\,{\sc xxv}}

\newcommand{\bc}{\color{blue}}

\def\lessim{\raise-.5ex\hbox{$\buildrel<\over{\scriptstyle\mathtt{\sim}}$}}
\def\grtsim{\raise-.5ex\hbox{$\buildrel>\over{\scriptstyle\mathtt{\sim}}$}}

\title[X-ray Variability of RT\,Cru]{Long-term X-ray Variability of the Symbiotic System RT\,Cru based on \textit{Chandra} Spectroscopy}
\author[A.~Danehkar et al.]{A.~Danehkar$^{{\bc 1},{\bc 2}}$\thanks{E-mail: danehkar@umich.edu}, 
M.~Karovska$^{{\bc 2}}$, 
J.~J.~Drake$^{{\bc 2}}$ and 
V.~L.~Kashyap$^{{\bc 2}}$
\\
$^{1}$Department of Astronomy, University of Michigan, 1085 S. University Ave., Ann Arbor, MI 48109, USA\\
$^{2}$Harvard-Smithsonian Center for Astrophysics, 60 Garden St., Cambridge, MA 02138, USA
}

\date{Accepted 2020 November 12. Received 2020 November 12; in original form 2020 July 27}

\pubyear{2020}

\begin{document}
\label{firstpage}

\pagerange{\pageref{firstpage}--\pageref{lastpage}} \pubyear{2020}

\maketitle

\begin{abstract}
RT\,Cru belongs to the rare class of hard X-ray emitting symbiotics, whose origin is not yet fully understood. In this work, we have conducted a detailed spectroscopic analysis of X-ray emission from RT\,Cru based on observations taken by the \textit{Chandra} Observatory using the Low Energy Transmission Grating (LETG) on the High-Resolution Camera Spectrometer (HRC-S) in 2015 and the High Energy Transmission Grating (HETG) on the Advanced CCD Imaging Spectrometer S-array (ACIS-S) in 2005. Our thermal plasma modeling of the time-averaged HRC-S/LETG spectrum suggests a mean temperature of $kT \sim 1.3$\,keV, whereas $kT \sim 9.6$\,keV according to the time-averaged ACIS-S/HETG. The soft thermal plasma emission component ($\sim1.3$\,keV) found in the HRC-S is heavily obscured by dense materials ($> 5 \times 10^{23}$\,cm$^{-2}$). The aperiodic variability seen in its light curves could be due to changes in either absorbing material covering the hard X-ray source or intrinsic emission mechanism in the inner layers of the accretion disk. To understand the variability, we extracted the spectra in the ``low/hard'' and ``high/soft'' spectral states, which indicated higher plasma temperatures in the low/hard states of both the ACIS-S and HRC-S. The source also has a fluorescent iron emission line at 6.4 keV, likely emitted from reflection off an accretion disk or dense absorber, which was twice as bright in the HRC-S epoch compared to the ACIS-S. The soft thermal component identified in the HRC-S might be an indication of a jet that deserves further evaluations using high-resolution imaging observations. 
\end{abstract}




\section{Introduction}
\label{rtcru:introduction}

Symbiotic systems are interacting binary stars consisting of a \textit{hot} accreting compact object and a \textit{cool} red giant companion \citep[see e.g.][]{Allen1984,Kenyon1986,Kenyon1987,Muerset1991,Belczynski2000}. Many of them demonstrate long- and shorter-term variability \citep{Houk1963,Mikolajewski1990,Sokoloski2001,Sokoloski2003a,Hinkle2009}. 
Although symbiotic stars mostly display soft or supersoft X-ray emission \citep[see e.g.][]{Muerset1997}, a few of them 
show hard X-ray emission, namely RT\,Cru \citep{Luna2007,Kennea2009,Eze2014,Ducci2016,Luna2018}, 
CH\,Cyg \citep{Kennea2009,Eze2014}, T\,CrB \citep{Tueller2005,Luna2008,Kennea2009,Eze2014}, CD--57\,3057 (SS73\,17) \citep{Smith2008,Kennea2009,Eze2010,Eze2014}, and
MWC\,560 \citep{Stute2009}. These hard X-ray emitting symbiotic systems pose a challenge to our understanding of accreting white dwarfs (WDs). We have not yet clearly understood the origin of hard X-ray emitting symbiotics.
It was argued that hard X-ray emission could be an indication of massive WDs \citep{Luna2007}.
Moreover, \citet{Kennea2009} proposed that these hard X-ray emitting symbiotics with massive WDs could be potentially progenitors of type Ia supernovae. 

\begin{table*}
\begin{center}
\caption[]{\textit{Chandra}  Observations of RT\,Cru}
\label{tab:obs:log}
\begin{tabular}{llllccc}
  \hline\hline\noalign{\smallskip}
Instrument & Gratings  & Seq.\,No & Obs.\,ID \& PI &  UT Start  &  UT End 
            & Time (ks) \\
\noalign{\smallskip}
   \hline\noalign{\smallskip}
HRC-S & LETG & 300332 & 16688, Karovska & 2015 Nov 23, 02:01 & 2015 Nov 23, 09:38 
         & 25.15 \\
\noalign{\smallskip}
HRC-S & LETG & 300332 & 18710, Karovska & 2015 Nov 23, 22:42 & 2015 Nov 24, 14:13 
         & 53.73 \\
\noalign{\smallskip}
ACIS-S & HETG & 300171 & 7186, Sokoloski & 2005 Oct 19, 10:21 & 2005 Oct 20, 16:31
         & 49.34 \\      
\noalign{\smallskip}\hline
\end{tabular}
\end{center}
\end{table*}

RT\,Cru was first classified as a symbiotic star by \citet{Cieslinski1994} based on its optical light curve features. Previously, it was identified as a visible star \citep{Leavitt1906,Duner1911}, a L-type variable star \citep{Kukarkin1969} and an early (O-A) type irregular variable star \citep{Kholopov1987}. It was observed to contain rapid brightness variations with scales of 10-30 minutes, so called flickering \citep{Cieslinski1994}. 
It was also found that it has the hard X-rays emission using the IBIS (Imager on Board the \textit{INTEGRAL} Satellite) instrument of the \textit{International Gamma-Ray Astrophysics Laboratory} (\textit{INTEGRAL}) satellite in 2003 and 2004 \citep{Chernyakova2005}, and again in 2012 and 2015 \citep{Sguera2012,Sguera2015}. The X-Ray Telescope (XRT) on the \textit{Swift} satellite also confirmed the presence of hard X-ray emission from its accreting WD, while also showing a large variation in X-ray flux of a factor of 3 over a period of 5 days \citep{Kennea2009}. More recently, \textit{NuSTAR} observations of RT\,Cru pointed to a reflection in the hard X-ray spectrum  from the boundary layer of the accretion disk \citep{Luna2018}. 

Several X-ray observations of RT\,Cru revealed the presence of high plasma temperatures associated with hard X-ray emission. Using \textit{Chandra} observations (0.3--8\,keV), \cite{Luna2007} determined a plasma temperature of $kT=8.6^{+3.8}_{-2.5}$\,keV from a highly absorbed, optically thin thermal plasma model, and a 
temperature 
of $kT_{\rm max}=80^{+\cdots}_{-24}$\,keV by fitting a cooling flow model with absorbers. Moreover, \cite{Kennea2009} derived 
$kT=37^{+7}_{-6}$\,keV with a partially covered absorbed thermal bremsstrahlung model from combined the Burst Alert Telescope (BAT) and X-ray telescope (XRT) on the \textit{Swift} satellite. A \textit{Suzaku} observation of RT\,Cru was also modeled with a thermal bremsstrahlung continuum having a temperature of $kT=29^{+9}_{-5}$\,keV \citep{Eze2014}. 
Furthermore, a high cooling flow temperature of $kT_{\rm max}=51^{+3}_{-2}$\,keV was estimated from combined archival INTEGRAL $\gamma$-ray satellite and \textit{Swift}/XRT data, in addition to $kT_{\rm max}=79.9^{+\cdots}_{-19.9}$\,keV from \textit{Suzaku} data \citep{Ducci2016}. More recently, \citet{Luna2018} obtained a maximum post-shock temperature of $kT=53\pm 4$\,keV from spectral fitting analysis of \textit{Suzaku} and \textit{NuSTAR}+\textit{Swift} observations, and suggested a WD mass of $1.25 \pm 0.02 M_{\odot}$. 

Previous X-ray observations of RT\,Cru have effectively measured its spectral features in the hard X-ray excess ($\gg 1$\,keV), except for the recent \textit{Chandra} observations in 2015 that we analyze here. 
The X-ray observations prior to 2012 implied that RT\,Cru has X-ray emission produced by plasmas having high temperatures ($kT> 9$\,keV).
However, our observations in 2012 taken simultaneously by the \textit{Chandra} high-resolution camera imager (HRC-I) and \textit{Swift}/XRT indicated the presence of 
flare-like variations in the soft band below 4\,keV, in addition to 
a possible soft thermal plasma component by \textit{Chandra}/HRC-I and an anticorrelation between the intensity and spectral hardness by \textit{Swift}/XRT \citep{Kashyap2013}. 
The flare-like variations have been detected previously \citep{Luna2007,Kennea2009}, and again recently by \citet{Ducci2016}.
Our knowledge of RT\,Cru was therefore limited to its X-ray features in the hard X-ray domain and flare-like variations prior to its recent \textit{Chandra} spectroscopy. 

In this study, we conducted new spectral analysis of \textit{Chandra} 
spectroscopic observations of RT\,Cru taken in 2005 and 2015.
The first \textit{Chandra} spectroscopic observation in 2005 was taken using the Advanced CCD Imaging Spectrometer S-array (ACIS-S) with the High Energy Transmission Grating (HETG; see \S\,\ref{rtcru:observations:acis}) that was  more sensitive in the ``hard band'' above 4\,keV.
The recent \textit{Chandra} observations in 2015 allowed us to have a new look into the X-ray characteristics of this peculiar symbiotic star after 10 years with the same space telescope, but the High-Resolution
Camera Spectrometer (HRC-S) with the Low Energy Transmission Grating (LETG; see \S\,\ref{rtcru:observations:hrcs}) that are more sensitive in the ``soft band'' below 4\,keV.
We used these observations in two different epochs to determine the plasma temperatures and absorbing column densities in the \textit{soft} and \textit{hard} energy bands, and \textit{low/hard} and \textit{high/soft} spectral states, as well as the mass-accretion rates in different epochs and states. In Section~\ref{rtcru:observations}, we describe the
observations and data reduction. Section~\ref{rtcru:analysis} presents the timing and hardness ratio analyses, followed by the spectral analysis and modeling results. 
A discussion is presented in Section~\ref{rtcru:discussion}, and a conclusion is given in Section~\ref{rtcru:conclusion}.

\section{Observations and Data Reduction}
\label{rtcru:observations}

X-ray spectroscopic observations of RT\,Cru were taken by the \textit{Chandra} Observatory using the ACIS-S \citep{Garmire2003} in 2005 and the HRC-S \citep{Murray1997} in 2015. The ACIS-S observation was previously analyzed by \citet{Luna2007}. In our study, we employed a Markov chain Monte Carlo (MCMC) based Bayesian statistic method to determine 
the best-fitted parameters in our spectral modeling (see \S\,\ref{rtcru:analysis:spectral}), and a novel approach to study the HRC-S data by decomposing the observational events according to the \textit{low/hard} and \textit{high/soft} spectral states (see \S\,\ref{rtcru:analysis:hardness}) by building Good Time Interval (GTI) tables for the data reduction described in this section. We utilized the same method and approach to further study the ACIS-S data that were not implemented previously by \citet{Luna2007}. 
 
\subsection{HRC-S/LETG Data Reduction}
\label{rtcru:observations:hrcs}

RT\,Cru was observed over two visits (Proposal 16300574, Obs.ID 16688 \& 18710, PI: Karovska) during 2015 November 23--24  with the LETG \citep{Brinkman2000} on the HRC-S. 
The observation details are given in Table~\ref{tab:obs:log}. 
Useful exposure times of 25.15 and 53.73\,ks were obtained for the two visits, respectively, which yield a total exposure time of about 79\,ks over a period of $\sim 1.5$ days.

The data reduction was performed according to standard procedures using the software package \textsc{ciao} v4.12 ``Chandra Interactive Analysis of 
Observation''\footnote{\url{https://cxc.harvard.edu/ciao/}} 
\citep{Fruscione2006} and the calibration files from CALDB v4.9.1. 
All events were manually processed using the \textsc{ciao} functions. The dispersed ($m=\pm 1$) source spectra were extracted using the \textsc{ciao} tools \textsf{tgextract} and \textsf{dmtype2split}.  The background spectra were also produced using the \textsf{tg\_bkg} script. The response matrix function (RMF) and the ancillary response matrix (ARF) files of the first-orders ($m=\pm 1$) were generated using the tool \textsf{mktgresp}. 

For the hardness ratio analysis, the \textsc{ciao} function \textsf{dmcopy} was used to filter the level 1 (L1) events (reprocessed by \textsf{hrc\_process\_events}) based on GTI tables made for the high/soft and low/hard states (see \S\,\ref{rtcru:analysis:hardness}), while the dead time correction factor, the exposure time, and the live time (that excludes the dead time associated with the image transfer) were manually corrected in the header keywords of the event files during the standard reduction after the event filtering procedure (see 
``CIAO Manual: LETG/HRC-S Grating Spectra''\footnote{\href{https://cxc.cfa.harvard.edu/ciao/threads/spectra_letghrcs}{cxc.cfa.harvard.edu/ciao/threads/spectra\_letghrcs}}). 
The grating events were manually assigned to spectral orders using the \textsc{ciao} function \textsf{tg\_resolve\_events}, 
and the source \textsf{pha} files were then extracted from the event files using the \textsc{ciao} tool \textsf{tgextract}. 
To have high signal-to-noise ratios in the time-averaged analysis, we used the \textsc{ciao} tool \textsf{combine\_grating\_spectra} to merge the dispersed spectra orders from the two observations. 
To produce the low/hard- and high/soft-state spectra, we similarly merged the dispersed spectra from events filtered based on GTI tables.

\subsection{ACIS-S/HETG Data Reduction}
\label{rtcru:observations:acis}

RT\,Cru was observed on 2005 October 19 (Proposal 07300767, Obs.ID 7186, PI: Sokoloski) with the HETG 
\citep{Weisskopf2002,Canizares2005} on the ACIS-S. 
The observation details are presented in Table \ref{tab:obs:log}. 
The useful exposure time was about $49$\,ks. 
The HETG instrument has two sets of gratings: the medium energy grating
(MEG) and the high energy grating (HEG). The MEG covers 0.4--7\,keV with a full-width at half maximum (FWHM) of 0.023 {\AA},
whereas the HEG covers 0.8--10\,keV with an FWHM of 0.012 {\AA} (See ``\textit{Chandra} Proposers' Observatory Guide'' 
 Rev.22.0, 2019). 
The MEG with a resolving power of $E/\triangle E = 660$ at 0.826\,keV can effectively measure the soft band, while
the HEG with a resolving power of $E/\triangle E = 1000$ at 1\,keV was designed to be more efficient in the hard band. 

We reprocessed the ACIS-S/HETG data of RT\,Cru with the recent calibration data (CALDB v4.9.1), and manually applied the \textsc{ciao} standard functions to the HETG data to generate the plus and
minus first-order ($m=\pm1$) MEG and HEG data and their corresponding response files. 
We used the \textsc{ciao} task \textsf{tgextact} to extract the pulse height amplitude (\textsf{pha}) files from the MEG and HEG $m=\pm1$ orders, 
and the \textsc{ciao} task \textsf{mkgrmf} and  \textsf{mkgarf} to produce the redistribution matrix (\textsf{rmf}) and auxiliary response (\textsf{arf}) files.

\begin{figure}
\begin{center}
\includegraphics[width=2.9in, trim = 0 20 10 10, clip, angle=0]{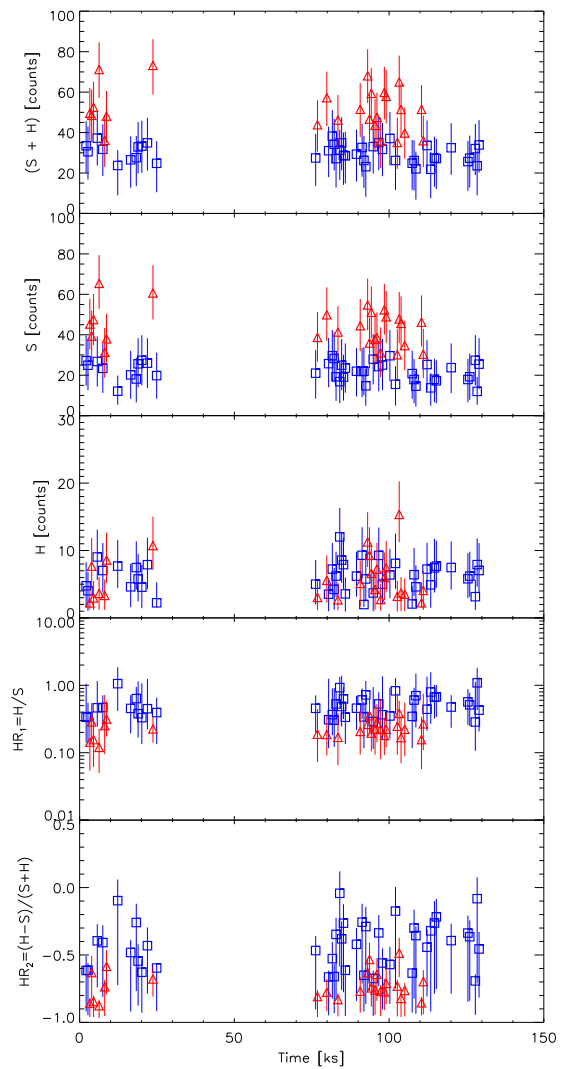}%
\caption{From top to bottom: the HRC-S/LETG light curves of RT\,Cru binned at 600\,sec in the 0.3--8 keV
  broad band ($S+H$), 0.3--4\,keV energy ($S$), 4--8\,keV energy ($H$), and the hardness ratios $\mathrm{HR_{1}}=H/S$ and $\mathrm{HR_{2}}=(H-S)/(S+H)$, obtained from the source and background light curves using the BEHR. The time unit is kilosecond (starts at 2015-11-23 02:01). The energy band unit is counts. Plots show only bins with both statistically significant $S$ and $H$ bands. 
The X-ray source denoted by blue squares and red triangles are classified under the low/hard and high/soft states, respectively (see  Figure~\ref{fig:rtcru:hard} ).
\label{fig:rtcru:light}%
}
\end{center}
\end{figure}
\begin{figure}
\begin{center}
\includegraphics[width=2.9in, trim = 0 20 10 10, clip, angle=0]{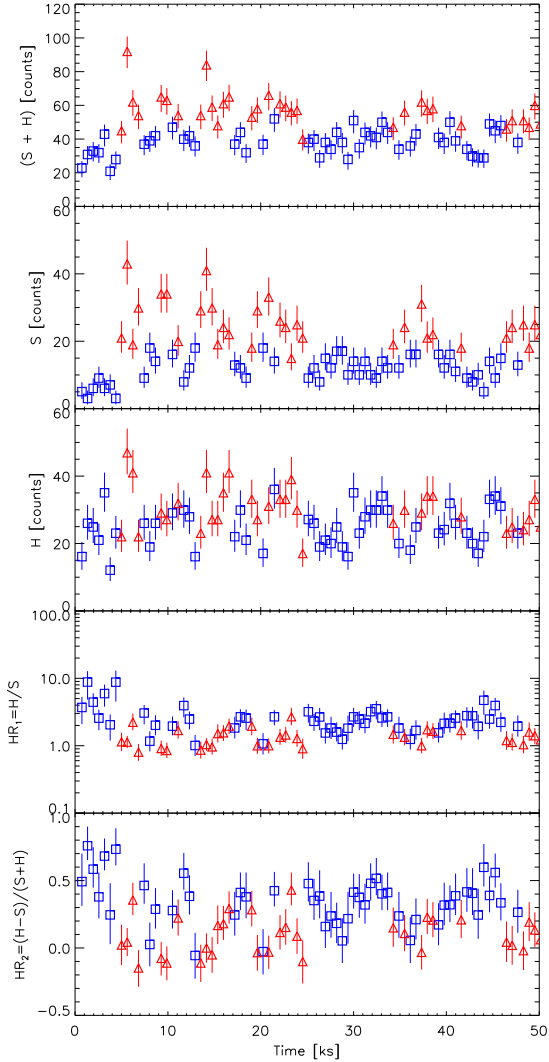}%
\caption{The same as Figure~\ref{fig:rtcru:light}, but for the ACIS-S/HETG light curves of RT\,Cru with a bin size of 600\,sec. 
The hardness ratios $\mathrm{HR_{1}}$ and $\mathrm{HR_{2}}$ were derived from the source light curves using the BEHR.
\label{fig:rtcru:light2}%
}
\end{center}
\end{figure}

To implement our hardness ratio analysis, we employed the \textsc{ciao} function \textsf{dmcopy} to copy events from the L1-event file (reprocessed by \textsf{acis\_process\_events}) based on a GTI table made for each state (high/soft and low/hard states defined in \S\,\ref{rtcru:analysis:hardness}). We then reprocessed the copied L1-event file with the \textsc{ciao} standard functions (\textsf{tg\_resolve\_events} and \textsf{tgextract}) to create the \textsf{pha} file. We merged the $m=\pm1$ order MEG and HEG data using the \textsc{ciao} tool \textsf{combine\_grating\_spectra} to make the time-averaged spectrum, low/hard-state and high/soft-stat spectra. 

\section{Data Analysis and Results}
\label{rtcru:analysis}

\subsection{Light Curves}

\begin{figure}
\begin{center}
\includegraphics[width=2.9in, trim = 0 25 10 10, clip, angle=0]{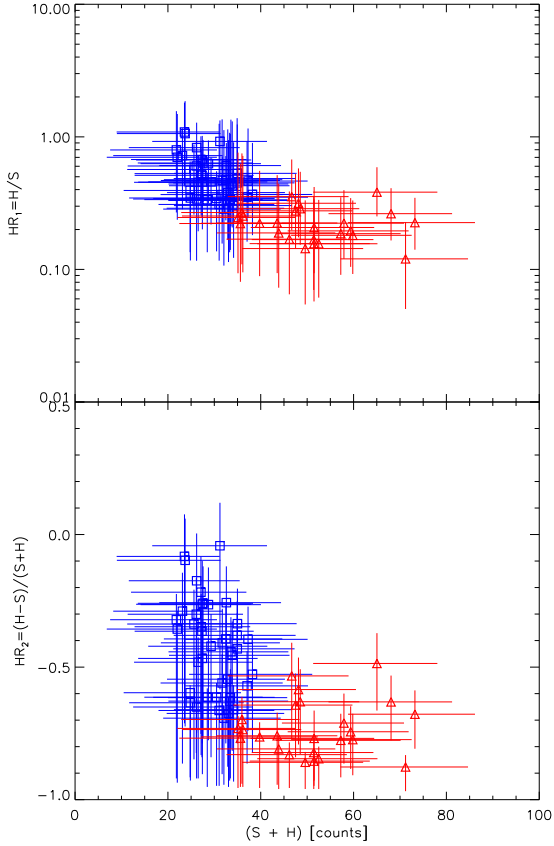}%
\caption{Hardness ratio diagrams of the HRC-S/LETG data: the hardness ratios $\mathrm{HR_{1}}=H/S$ (top
  panel), and $\mathrm{HR_{2}}=(H-S)/(S+H)$ (bottom panel) plotted
  against the energy band 0.3--8\,keV ($S+H$; in counts) binned at 600 sec. The hardness ratios were calculated using the BEHR from the source and background light curves. The X-ray source in the \textit{low/hard} and \textit{high/soft} states is denoted by blue squares and red triangles, respectively. Plots show only bins with both statistically significant $S$ and $H$ bands. 
\label{fig:rtcru:hard}%
}
\end{center}
\end{figure}

To investigate the variability and hardness ratios of RT\,Cru during the HRC-S/LETG observations in 2015, we utilized the \textsc{ciao} tool \textsf{dmextract} to create light curves in different energy bands and with different binning. 
For these hardness ratios, we chose the \textit{soft} band ($S$: 0.3--4\,keV) and the \textit{hard} band ($H$: 4--8\,keV).
We binned the time series at 60 and 600~sec intervals in order to detect any flickering variations and possible periodic brightness modulations of binary orbital periods or WD rotation periods. We used the tools \textsf{efsearch} and \textsf{powspec} 
from \textsc{xronos} 
 timing analysis software package v5.22 \citep{Stella1992} to search for periodic modulations in the light curves binned at 60\,sec in different energy bands. Although the light curves show aperiodic flickering variations, we did not detect any periodic modulations. This agrees with the previous studies  \citep{Luna2007,Kennea2009,Ducci2016}. We used the HRC-S/LETG light curves binned at 600\,sec for our hardness ratio analysis, and to create GTI tables for filtering observational events according to the HR diagrams shown in Figure~\ref{fig:rtcru:hard}.

Figure~\ref{fig:rtcru:light} shows the HRC-S/LETG light curves of RT\,Cru binned at 600-seconds intervals in the 0.3--8\,keV energy band ($S+H$), whereas the $S$ band (0.3--4\,keV) and the $H$ band (4--8\,keV). 
The light curves present the net counts and were background subtracted. We notice that the light curves are dominated by the background noise in several intervals, so it was necessary to carefully handle both, the source and background, as different data sets and simultaneously model both of them in the spectral analysis. In Figure~\ref{fig:rtcru:light}, we also plotted the hardness ratios $\mathrm{HR_{1}}$ and $\mathrm{HR_{2}}$ described in \S\,\ref{rtcru:analysis:hardness}. The binned light-curve points plotted by blue squares and red triangles are according to the \textit{low/hard} and \textit{high/soft} spectral states classified on the HR diagrams in \S\,\ref{rtcru:analysis:hardness}, respectively.

\begin{figure}
\begin{center}
\includegraphics[width=2.9in, trim = 0 25 10 10, clip, angle=0]{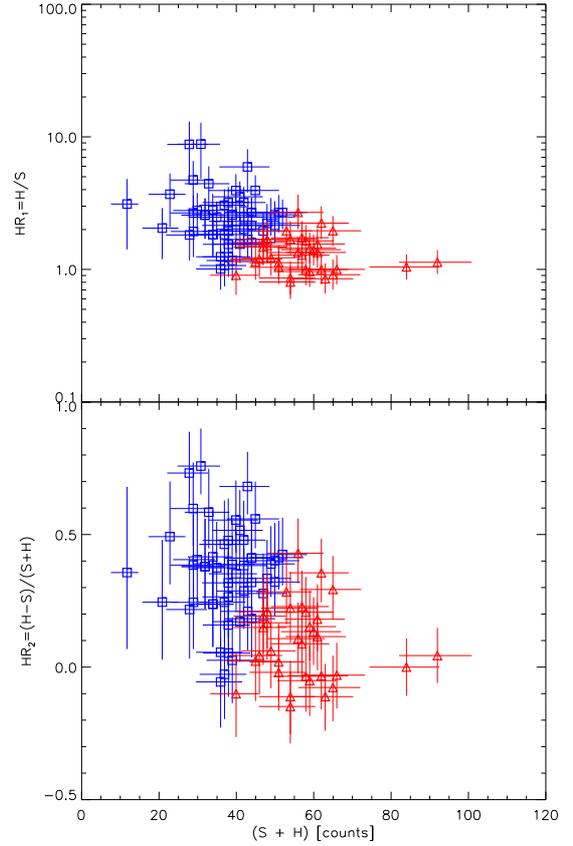}%
\caption{The same as Figure~\ref{fig:rtcru:hard}, but for the ACIS-S/HETG data. 
\label{fig:rtcru:hard2}%
}
\end{center}
\end{figure}

To analyze the hardness ratios of the ACIS-S/HETG observation of RT\,Cru in 2005, we also employed the ``ACIS Grating Light Curve'' (\textsf{aglc}) program\footnote{\url{http://space.mit.edu/cxc/analysis/aglc/}} developed by \citet{Huenemoerder2011} in the ``Interactive Spectral Interpretation System''\footnote{\url{http://space.mit.edu/asc/isis/}} \citep[\textsc{isis};][]{Houck2000} to generate the light curves in 60-s and 600-s bins for three bands: the soft band ($S$: 0.3--4
keV), the hard band ($H$: 4--8 keV), and 0.3--8 keV broad band ($S+H$: 0.3--8.0
keV). We did not detect any periodic modulations in the light curves binned at 60\,sec. 
The ACIS-S/HETG light curves binned at 600\,sec were used to conduct the hardness ratio analysis and generate GTI tables for separating observational events according to the low/hard and high/soft spectral states in the HR diagrams (see Figure~\ref{fig:rtcru:hard2}).

In Figure~\ref{fig:rtcru:light2} we plotted the light curves of RT\,Cru derived from the ACIS-S/HETG data with a bin size of 600 sec for the broad $S+H$ energy band (0.3--8\,keV), the $S$ band (0.3--4\,keV), and the $H$ band (4--8\,keV). 
The light curve units are in counts. The ACIS-S, in contrast to the HRC-S, has a negligible background noise in extracted spectra for the exposures analyzed here. 
Figure~\ref{fig:rtcru:light2} also shows the hardness ratios $\mathrm{HR_{1}}$ and $\mathrm{HR_{2}}$ defined in \S\,\ref{rtcru:analysis:hardness}. Similarly, we also plotted the binned light curves in the low/hard and high/soft states according to the HR diagrams (see \S\,\ref{rtcru:analysis:hardness}) using blue squares and red triangles, respectively.

\subsection{Hardness Ratios}
\label{rtcru:analysis:hardness}

We conducted the hardness ratio analysis based on the light curves binned at 600 sec extracted from the soft and hard energy bands. Similar hardness ratio studies of RT\,Cru have been carried out by \citet{Kennea2009}, \citet{Ducci2016}, and \citet{Luna2018}. 
Hardness ratio spectral analyses have been also used to separate the soft thermal components and the hard nonthermal components in \textit{Chandra} X-ray surveys \citep[see e.g.][]{Brassington2008,Plucinsky2008}, 
as well as X-ray studies of active galactic nuclei \citep[e.g.][]{Worrall2010,Danehkar2018}.

We calculated the hardness ratios ($\mathrm{HR_{1}}$ and $\mathrm{HR_{2}}$) using the following definitions and the light-curves binned at 600\,sec in the \textit{soft} ($S$: 0.3--4\,keV) and \textit{hard} ($H$: 4--8\,keV) bands:
\begin{align}
\mathrm{HR_{1}} & =H/S,\label{eq:hr:1}\\
\mathrm{HR_{2}} & =(H-S)/(S+H).\label{eq:hr:2}
\end{align}
To propagate uncertainties of the source and background counts, we employed the Bayesian Estimator for Hardness Ratios \citep[BEHR;][]{Park2006}. 

The hardness ratios $\mathrm{HR_{1}}$ and $\mathrm{HR_{2}}$ of the HRC-S/LETG light curves binned at 600-seconds intervals are shown in Figure~\ref{fig:rtcru:light} (two bottom panels).  For the HRC-S/LETG observations, we also plotted the hardness ratios $\mathrm{HR_{1}}$ and $\mathrm{HR_{2}}$ versus the 0.3--8\,keV energy band ($H+S$) in Figure~\ref{fig:rtcru:hard}. It is seen that there is an anticorrelation between the hardness ratio $\mathrm{HR_{1}}$ and the brightness, which is in agreement with \citet{Kennea2009}, \citet{Kashyap2013} and \citet{Ducci2016}. The light curves shown in Figure~\ref{fig:rtcru:light} were also classified under the \textit{low} (hard) and \textit{high} (soft) states  based on the $\mathrm{HR_{2}}$ versus $S+H$ diagram of Figure~\ref{fig:rtcru:hard}. The division in the HR diagrams of the HRC-S/LETG data was selected according to $\mathrm{HR_{2}}=0.025(S+H)-1.54$ that yields an approximately equal total net counts in each state (low and high). 
Similarly, we divided the HR diagrams of the ACIS-S/HETG data based on $\mathrm{HR_{2}}=0.025(S+H)-0.96$ that provides roughly equal total net counts in the low/hard and high/soft states (see Figure~\ref{fig:rtcru:hard2}). 
The $S$ band light curves of the HRC-S/LETG demonstrate that the X-ray source in the high/soft state emerges as approximately hourly aperiodic flickering (brightening) events.
However, these brightening events were also visible in the $H$ band light curves of the ACIS-S/HETG, in addition to those in its $S$ band. In Figures \ref{fig:rtcru:hard} and \ref{fig:rtcru:hard2}, we plotted the low/hard and high/soft state events by blue squares and red triangles, respectively. 

\subsection{Spectral Analysis}
\label{rtcru:analysis:spectral}

The boundary layer of an accretion disk could be optically thin or thick to its X-ray radiation in different regions, while obscuring clumpy materials moving within the line of sight around the X-ray source are thought to produce X-ray variations in RT\,Cru \citep{Kennea2009}. The low/hard (high/soft) state that is associated with lower (higher) brightening events could originate from
either the thermal variability in the accretion flows or obscuring clumpy medium. 
To understand better the X-ray variability of RT\,Cru, we therefore modeled the spectra using 
a thermal plasma model covered by a local absorber that could produce X-ray variations made by obscuring clumpy materials.

We conducted the spectral analysis on the time-averaged spectra of the merged HRC-S/LETG observations and the simultaneous spectral fitting of the MEG and HEG $m=\pm1$ orders of the ACIS-S/HETG data.
We used the CXC's Modeling and Fitting Package \textsf{Sherpa}\footnote{\url{https://cxc.cfa.harvard.edu/sherpa/}} \citep{Freeman2001}, from \textsc{ciao} v4.12, running under Python v3.5.4, together with \textsc{xspec} X-ray spectral-fitting models \citep{Arnaud1996}. 
The time-averaged spectrum analysis allowed us to study the overall physical properties of this X-ray source without considering 
them in the low/hard to high/soft states.  
To model the background in each data set, we employed an unabsorbed broken powerlaw component (\textsc{xspec} model \textsf{bknpower}) without setting any auxiliary matrix over the 0.3--8\,keV energy range. 
For the background modeling, we used the ``cstat'' statistic maximum-likelihood function \citep{Cash1979}, 
together with the Levenberg--Marquardt optimization method \citep{More1978}.
As seen in Table~\ref{rtcru:mekal:parameters}, the powerlaw indexes and normalization factors of the background models are roughly similar
in the HRC-S time-averaged, low/hard-state, and high/soft-state spectra. For the ACIS-S observations (see Table~\ref{rtcru:mekal:parameters2}),  
the backgrounds are not as high as those in the HRC-S. It can be noticed that the normalization factors  
and the powerlaw indexes of the HRC-S background models in the time-averaged, low/hard-state, and high/soft-state spectra are similar to each other. 
We then fixed the broken powerlaw model of the background spectrum, and we fitted the source spectrum for each data set.  
As seen in Figures~\ref{fig:rtcru:hard} and \ref{fig:rtcru:hard2}, there are hardness--brightness anticorrelations in both  the HRC-S and ACIS-S observations due to some brightness changes on hour timescales mostly in the soft excess below 4 keV. 
To exclude any effect of the soft band variations on our fitting, we first constructed a phenomenological model of the hard band (4--8\,keV) in \S\,\ref{rtcru:analysis:spectral:hard}, and then extended our fitting model to the soft band (0.3--4\,keV) in \S\,\ref{rtcru:analysis:spectral:soft}. 

\begin{figure}
\begin{center}
\includegraphics[width=2.6in, trim = 20 10 20 20, clip, angle=270]{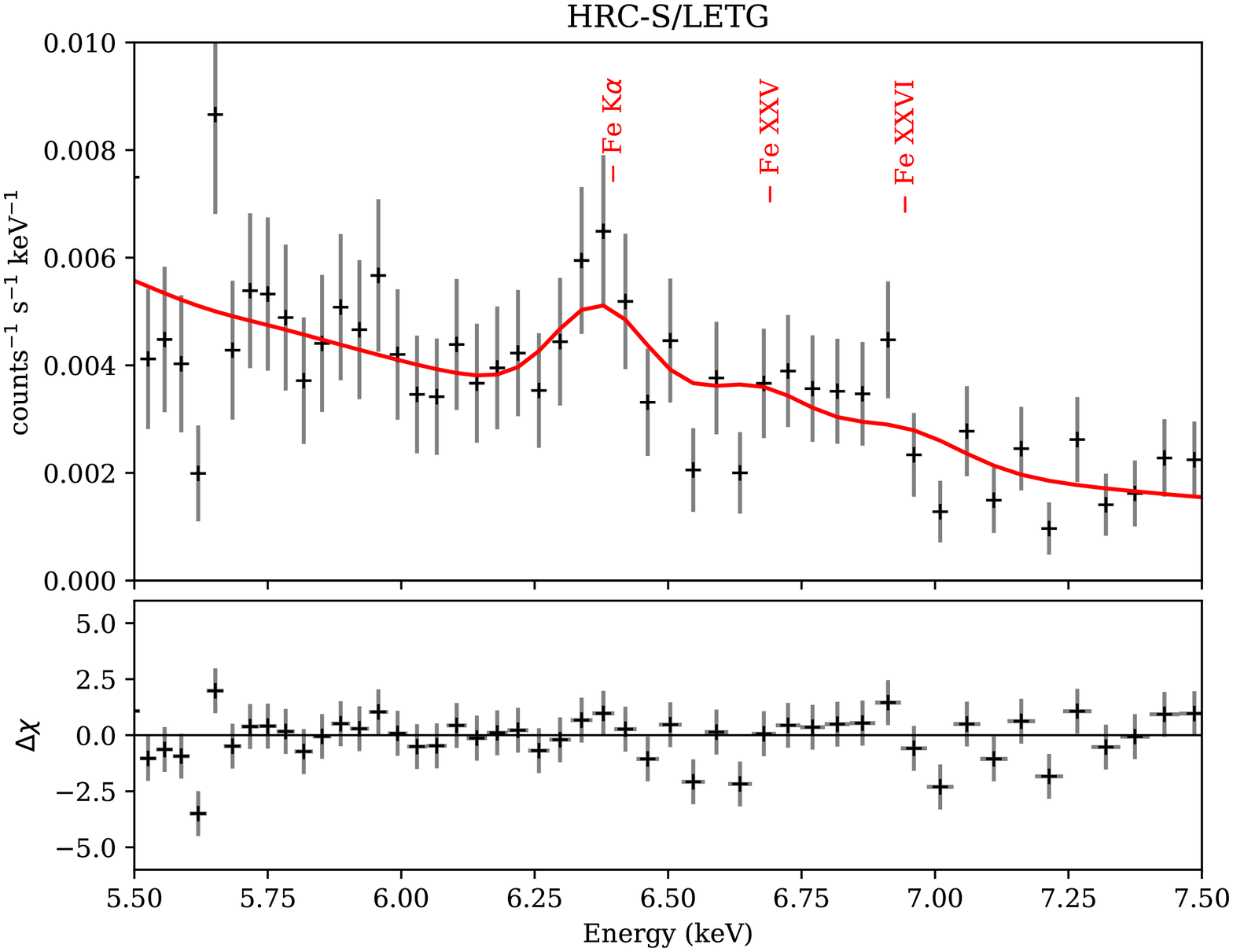}\\
\includegraphics[width=2.6in, trim = 20 10 20 20, clip, angle=270]{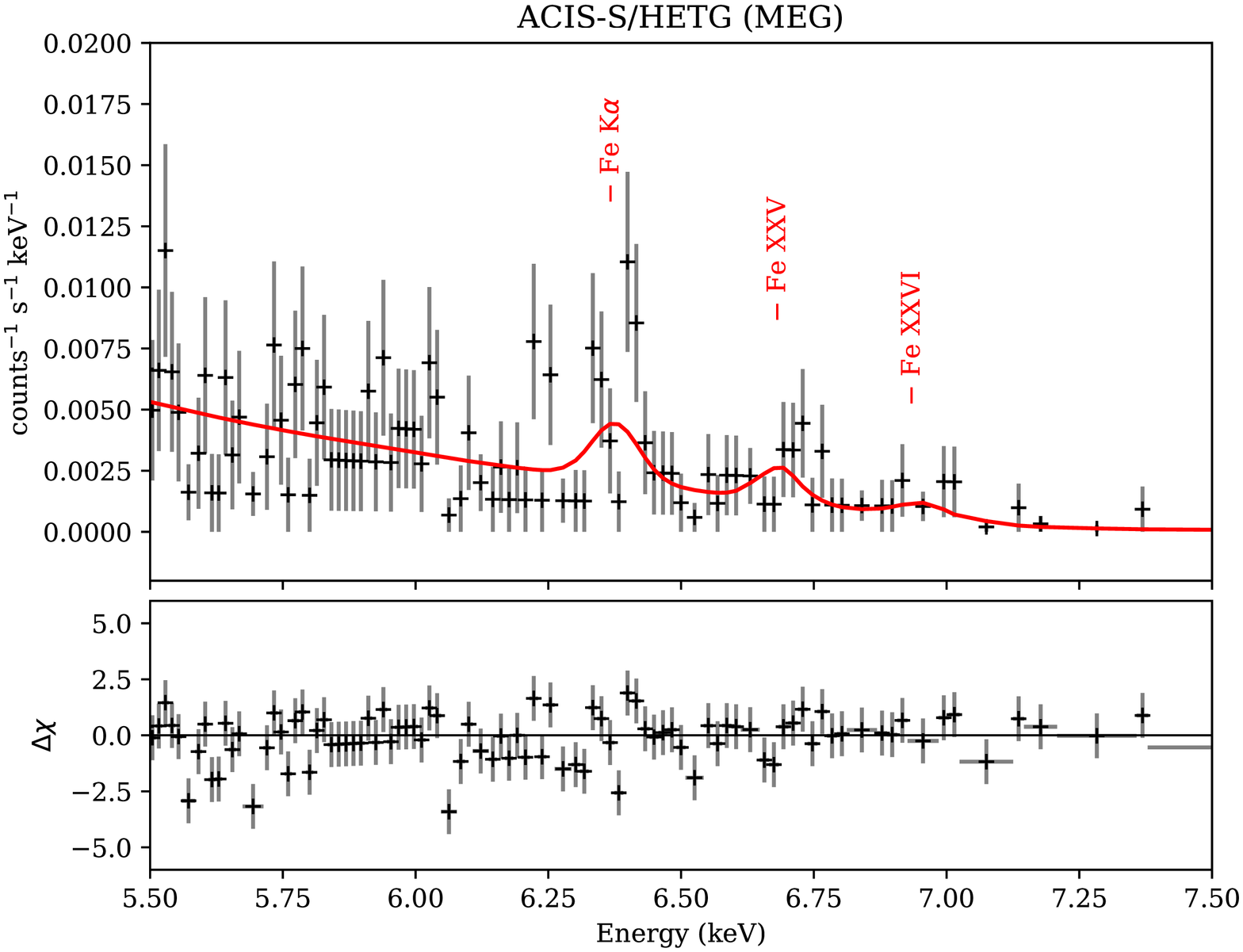}\\
\includegraphics[width=2.6in, trim = 20 10 20 20, clip, angle=270]{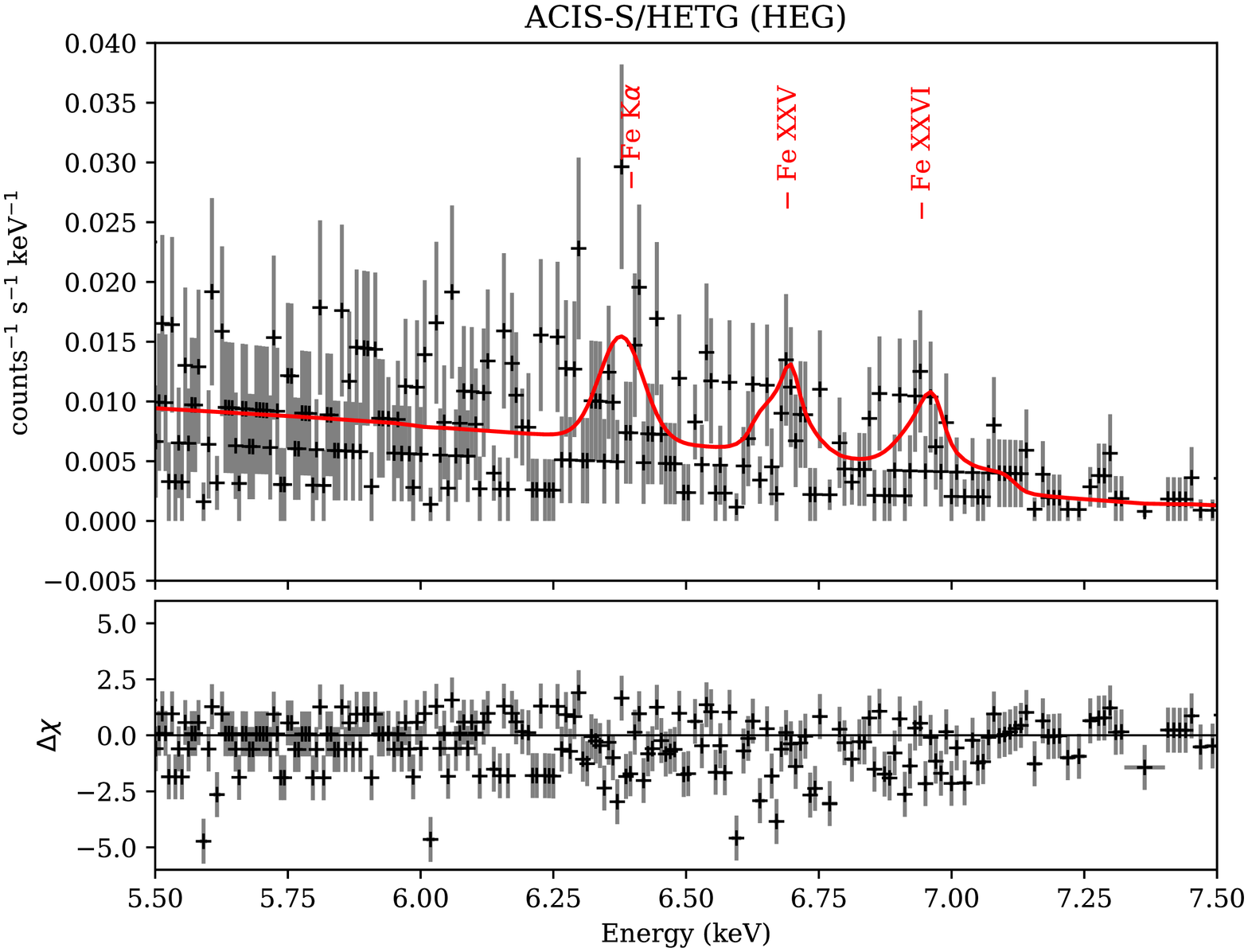}
\caption{The Fe\,K$\alpha$, \fexxv, and \fexxvi\ lines in the time-averaged HRC-S/LETG spectrum (top panel) and ACIS-S/HETG MEG and HEG-grating spectra (bottom panels), modeled
by the Gaussian functions (\textsc{xspec} model \textsf{gauss}) added to the \textsf{powerlaw} model with the parameters listed in Tables~\ref{rtcru:mekal:parameters} and \ref{rtcru:mekal:parameters2}, respectively. 
The ACIS-S MEG- and HEG-grating spectra were simultaneously fitted. 
The spectra are shown without any grouping scheme.
\label{fig:rtcru:ironlines}%
}
\end{center}
\end{figure}

We conducted the ``Bayesian Low-Count X-ray Spectral Analysis in Python''\footnote{\url{https://hea-www.harvard.edu/AstroStat/pyBLoCXS/}}  \citep[pyBLoCXS;][]{vanDyk2001,Protassov2002,Lee2011,Xu2014} that yielded the best-fitting parameters with their corresponding uncertainties using an MCMC-based algorithm. The sampler and random walker method chosen in the pyBLoCXS analysis was the Metropolis and Metropolis-Hastings (``MetropolisMH'') algorithm \citep{Metropolis1953,Hastings1970}. 
For our MCMC fitting, we utilized the ``Cash'' Poisson logarithm-likelihood function \citep{Cash1979}, together with 
the Nelder-Mead Simplex optimization method \citep{Lagarias1998,Wright1996}. 
We ran 20,000 iterations to produce the MCMC chain in each spectral modeling case. 
The posterior probability distributions of the best-fitted parameters were extracted from the MCMC chain, which were also plotted in multi-dimensional projected 2-D histograms together with probability density function (PDF) plots
using the Python module \textsf{corner} \citep{Foreman-Mackey2016}.\footnote{The module \textsf{corner} uses the Python packages \textsf{matplotlib} \citep{Hunter2007} and \textsf{numpy} \citep{vanderWalt2011,Harris2020}.}

\begin{figure*}
\begin{center}
\includegraphics[width=4.60in, trim = 0 0 0 0, clip, angle=270]{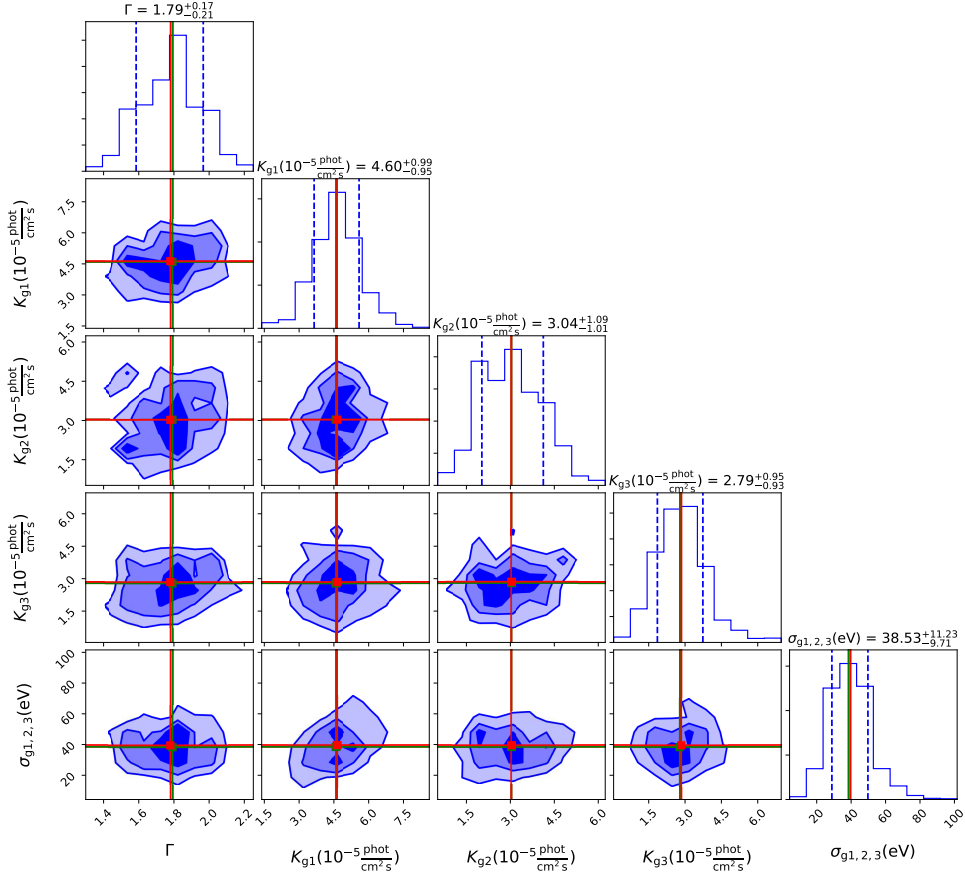}%
\caption{The posterior probability distributions of the model parameters fitted to the ACIS-S/HETG time-averaged spectrum ($m=\pm1$ orders; MEG and HEG) over the hard excess (4--8\,keV). Contours are shown at the 1-$\sigma$ (68\%), 
2-$\sigma$ (95\%), and 3-$\sigma$ (99.7\%) confidence levels. 
The model parameters are as follows: the \textsf{powerlaw} index ($\Gamma$), the normalization factor ($K_{\rm g1}$, $K_{\rm g2}$, and $K_{\rm g3}$) of the Gaussian components (\textsc{xspec} model \textsf{gauss}) associated with the Fe K$\alpha$, Fe\,{\sc xxv}, and Fe\,{\sc xxvi} lines, respectively, and 
the line width ($\sigma_{\rm g1,2,3}$) of the Gaussian components. 
The best-fitting parameters located at the true means drawn by the solid red lines are listed in Table~\ref{rtcru:mekal:parameters2}.
The 1-$\sigma$ (68\%) confidence levels are plotted by the dashed blue lines in the PDF plots, and the empirical means shown by the solid green lines. 
\label{fig:rtcru:mcmc1}%
}
\end{center}
\end{figure*}

\subsubsection{Hard Excess and Iron Lines above 4\,keV}
\label{rtcru:analysis:spectral:hard}

To model the hard excess, we considered the combined first-order ($m = \pm 1$), time-averaged spectrum in the 4--8\,keV energy region. 
To model the continuum, we initially used a fully absorbed powerlaw component, implemented by using the \textsc{xspec} model \textsf{phabs} $\times$ \textsf{powerlaw} that describes X-ray hard non-thermal emission of the accretion disk (\textsf{powerlaw}) and local absorption from possible moving dense material along the line of sight partially covering the source (\textsf{pcfabs}). As the soft excess is highly variable, we matched the absorbed powerlaw model to the spectrum above 4\,keV. Three Gaussian components (\textsc{xspec} model \textsf{gauss}) were also included at 6.379, 6.693, and 6.946\,keV (at the rest frame, they are not redshifted or blueshifted), which could be associated with the Fe K$\alpha$, Fe\,{\sc xxv}, and Fe\,{\sc xxvi} lines, respectively. Figure~\ref{fig:rtcru:ironlines} shows the iron lines in the time-averaged spectra of HRC-S and ACIS-S (MEG and HEG gratings). It can be seen that the Fe K$\alpha$ are stronger than the Fe\,{\sc xxv}, and Fe\,{\sc xxvi} lines.
As the Fe\,{\sc xxv} and Fe\,{\sc xxvi} lines in both the HRC-S and ACIS-S spectra had insufficient count rates for constraining their upper and lower uncertainties using our MCMC-based method, we linked their line widths to the line width ($\sigma$ in Tables~\ref{rtcru:mekal:parameters} and \ref{rtcru:mekal:parameters2}) of the Fe K$\alpha$ line with higher count rates in our spectral fitting analysis (using the \textsf{link} function in \textsf{Sherpa}). 
However, the equivalent widths of the Fe\,{\sc xxv} and Fe\,{\sc xxvi} lines were reported to be lower than that of the Fe K$\alpha$ line \citep{Eze2014}. 
To achieve a better fit, we then used a powerlaw component partially covered by an absorption model, so the phenomenological model of the hard excess (4--8\,keV) with iron lines is \textsf{pcfabs}  $\times$ \textsf{powerlaw} $+\sum$\,\textsf{gauss}. 

Using the Levenberg--Marquardt optimization method and cstat maximum-likelihood function (without pyBLoCXS), the fully absorbed powerlaw modeling (\textsf{phabs} $\times$ \textsf{powerlaw}) of the HRC-S/LETG time-averaged spectrum yielded a photon index of $\Gamma = 1.84^{+0.06}_{-0.06}$. Similarly, for the combined MEG and HEG fitting of the ACIS-S first-order ($m=\pm 1$) data we found $1.27^{+0.45}_{-0.42}$, which roughly agrees with $\Gamma = 1.05^{+0.24}_{-0.27}$ found by \citet{Luna2007}.  
For the absorption component, we also obtained absorbing column densities $N_{\rm H}= 15 \pm{6}\times 10^{22}$\,cm$^{-2}$ for the HRC-S data and $N_{\rm H}= 22 \pm 7 \times 10^{22}$\,cm$^{-2}$ for ACIS-S data, which are higher than $N_{\rm H}= 7 \pm 1 \times 10^{22}$\,cm$^{-2}$ found by \citet{Luna2007}. 
Using the Nelder--Mead Simplex optimization method and Cash logarithm-likelihood function (with pyBLoCXS), 
the best-fitting model of the HRC-S time-averaged data with a partially covered absorbed powerlaw component (\textsf{pcfabs} $\times$ \textsf{powerlaw}) has a powerlaw index $\Gamma = 1.70^{+ 0.21}_{-0.39}$, 
while for the ACIS-S data $\Gamma = 1.79^{+ 0.17}_{-0.21}$. 
The best-fitting photon indexes of the partially covered absorbed powerlaw model and the best-fitting parameters of the three Gaussian components (\textsf{pcfabs}$\times$\textsf{powerlaw}$+\sum$\,\textsf{gauss}) obtained for the HRC-S and ACIS-S data are presented in Tables \ref{rtcru:mekal:parameters} and \ref{rtcru:mekal:parameters2}, respectively. These parameters were determined by restricting the spectral fitting analysis to the hard excess (4--8 keV).

Figure~\ref{fig:rtcru:mcmc1} shows the posterior probability distributions of the parameters plotted by the module \textsf{corner} for the partially absorbed powerlaw model with three Gaussian components 
simultaneously fitted to the MEG and HEG first-orders ($m=\pm 1$) of the ACIS-S/HETG time-averaged data over the hard excess (4--8\,keV). 
The multi-dimensional probability distributions of the parameters for the HRC-S/LETG time-averaged data are presented in Figure~A1 in Appendix~A (see Supplementary Material). These probability distributions of the model parameters were obtained from the MCMC chains built by pyBLoCXS in \textsf{Sherpa}. 
It can be seen that the powerlaw index and the parameters of the Gaussian components were properly fitted in both the ACIS-S and HRC-S observations, while we obtained a much better fit for the ACIS-S data. The 68\% confidence uncertainties of the model parameters fitted to the HRC-S data  are much higher than those derived for the ACIS-S data. 

\begin{table*}
\begin{center}
\caption{Partially covered absorbed powerlaw and thermal modeling of the HRC-S/LETG data.  
}\label{rtcru:mekal:parameters}
\begin{tabular}{llC{3.5cm}ccc}
\hline\hline
\noalign{\smallskip}
& {Component} & {Parameter} & {Time-Averaged}  & {Low/Hard-State}   & {High/Soft-State}  \\
\noalign{\smallskip}
\hline 
\noalign{\smallskip}
&  & \multicolumn{4}{c}{\textit{Source}: \textsf{phabs}(\hi$)^{\rm \bf a}\times$(\textsf{pcfabs}$\times$\textsf{mekal}$+$\textsf{pcfabs}$\times$\textsf{powerlaw}$+\sum$\,\textsf{gauss})} \\
\noalign{\smallskip}
\hline 
\noalign{\smallskip}
&  & ${\rm cstat}/{\rm d.o.f}$ \dotfill &  $ 1.03$ &  $1.09$ & $1.08$ \\ 
\noalign{\bigskip}  
\multirow{4}{*}{\rotatebox{90}{Soft Excess}}  & \textsf{phabs}(\hi)\,$^{\rm \bf a}$ & $N_{\rm H}$(cm$^{-2}$ ) \dotfill & $     1.8 \times 10^{21}$ & $     1.8 \times 10^{21}$  &  $     1.8 \times 10^{21}$ \\ 
\noalign{\medskip} 
& \textsf{phabs} & $N_{\rm H,mk}$(cm$^{-2}$) \dotfill & $     75.88^{+17.92}_{-21.64}\times 10^{22}$ &  $     63.08^{+17.81}_{-11.40}\times 10^{22}$ & $     46.65^{+12.63}_{-21.88}\times 10^{22}$  \\ 
\noalign{\medskip}
& \textsf{mekal}  & $kT$(keV) \dotfill  & $     1.28^{+0.93}_{-0.26}$   &  $     1.53^{+0.14}_{-0.12}$&  $     0.57^{+0.34}_{-0.39}$ \\ 
\noalign{\smallskip}
&                  & $A_{\rm Fe} (A_{\rm Fe_{\odot}})$ \dotfill  & $0.15^{+0.40}_{-0.10}$   &  $0.15$ & $0.15$  \\ 
\noalign{\smallskip}  
&                  & $K_{\rm mk}$($10^{-14}$\,cm$^{-5}$)\,$^{\rm \bf b}$  \dotfill  &  $0.86^{+1.32}_{-0.72}$   &  $0.86$ &  $0.86$ \\   
\noalign{\medskip}
& \textsf{apec}\,$^{\rm \bf c}$    & $kT$(keV) \dotfill  & $     1.13^{+0.41}_{-0.23}$   &  $     1.26^{+0.05}_{-0.05}$ &  $     0.68^{+0.13}_{-0.33}$ \\ 
\noalign{\smallskip}  
&                  & $K_{\rm ap}$($10^{-14}$\,cm$^{-5}$)\,$^{\rm \bf b}$  \dotfill  &  $1.57^{+2.95}_{-1.08}$   & $1.57$  &  $1.57$ \\   
\noalign{\bigskip}               
\multirow{4}{*}{\rotatebox{90}{Hard Excess}} & \textsf{pcfabs} & $N_{\rm H,p}$(cm$^{-2}$ ) \dotfill & $     4.47^{+0.70}_{-0.51} \times 10^{22}$ & $      5.42^{+1.03}_{-1.04} \times 10^{22}$  & $      7.35^{+1.74}_{-1.44} \times 10^{22}$  \\ 
\noalign{\smallskip}
&                 & $f_{\rm p}$ \dotfill & $      0.83^{+0.02}_{-0.02} $ &  $      0.93^{+0.01}_{-0.02} $  &   $      0.78^{+0.03}_{-0.04} $ \\ 
\noalign{\medskip} 
& \textsf{powerlaw} & $\Gamma$\,$^{\rm \bf d}$ \dotfill  & $     1.70^{+0.21}_{-0.39}$   & $    2.40^{+0.23}_{-0.52}$  &   $     1.69^{+0.10}_{-0.08}$ \\ 
\noalign{\smallskip}
&                   & $K_{\rm p}$($\frac{\mathrm{photons}}{\mathrm{keV\,cm^{2}\,s}}$) \dotfill & $    5.56^{+0.77}_{-0.56} \times 10^{-3}$    &  $    13.65^{+3.71}_{-3.21} \times 10^{-3}$ &  $    9.61^{+1.29}_{-1.42} \times 10^{-3}$ \\ 
\noalign{\bigskip}  
 & \textsf{gauss}(g1) & $E$(keV) \dotfill & $     6.379$   & $     6.379$  & $     6.379$  \\ 
\noalign{\smallskip}
\multirow{4}{*}{\rotatebox{90}{Iron Lines}} & (Fe\,K$\alpha$)   & $K_{\rm g1}$($\frac{\mathrm{photons}}{\mathrm{cm^{2}\,s}}$) \dotfill  & $     9.95^{+4.23}_{-2.72}\times10^{-5}$ & $     12.18^{+13.35}_{-8.87}\times10^{-5}$  & $     21.38^{+15.55}_{-9.58}\times10^{-5}$  \\ 
\noalign{\medskip}       
& \textsf{gauss}(g2) & $E$(keV) \dotfill & $     6.693$   & $     6.693$  &  $     6.693$ \\ 
\noalign{\smallskip}
& (\fexxv)         & $K_{\rm g2}$($\frac{\mathrm{photons}}{\mathrm{cm^{2}\,s}}$) \dotfill  & $      4.82^{+3.77}_{-2.42}\times10^{-5}$ & $      12.24^{+17.09}_{-7.69}\times10^{-5}$  & $      13.28^{+10.11}_{-10.24}\times10^{-5}$  \\ 
\noalign{\medskip}  
& \textsf{gauss}(g3) & $E$(keV) \dotfill & $     6.946$   &  $     6.946$ &  $     6.946$ \\ 
\noalign{\smallskip}
& (\fexxvi)        & $K_{\rm g3}$($\frac{\mathrm{photons}}{\mathrm{cm^{2}\,s}}$) \dotfill  & $     5.70^{+3.58}_{-3.27}\times10^{-5}$ &  $      22.46^{+22.31}_{-15.04}\times10^{-5}$ &  $      13.28^{+10.11}_{-10.24}\times10^{-5}$ \\ 
\noalign{\smallskip}
& {g1--3}\,$^{\rm \bf e}$         & $\sigma_{\rm g1,2,3}$(eV) \dotfill & $     38.08^{+103.49}_{-24.99}$   &  $     472.99^{+119.48}_{-224.38}$&  $     169.98^{+182.81}_{-100.17}$ \\  
\noalign{\bigskip}  
&     & $F_{\mbox{\scriptsize 0.3-8}}$($\mathrm{ergs}\,\mathrm{s}^{-1}\,\mathrm{cm}^{-2}$)\,$^{\rm \bf f}$  \dotfill &  $     23.2^{+0.5}_{-0.4}\times 10^{-12}$  & $     29.0^{+0.2}_{-0.2}\times 10^{-12}$ & $     32.9^{+0.4}_{-0.4}\times 10^{-12}$  \\  
\noalign{\smallskip}
&     & $L_{\mbox{\scriptsize 0.3-8}}$($\mathrm{ergs}\,\mathrm{s}^{-1}$)\,$^{\rm \bf g}$
\dotfill &  $     219.3^{+1.8}_{-2.0}\times 10^{33}$  & $     247.3^{+10.8}_{-9.3}\times 10^{33}$ & $     173.0^{+3.2}_{-2.7}\times 10^{33}$  \\ 
\noalign{\smallskip}
\hline
\noalign{\smallskip}
& & \multicolumn{4}{c}{\textit{Background}: \textsf{bknpower}} \\
\noalign{\smallskip}
\hline 
\noalign{\smallskip}
&  & ${\rm cstat}/{\rm d.o.f}$ \dotfill &  $ 1.03$ &  $1.01$ &  $1.01$ \\ 
\noalign{\medskip} 
& \textsf{\textsf{bknpower}} & $\Gamma_1$ \dotfill  & $    1.88^{+0.01}_{-0.05}$   & $    1.87^{+0.02}_{-0.05}$  &    $    1.88^{+0.03}_{-0.01}$ \\ 
\noalign{\smallskip}
&  & $E_{\rm break}$(keV) \dotfill  & $    3.10^{+1.45}_{-2.02}$   & $    2.77^{+1.44}_{-1.75}$   &  $    2.89^{+1.22}_{-2.89}$  \\ 
\noalign{\smallskip}
&  & $\Gamma_2$ \dotfill  & $    2.01^{+0.21}_{-0.10}$   &  $    2.01^{+0.17}_{-0.10}$  &  $    1.98^{+0.08}_{-0.07}$ \\ 
\noalign{\smallskip}
&                   & $K$($\frac{\mathrm{photons}}{\mathrm{keV\,cm^{2}\,s}}$) \dotfill & $    69.51^{+2.93}_{-0.05} \times 10^{-2}$    &  $    69.80^{+2.87}_{-0.10} \times 10^{-2}$  &  $    68.64^{+0.53}_{-68.64} \times 10^{-2}$ \\ 
\noalign{\smallskip}
\hline
\noalign{\smallskip}
\end{tabular}
\end{center}
\begin{tablenotes}
\item[1]\textbf{Notes.} The uncertainties of the fitted parameters are at 68\% confidence level. The fully absorbed thermal plasma model was fitted to the low/hard- and high/soft-state spectra assuming
the metal abundance $A_{\rm Fe}$ and normalization factors $K_{\rm mk}$ and $K_{\rm ap}$ derived from
the time-averaged spectrum. 
$^{\rm \bf a}$~The Galactic \hi\ foreground absorption.
$^{\rm \bf b}$~Normalization unit in $(10^{-14}/4\pi D^2) \int n_{\rm e} n_{\rm H} dV$, 
where $D$(cm) is the angular diameter distance, $n_{\rm e}$(cm$^{-3}$) the electron density, $n_{\rm H}$ the hydrogen density, and $V$ the volume.
$^{\rm \bf c}$~The \textsf{apec} model substituted for the \textsf{mekal} model was constrained by adopting all the parameters determined from the absorbed thermal plasma model plus the partially absorbed \textsf{powerlaw} and \textsf{gauss} models, and assuming 
the same metal abundance $A_{\rm Fe}$ found by the \textsf{mekal} model. The model made with \textsf{apec} yields the same ${\rm cstat}/{\rm d.o.f}$.
$^{\rm \bf d}$~The powerlaw index $\Gamma$ is associated with the phenomenological model that describes the continuum 
over the 4--8\,keV energy range. 
$^{\rm \bf e}$~The Fe\,{\sc xxv} and Fe\,{\sc xxvi} line widths were linked to the Fe K$\alpha$ line width  (having higher count rates).
$^{\rm \bf f}$~The absorbed X-ray flux over the 0.3--8 keV energy band. 
$^{\rm \bf g}$~The unabsorbed X-ray luminosity over the 0.3--8 keV energy band ($D=1.64$\,kpc).
\end{tablenotes}
\end{table*}

\begin{table*}
\begin{center}
\caption{The same as Table~\ref{rtcru:mekal:parameters}, but for the ACIS-S/HETG data. 
}\label{rtcru:mekal:parameters2}
\begin{tabular}{llC{3.5cm}ccc}
\hline\hline
\noalign{\smallskip}
& {Component} & {Parameter} & {Time-Averaged}  & {Low/Hard-State}   & {High/Soft-State}  \\
\noalign{\smallskip}
\hline 
\noalign{\smallskip}
&  & \multicolumn{4}{c}{\textit{Source}: \textsf{phabs}(\hi$)\times$(\textsf{phabs}$\times$\textsf{mekal}$+$\textsf{pcfabs}$\times$\textsf{powerlaw}$+\sum$\,\textsf{gauss})} \\
\noalign{\smallskip}
\hline 
\noalign{\smallskip}
&  & ${\rm cstat}/{\rm d.o.f}$ \dotfill &  $ 0.31$ &  $0.14$ & $0.15$ \\ 
\noalign{\bigskip}  
\multirow{4}{*}{\rotatebox{90}{Soft Excess}}  & \textsf{phabs}(\hi) & $N_{\rm H}$(cm$^{-2}$ ) \dotfill & $     1.8 \times 10^{21}$ & $     1.8 \times 10^{21}$  &  $     1.8 \times 10^{21}$ \\ 
\noalign{\medskip} 
& \textsf{phabs} & $N_{\rm H,mk}$(cm$^{-2}$) \dotfill & $     12.44^{+0.71}_{-0.84}\times 10^{22}$ & $     7.56^{+1.38}_{-1.09}\times 10^{22}$ & $     12.33^{+1.67}_{-1.83}\times 10^{22}$ \\ 
\noalign{\medskip}
& \textsf{mekal}  & $kT$(keV) \dotfill  & $     9.64^{+2.69}_{-3.96}$   &  $     45.86^{+12.74}_{-17.48}$ &  $     4.37^{+2.25}_{-2.38}$ \\ 
\noalign{\smallskip}
&                  & $A_{\rm Fe} (A_{\rm Fe_{\odot}})$ \dotfill  & $0.31^{+1.96}_{-0.31}$    &  $0.31$ & $0.31$  \\ 
\noalign{\smallskip} 
&                  & $K_{\rm mk}$($10^{-14}$\,cm$^{-5}$) \dotfill  &  $1.62^{+0.13}_{-0.17}\times 10^{-2}$  & $0.59^{+0.21}_{-0.16}\times 10^{-2}$ &  $2.06^{+1.28}_{-0.45}\times 10^{-2}$ \\    
\noalign{\medskip}
& \textsf{apec}    & $kT$(keV) \dotfill  & $     10.25^{+5.11}_{-1.35}$   &  $     50.94^{+\cdots}_{-41.70}$ &  $     3.81^{+1.36}_{-0.78}$ \\ 
\noalign{\smallskip}  
&                  & $K_{\rm ap}$($10^{-14}$\,cm$^{-5}$)  \dotfill  &  $1.59^{+0.05}_{-0.06}\times 10^{-2}$   &  $0.61^{+0.09}_{-0.14}\times 10^{-2}$ & $2.32^{+0.46}_{-0.39}\times 10^{-2}$  \\   
\noalign{\bigskip}               
\multirow{4}{*}{\rotatebox{90}{Hard Excess}} & \textsf{pcfabs} & $N_{\rm H,p}$(cm$^{-2}$ ) \dotfill & $     105.27^{+46.08}_{-46.47} \times 10^{22}$ & $      30.77^{+5.22}_{-4.96} \times 10^{22}$  & $      43.28^{+12.52}_{-19.62} \times 10^{22}$  \\ 
\noalign{\smallskip}
&                 & $f_{\rm p}$ \dotfill & $      1.00$ &  $      1.00$  &   $      0.98^{+0.01}_{-0.01} $ \\ 
\noalign{\medskip} 
& \textsf{powerlaw} & $\Gamma$ \dotfill  & $     1.79^{+0.17}_{-0.21}$   & $    2.04^{+0.23}_{-0.24}$  &   $     0.92^{+0.39}_{-0.15}$ \\ 
\noalign{\smallskip}
&                   & $K_{\rm p}$($\frac{\mathrm{photons}}{\mathrm{keV\,cm^{2}\,s}}$) \dotfill & $    15.68^{+11.79}_{-5.68} \times 10^{-3}$    &  $    10.20^{+2.50}_{-2.63} \times 10^{-3}$ &  $    1.88^{+0.89}_{-0.50} \times 10^{-3}$ \\ 
\noalign{\bigskip}
 & \textsf{gauss}(g1) & $E$(keV) \dotfill & $     6.379$   & $     6.379$  & $     6.379$  \\ 
\noalign{\smallskip}
\multirow{4}{*}{\rotatebox{90}{Iron Lines}} & (Fe\,K$\alpha$)   & $K_{\rm g1}$($\frac{\mathrm{photons}}{\mathrm{cm^{2}\,s}}$) \dotfill  & $     4.60^{+0.99}_{-0.95}\times10^{-5}$ & $     10.46^{+3.42}_{-2.61}\times10^{-5}$  & $     5.30^{+2.56}_{-2.14}\times10^{-5}$  \\ 
\noalign{\medskip}       
& \textsf{gauss}(g2) & $E$(keV) \dotfill & $     6.693$   & $     6.693$  &  $     6.693$ \\ 
\noalign{\smallskip}
& (\fexxv)         & $K_{\rm g2}$($\frac{\mathrm{photons}}{\mathrm{cm^{2}\,s}}$) \dotfill  & $      3.04^{+1.09}_{-1.01}\times10^{-5}$ & $      2.46^{+3.01}_{-1.56}\times10^{-5}$  & $      6.91^{+3.58}_{-4.35}\times10^{-5}$  \\ 
\noalign{\medskip}  
& \textsf{gauss}(g3) & $E$(keV) \dotfill & $     6.946$   &  $     6.946$ &  $     6.946$ \\ 
\noalign{\smallskip}
& (\fexxvi)        & $K_{\rm g3}$($\frac{\mathrm{photons}}{\mathrm{cm^{2}\,s}}$) \dotfill  & $     2.79^{+0.95}_{-0.93}\times10^{-5}$ &  $      1.43^{+2.74}_{-1.04}\times10^{-5}$ &  $      5.52^{+4.26}_{-3.18}\times10^{-5}$ \\ 
\noalign{\smallskip}
& {g1--3}         & $\sigma_{\rm g1,2,3}$(eV) \dotfill & $     38.53^{+11.23}_{-9.71}$   &  $     71.65^{+48.32}_{-20.67}$&  $     174.73^{+14.85}_{-103.58}$ \\ 
\noalign{\bigskip}  
&     & $F_{\mbox{\scriptsize 0.3-8}}$($\mathrm{ergs}\,\mathrm{s}^{-1}\,\mathrm{cm}^{-2}$) \dotfill &  $     14.5^{+0.4}_{-0.4}\times 10^{-12}$  & $     12.3^{+0.6}_{-1.1}\times 10^{-12}$  & $     16.4^{+0.9}_{-1.0}\times 10^{-12}$  \\ 
\noalign{\smallskip}
&     & $L_{\mbox{\scriptsize 0.3-8}}$($\mathrm{ergs}\,\mathrm{s}^{-1}$)
\dotfill &  $    38.5^{+1.0}_{-1.4}\times 10^{33}$  & $     20.4^{+1.3}_{-1.1}\times 10^{33}$ & $     18.0^{+1.1}_{-1.1}\times 10^{33}$  \\ 
\noalign{\smallskip}
\hline
\noalign{\smallskip}
& & \multicolumn{4}{c}{\textit{Background}: \textsf{bknpower}} \\
\noalign{\smallskip}
\hline 
\noalign{\smallskip}
&  & ${\rm cstat}/{\rm d.o.f}$ \dotfill &  $ 0.35$ &  $0.15$ &  $0.15$ \\ 
\noalign{\medskip} 
& \textsf{\textsf{bknpower}} & $\Gamma_1$ \dotfill  & $    6.00^{+1.22}_{-1.05}$   & $    6.25^{+0.87}_{-0.89}$  &    $    5.74^{-1.54}_{-2.05}$ \\ 
\noalign{\smallskip}
&  & $E_{\rm break}$(keV) \dotfill  & $    1.37^{+0.12}_{-0.34}$   & $    1.38^{+0.07}_{-0.27}$   &  $    1.45^{+0.20}_{-0.56}$  \\ 
\noalign{\smallskip}
&  & $\Gamma_2$ \dotfill  & $    0.71^{+1.08}_{-0.30}$   &  $    0.44^{+0.91}_{-0.25}$  &  $    0.84^{+1.72}_{-0.52}$ \\ 
\noalign{\smallskip}
&                   & $K$($\frac{\mathrm{photons}}{\mathrm{keV\,cm^{2}\,s}}$) \dotfill & $    28.54^{+14.61}_{-12.71} \times 10^{-5}$    &  $    27.30^{+11.48}_{-9.62} \times 10^{-5}$  &  $    35.50^{+27.01}_{-22.78} \times 10^{-5}$ \\ 
\noalign{\smallskip}
\hline
\noalign{\smallskip}
\end{tabular}
\end{center}
\begin{tablenotes}
\item[1]\textbf{Notes.} The uncertainties of the fitted parameters are at 68\% confidence level. The background model parameters, the \textsf{powerlaw} photon index $\Gamma$, the \textsf{gauss} parameters, and the \textsf{mekal} metal abundance $A_{\rm Fe}$ were determined from the simultaneously fitted MEG and HEG data, while the remaining parameters were derived from the MEG spectrum. 
The fully absorbed thermal plasma model was fitted to the low/hard- and high/soft-state spectra assuming
the metal abundance $A_{\rm Fe}$ derived from the time-averaged spectrum. The \textsf{apec} model parameters were obtained by fixing the parameters derived from the full model with \textsf{mekal}.
\end{tablenotes}
\end{table*}

\subsubsection{Extension to Soft Excess below 4\,keV}
\label{rtcru:analysis:spectral:soft}

To describe the emission from optically thin thermal plasma in the broad band (0.3--8\,keV), we included a fully absorbed \textsc{xspec} model \textsf{mekal} that describes a Mewe--Kaastra--Liedahl thermal plasma \citep[][]{Mewe1985,Mewe1986,Kaastra1992,Liedahl1995,Phillips1999} using ionization balances from \citet{Arnaud1985} and \citet{Arnaud1992}. 
Moreover, we reproduced the thermal emission by replacing \textsf{mekal} with the \textsf{apec} model \citep[Astrophysical Plasma Emission Database;][]{Smith2001} that calculates emission spectra using the AtomDB atomic database \citep{Smith2001,Foster2012}.
We matched the \textsc{xspec} models \textsf{phabs} $\times$ \textsf{mekal} to the 0.3--8\,keV energy range by fixing 
the \textsf{powerlaw} index $\Gamma$ and \textsf{gauss} parameters previously fitted to the hard energy band above 4\,keV, but 
constraining again the \textsf{powerlaw} normalization factor $K_{\rm p}$, as well as the \textsf{pcfabs} column density $N_{\rm H,p}$ and covering fraction $f_{\rm p}$. 
The thermal plasma modeling was implemented by computing the \textsf{mekal} model in each fitting iteration. 
The \textsf{mekal} normalization factor is defined as $K_{\rm mk}=(10^{-14}/4\pi D^2) \int n_{\rm e} n_{\rm H} dV$, which is a function of
the angular diameter distance $D$(cm), electron density $n_{\rm e}$(cm$^{-3}$), hydrogen density $n_{\rm H}$(cm$^{-3}$), and gas volume $V$.
To make the \textsf{mekal} normalization factor consistent with that in the \textsf{apec} model, we set a hydrogen density of $n_{\rm H}=1$\,cm$^{-3}$ that is used by default in \textsf{mekal}. Different values of $n_{\rm H}$ result in different $K_{\rm mk}$. We should note that the number density of a dense accretion disk in the symbiotic system could be $n_{\rm H}\sim 10^8$--$10^9$\,cm$^{-3}$, however, we could not constrain $n_{\rm H}$ with the current available data. 
The free parameters are
the plasma temperature ($kT$), metal abundance ($A_{\rm Fe}$), normalization factor ($K_{\rm mk}$), and absorbing column density ($N_{\rm H,mk}$) of the fully absorbed thermal model (\textsf{phabs}$\times$\textsf{mekal}). 

We adopted the ISM composition of \citet{Wilms2000} as the baseline abundances for all the models. We determined
a metal abundance of $A_{\rm Fe}=0.15^{+0.40}_{-0.10}$ (with respect to \citeauthor{Wilms2000}) from the initial MCMC analysis of the \textsf{mekal} model fitted to the time-averaged HRC-S/LETG spectrum.
Although this is ostensibly smaller than the value found by \citet{Luna2007} -- $A_{\rm Fe} = 0.30^{+0.15}_{-0.28}$ at 1-$\sigma$ confidence relative to a baseline composition of \citet{Anders1989} -- 
but the uncertainty range overlap. 
For the ACIS-S/HETG data, we initially fitted the fully absorbed thermal plasma model simultaneously to both the MEG and HEG gratings time-averaged spectra, which provided $A_{\rm Fe}=0.31^{+1.96}_{-0.31}$ (at 68\% confidence).
As the metal abundance has a high uncertainty, we fixed it to this value to conduct further the MCMC-based analysis.
While the HEG grating observations allowed us to measure the iron lines in the hard band, the resulting spectrum 
has fewer counts in the soft band, leading to weak constraints for the \textsf{mekal} model.
As the HEG was less efficient in the soft band (resulting in poor constraints), we excluded the HEG grating and fitted the absorbed thermal model to the MEG grating ($m=\pm 1$) with the same values of the Gaussian parameters, powerlaw index $\Gamma$, and abundance $A_{\rm Fe}$ determined from the simultaneously fitted MEG and HEG data.

\begin{figure}
\begin{center}
\includegraphics[width=2.6in, trim = 20 10 20 20, clip, angle=270]{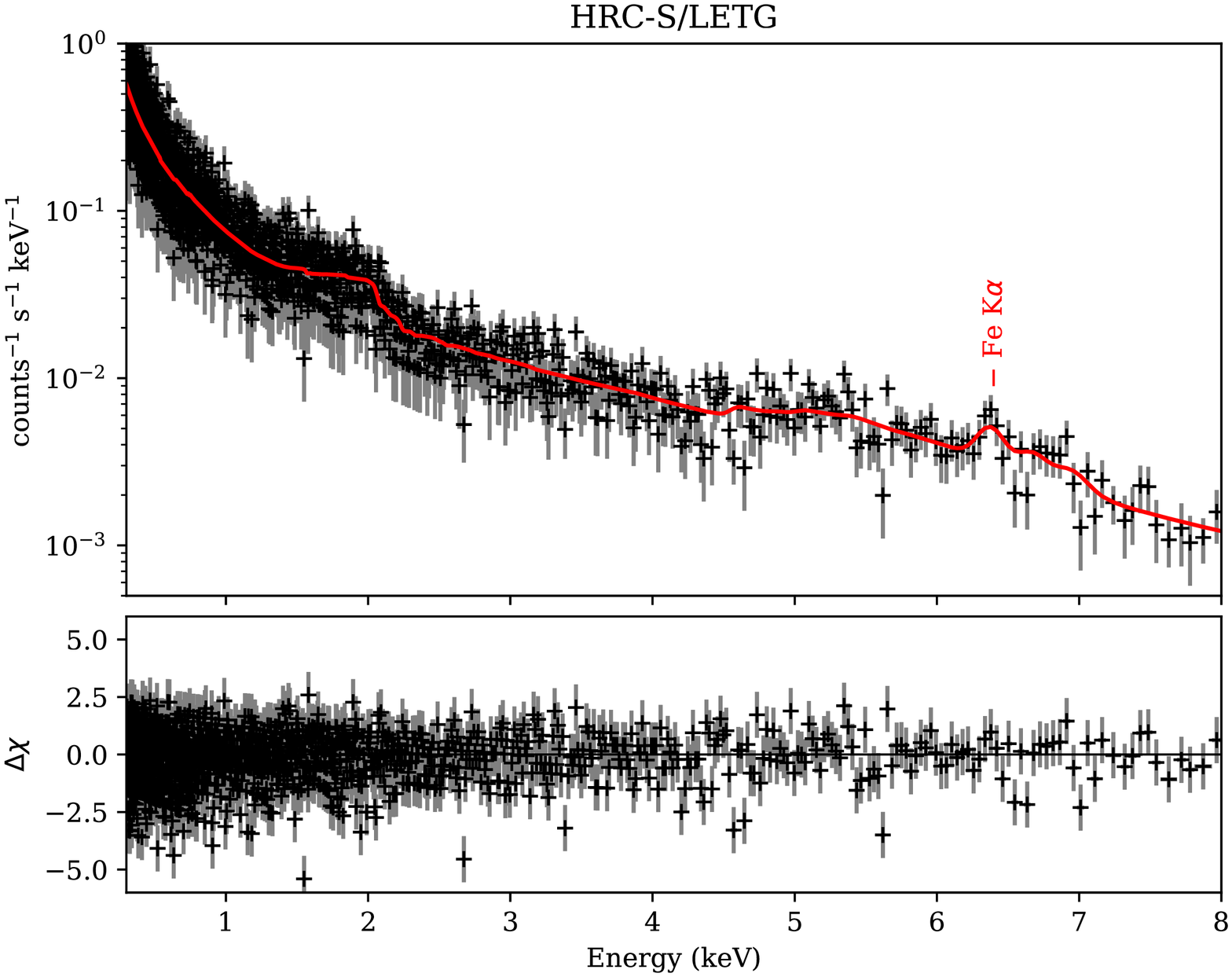}\\
\includegraphics[width=2.6in, trim = 20 10 20 20, clip, angle=270]{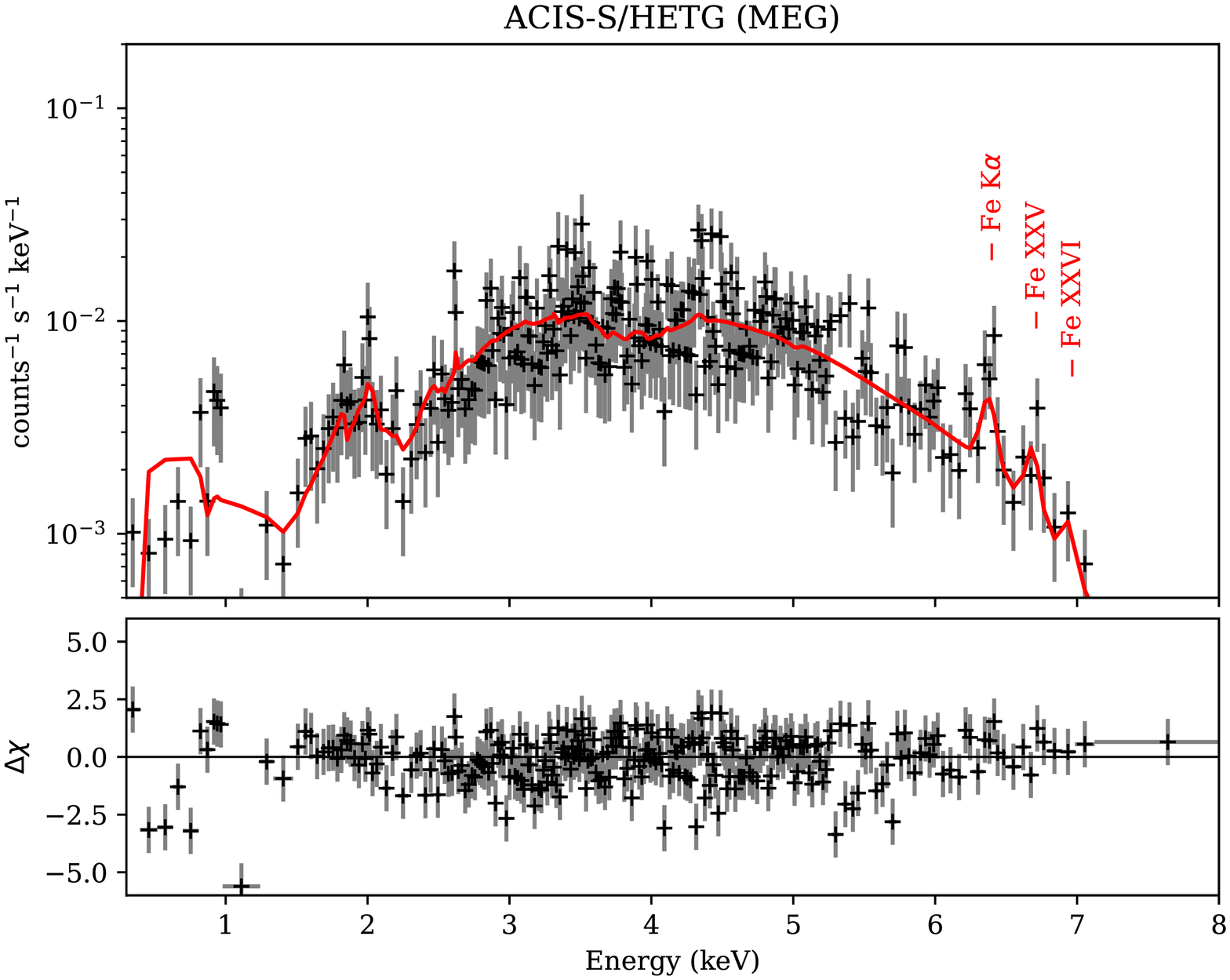}\\
\includegraphics[width=2.6in, trim = 20 10 20 20, clip, angle=270]{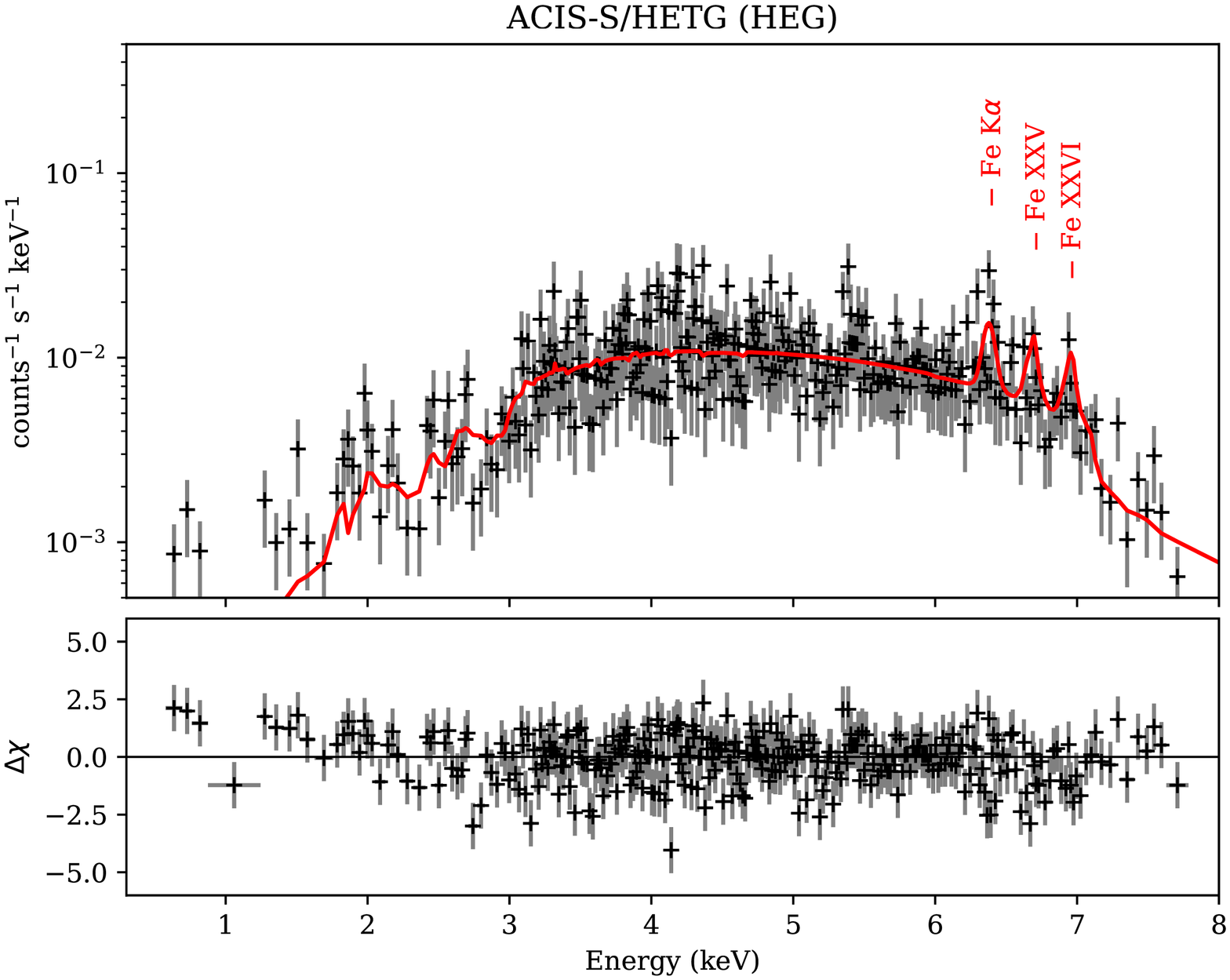}
\caption{The dispersed ($m= \pm 1$) time-averaged HRC-S/LETG spectrum (top panel) and the ACIS-S/HETG ($m= \pm 1$) MEG and HEG grating spectra (bottom panels) of RT\,Cru. 
The source model is a partially covered absorbed powerlaw plus fully absorbed thermal plasma together with Gaussian iron lines, \textsf{phabs} $\times$ (\textsf{phabs} $\times$ \textsf{mekal} $+$ \textsf{pcfabs} $\times$ \textsf{powerlaw} $+\sum$\,\textsf{gauss}). 
The best-fitting parameters for the HRC-S and ACIS-S 
are listed in Tables~\ref{rtcru:mekal:parameters} and ~\ref{rtcru:mekal:parameters2}, respectively. 
The plotted spectra were grouped into bins containing $\geq5$ counts each after the spectral fitting analysis. 
\label{fig:rtcru:fitting}%
}
\end{center}
\end{figure}

To account for the Galactic \hi\ foreground absorption, we considered the total Galactic mean column density of $N_{\rm H\,I}= 9.76\times 10^{21}$\,cm$^{-2}$ and $E(B-V)=2.052$ from the UK Swift Science Data Centre\footnote{\url{https://www.swift.ac.uk/analysis/nhtot/}} \citep{Willingale2013}. 
As this value corresponds to the total Galactic column density of hydrogen in the line of sight, we need to scale it based on the Galactic location of RT\,Cru. 
From the parallax of $0.6096 \pm 0.0582 \times 10^{-3}$ arcsec reported by the second Gaia data release
 \citep{GaiaCollaboration2016,Luri2018,GaiaCollaboration2018}, we derived a distance of $1.64^{+0.17}_{-0.14}$ kpc to RT\,Cru. \cite{Bailer-Jones2018} also estimated a distance of $1.58_{-0.14}^{+0.17}$ by correcting the parallax for the nonlinearity of the transformation. 
Assuming $D=1.64$ kpc, we adopted  $N_{\rm H\,I}= 1.8\times 10^{21}$\,cm$^{-2}$ that is also associated with the RT\,Cru interstellar extinction of $E(B-V)=0.374$ measured from optical observations \citep{Luna2018}. We note that this column density can be also calculated using the empirical correlation $N$(\hi)[cm$^{-2}] = E(B-V)[{\rm mag}] \times 4.8 \times 10^{21}$ estimated from Ly$\alpha$ absorption measurements \citep{Bohlin1978}. 
To model the Galactic foreground absorption affecting the X-ray soft band, 
we employed another \textsf{phabs} component with the fixed parameter $N_{\rm H\,I}= 1.8\times 10^{21}$\,cm$^{-2}$.  
The final phenomenological model over the broad band (0.3--8\,keV) is \textsf{phabs}(\hi$)\times$(\textsf{phabs}$\times$\textsf{mekal}$+$\textsf{pcfabs}$\times$\textsf{powerlaw}$+\sum$\,\textsf{gauss}).
We plotted the best-fitted models for the time-averaged spectra of the HRC-S and ACIS-S observations in Figure~\ref{fig:rtcru:fitting} along with $\Delta \chi \equiv ({\rm data}-{\rm model})/{\rm error}$ (residuals divided by uncertainties).  The spectra shown in this figure were grouped to obtain a minimum of 5 counts\,bin$^{-1}$, but our spectral fitting analysis of all the spectra was implemented  without any grouping scheme.

The HRC-S time-averaged spectrum was well matched to an absorbed thermal plasma model (\textsf{phabs}$\times$\textsf{mekal}) with a plasma temperature  of $kT=1.3^{+0.9}_{-0.3}$\,keV and
an absorbing column density of $N_{\rm H,mk}= 76^{+18}_{-22}\times 10^{22}$\,cm$^{-2}$. 
It can be seen that the plasma emission was heavily obstructed by the local presence of highly
dense clumps within the line of sight. 
To include the \textsf{apec} model, we fixed all fitted parameters and replaced \textsf{mekal} with \textsf{apec} having the same metal abundance derived from the \textsf{mekal} model. It was found that the \textsf{apec} model fits the HRC-S spectrum with $kT=1.1^{+0.4}_{-0.2}$\,keV that is roughly the same plasma temperature derived from \textsf{mekal}. 
The parameters of the thermal plasma model fitted to the HRC-S/LETG data are listed in Table~\ref{rtcru:mekal:parameters}.

\begin{figure*}
\begin{center}
\includegraphics[width=6.6in, trim = 0 0 0 0, clip, angle=270]{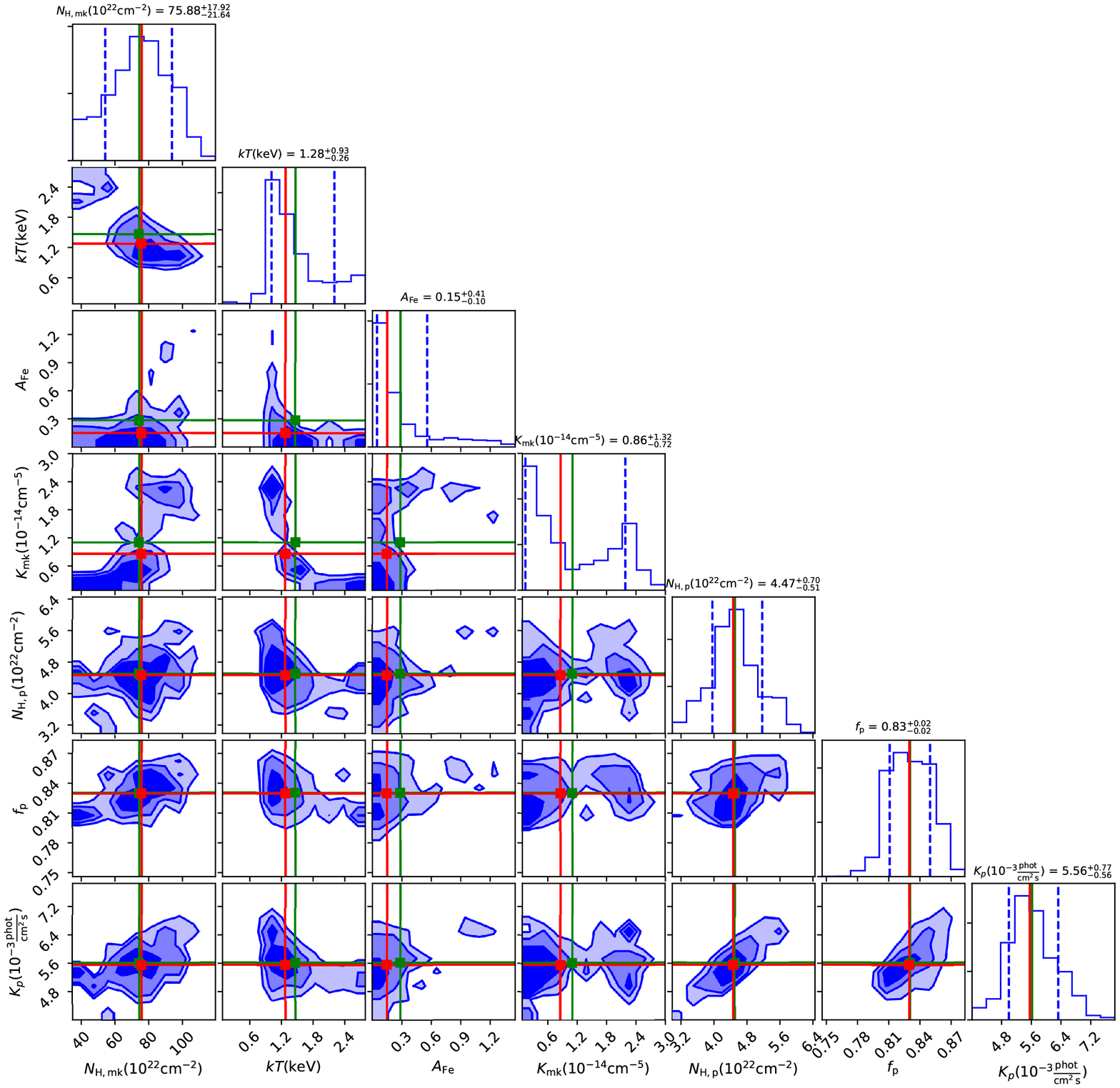}%
\caption{The posterior probability distributions of the parameters of the thermal plasma model fitted to the HRC-S/LETG time-averaged spectrum over the broad band (0.3--8\,keV). Contours are shown at the 1-$\sigma$ (68\%), 
2-$\sigma$ (95\%), and 3-$\sigma$ (99.7\%) confidence levels. 
The model parameters are as follows: the absorbing column density ($N_{\rm H,mk}$),  plasma temperature ($kT$), metal abundance ($A_{\rm Fe}$), and normalization factor ($K_{\rm mk}$) in $10^{-14}$ cm$^{-5}$) of the thermal plasma model (\textsf{phabs}$\times$\textsf{mekal}); and the column density ($N_{\rm H,p}$), covering fraction ($f_{\rm p}$), and normalization factor ($K_{\rm p}$) of the partially
absorbed powerlaw model (\textsf{pcfabs}$\times$\textsf{powerlaw}). 
The best-fitting parameters located at the true means drawn by the solid red lines are listed in Table~\ref{rtcru:mekal:parameters}.
The 1-$\sigma$ confidence levels are plotted by the dashed blue lines in the PDF plots, and the empirical means shown by the solid green lines. 
\label{fig:rtcru:mcmc7}%
}
\end{center}
\end{figure*}

Figure~\ref{fig:rtcru:mcmc7} shows the posterior probability distributions of the parameters of the absorbed thermal plasma model fitted together with the partially absorbed powerlaw model, while the powerlaw index $\Gamma$ and the Gaussian parameters were fixed at the values found in \S\,\ref{rtcru:analysis:spectral:hard}.  
As seen, there are strong constraints on the column density ($N_{\rm H,p}$), covering fraction ($f_{\rm p}$), and normalization factor ($K_{\rm p}$) of the partially absorbed powerlaw model. 
However, we notice extremely weak constraints on the metal abundance ($A_{\rm Fe}$) and normalization factor ($K_{\rm mk}$) of 
the \textsf{mekal} model. Although the plasma temperature ($kT$) is weakly constrained by the metal abundance ($A_{\rm Fe}$) and normalization factor ($K_{\rm mk}$), it is strongly restrained by the absorbing column density ($N_{\rm H,mk}$) of the thermal model, and the parameters of the partially absorbed powerlaw model.
The value of the absorbing column density $N_{\rm H,mk}= 76^{+18}_{-22}\times 10^{22}$\,cm$^{-2}$ indicates that the emission lines produced by the thermal plasma are strongly absorbed and are not easily distinguishable in the spectrum. 

For the ACIS-S time-averaged MEG spectrum, 
the best-fitting \textsf{mekal}  model has a plasma temperature  of $kT=9.6^{+2.7}_{-4.0}$\,keV, which is roughly similar to $kT=8.6^{+3.8}_{-2.5}$\,keV previously derived from the ACIS-S data by \citet{Luna2007}. 
For the fully absorbed thermal plasma model, 
we obtained an absorbing column density of $N_{\rm H,mk}= 12.4^{+0.7}_{-0.8}\times 10^{22}$\,cm$^{-2}$ in reasonable agreement with $N_{\rm H}= 9.3^{+3.2}_{-2.1}\times 10^{22}$\,cm$^{-2}$ found by \citet{Luna2007}.  
We should note that \citet{Luna2007} did not include the second \textsf{phabs} component with $N_{\rm H}= 0.18\times 10^{22}$\,cm$^{-2}$ to model the Galactic \hi\ foreground absorption. Moreover, we reprocessed the ACIS-S
data with the most recent calibration data and the updated reduction package \textsc{ciao}, and we still obtained 
similar results. The posterior probability distributions of the fitted parameters are shown in Figure~A7.
Table~\ref{rtcru:mekal:parameters2} presents the best-fitting model parameters for the ACIS-S/HETG data.

\begin{figure*}
\begin{center}
\includegraphics[width=5.in, trim = 0 0 0 0, clip, angle=270]{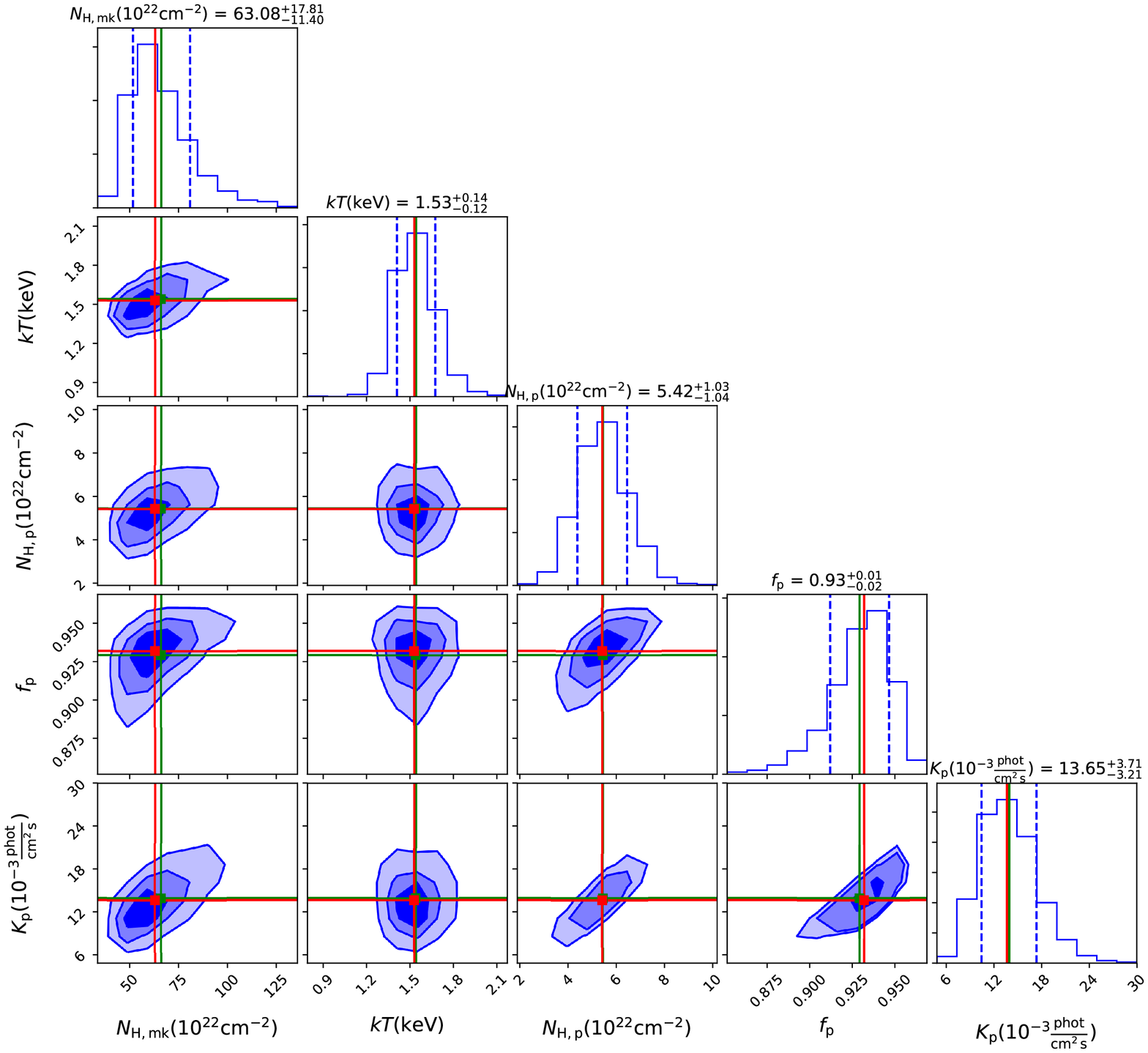}%
\caption{The same as Figure~\ref{fig:rtcru:mcmc7}, but for the HRC-S/LETG \textit{low/hard}-state spectrum. 
\label{fig:rtcru:mcmc:a6}%
}
\end{center}
\end{figure*}

\subsection{\textit{Low/Hard} and \textit{High/Soft} States}
\label{rtcru:analysis:spectral:low:high}
 
The hardness--brightness anticorrelations seen in Figures\,\ref{fig:rtcru:hard} and \ref{fig:rtcru:hard2} could be due to some flickering variations on hour timescales in the soft band. 
To understand the origin of such a variation, we extracted the spectra for both the low/hard and high/soft states according to criteria explained in \S\,\ref{rtcru:analysis:hardness}. 
The \textit{low/hard} and \textit{high/soft} state events are distinguished 
by blue squares and red triangles in Figures\,\ref{fig:rtcru:light}--\ref{fig:rtcru:hard2}, respectively. 
We utilized the first-order ($m = \pm 1$) spectra produced for the low/hard and high/soft states, and repeated the same spectral analysis performed for the time-averaged spectrum in \S~\ref{rtcru:analysis:spectral} for both the HRC-S/LETG and ACIS-S/HETG data. The parameters of the best-fitting models for the low/hard and high/soft states are listed in Tables~\ref{rtcru:mekal:parameters} and \ref{rtcru:mekal:parameters2}.

To produce the continuum over the 4--8\,keV energy range, we used a powerlaw component that could be a phenomenological model. 
It is seen that the powerlaw  photon index $\Gamma$ is higher when the source was in the low (harder) state, which is in contrast
to what we typically see for the powerlaw model (0.3--8\,keV) fitted to the low/hard and high/soft states in X-ray binaries. 
We obtained $\Gamma=2.40^{+0.23}_{-0.52}$ (low/hard-state) and $\Gamma=1.69^{+0.10}_{-0.08}$  (high/soft-state) for the HRC-S data,
and  $\Gamma=2.04^{+0.23}_{-0.24}$ (low/hard) and $\Gamma=0.92^{+0.39}_{-0.15}$  (high/soft) for the ACIS-S data. 
The powerlaw normalization factor $K_{\rm p}$ of the source in the low/hard state is around 1.4 time higher than that in the high/soft state in the HRC-S observations, while $K_{\rm p}$(low/hard) is about 5.4 time higher than $K_{\rm p}$(high/soft) in the ACIS-S observation. The iron line measurements in the low/hard and high/soft state spectra are extremely uncertain due to lower count rates compared to the time-average spectra. We notice that the Fe K$\alpha$ line at 6.4\,keV, which could originate from either the reflection off an accretion disk or neutral dense clumps, in the high/soft state is twice stronger than that in the low/hard state in the HRC-S observations. However, we see an opposite situation in the ACIS-S observation. 
Our Bayesian low-count analysis suggested that the highly ionized Ly$\alpha$ iron line at 6.95\,keV might be stronger in the HRC-S low/hard state, 
while this line and the highly ionized He$\alpha$ iron line at 6.67\,keV are likely stronger in the ACIS-S low/hard state, but with extremely high uncertainties in both the 2005 and 2015 observations. 
The probability distributions of the powerlaw index and the Gaussian parameters are shown in Figures~A2--A5 (Supplementary Material). We see that the powerlaw continuum and the Fe\,K$\alpha$ line were strongly constrained in the ACIS-S, but weakly in the HRC-S. Moreover, we do not see any robust constraints on the \fexxv\ and \fexxvi\ lines, and the line widths in the low/hard- and high/soft-state spectra.

We fitted the fully absorbed thermal plasma model to the low/hard- and high/soft-state spectra of the HRC-S data by assuming $A_{\rm Fe}=0.15$ deduced from the time-averaged spectra in \S\,\ref{rtcru:analysis:spectral}. 
As we could not determine the confidence level bounds of $K_{\rm mk}$ in the low/hard and high/soft states, we adopted the \textsf{mekal} normalization factor $K_{\rm mk}$ derived from the time-averaged HRC-S spectrum (having counts twice that in each state) to compute the covariance matrix for our MCMC statistic method. 
Fixing the metal abundance $A_{\rm Fe}$ and the normalization factor $K_{\rm mk}$ of the \textsf{mekal} model allowed us to properly constrain the confidence level bounds of other free parameters (see Figures~\ref{fig:rtcru:mcmc:a6} and A6). We derived a plasma temperature of $kT=1.5 \pm 0.1$ and $0.6^{+0.3}_{-0.4}$\,keV in the low/hard and high/soft states, respectively, while $kT=1.3^{+0.4}_{-0.2}$\,keV was obtained from the time-averaged spectrum containing both the low/hard- and high/soft-state events. 
Similarly, from the \textsf{apec} model we obtained $kT=1.3 \pm 0.1$ and $0.7^{+0.1}_{-0.3}$\,keV for the low/hard- and high/soft-state spectra, respectively.
It can be seen that the plasma was slightly hotter in the low (hard) state and cooler in the high (soft) state.
As seen in Figure~\ref{fig:rtcru:mcmc:a6}, the probability distributions of the plasma temperature ($kT$) and the column density ($N_{\rm H,mk}$) of the fully absorbed thermal plasma model were robustly fitted together with the column density ($N_{\rm H,p}$), covering fraction ($f_{\rm p}$), and normalization factor ($K_{\rm p}$) of the partially
absorbed powerlaw model in the low/hard state spectrum. However, we see in Figure~A6 that our fitting analysis provided an upper limit to the plasma temperature ($kT$) in the high/soft state spectrum, while we obtained solid constraints on the parameters of the absorbed powerlaw model, but an extremely weak constraint on the absorbing column density of the thermal plasma model.

The absorbed \textsf{mekal} model was also fitted to the low/hard- and high/soft-state ACIS-S spectra by taking $A_{\rm Fe}=0.31$ derived from the time-averaged data in \S\,\ref{rtcru:analysis:spectral}. 
We obtained $kT=4.4^{+2.3}_{-2.4}$ and $45.9^{+12.7}_{-17.5}$\,keV for the high/soft- and low/hard-state spectra, respectively, while we got $kT=9.6^{+2.7}_{-4.0}$\,keV from the time-averaged spectrum. 
The \textsf{apec} model provided $kT=3.8^{+1.4}_{-0.8}$ and $50.9^{+\cdots}_{-41.7}$\,keV  (the upper error cannot be constrained) for the high/soft- and low/hard-state spectra, respectively, which are similar to the \textsf{mekal} results.
As the low/hard-state spectrum did not have sufficient count rates, we could not determine its upper confidence limit using the \textsf{apec} model.
Again, similar to the HRC-S data, the thermal plasma was found to be hotter in the low/hard state and cooler in the high/soft state. However, the plasma temperatures derived from the ACIS-S/HETG data are higher than those from the HRC-S/LETG data.
We should note that the ACIS-S/HETG spectrometer is more efficient in the hard band, while the HRC-S/LETG spectrometer effectively records the soft band, so the discrepancy between the ACIS-S and HRC-S results could be related to their respective sensitivity energy ranges. Alternatively, the thermal plasma in RT\,Cru might be hotter in 2005 in the ACIS-S epoch, and cooler in 2015 during the HRC-S observations.
Figures~A8 and A9 present the probability distributions of the model parameters for the low/hard- and high/soft-state ACIS-S spectra. 
While the plasma temperature ($kT$) was robustly fitted in the time-averaged ACIS-S spectrum, the probability distributions of the \textsf{mekal} temperature are very wide in the low/hard and high/soft states.
However, the absorbing column density ($N_{\rm H,mk}$) and normalization factor ($K_{\rm mk}$) of the thermal model were well fitted in all the spectra (time-averaged, low and high). 
The covering fractions ($f_{\rm p}$) of the \textsf{powerlaw} model are about 1.0 in all the models (fully absorbed powerlaw) of the ACIS-S/HETG data .

\section{Discussion}
\label{rtcru:discussion}

\begin{table*}
\begin{center}
\caption{The unabsorbed \textit{hard} X-ray luminosity and mass-accretion rate from the HRC-S/LETG and ACIS-S/HETG data 
}\label{rtcru:mekal:lumin}
\begin{tabular}{C{3.5cm}ccc}
\hline\hline
\noalign{\smallskip}
\multicolumn{1}{l}{Parameter} & {Time-Averaged}  & {Low/Hard-State}   & {High/Soft-State}  \\
\noalign{\smallskip}
\hline 
\noalign{\smallskip}
\multicolumn{4}{c}{HRC-S/LETG: \textsf{powerlaw}} \\
\noalign{\smallskip}
\hline
\noalign{\smallskip}
$F_{\rm x}$($\mathrm{ergs}\,\mathrm{s}^{-1}\,\mathrm{cm}^{-2}$)\,$^{\rm \bf a}$ \dotfill  & $     34.90^{+0.38}_{-0.37}\times 10^{-12}$ &  $     64.55^{+1.67}_{-1.60}\times 10^{-12}$ &  $     60.01^{+1.33}_{-1.54}\times 10^{-12}$ \\ 
\noalign{\smallskip}      
$L_{\rm x}$($\mathrm{ergs}\,\mathrm{s}^{-1}$)\,$^{\rm \bf a}$ \dotfill  & $     11.23^{+0.12}_{-0.12}\times 10^{33}$ &  $      20.77^{+0.51}_{-0.54}\times 10^{33}$ &  $      19.31^{+0.43}_{-0.50}\times 10^{33}$ \\  
\noalign{\smallskip}      
$\dot{M}$($M_{\odot}\mathrm{yr}^{-1}$)\,$^{\rm \bf b}$ \dotfill  & $     1.52^{+0.02}_{-0.02}\times 10^{-9}$ & $     2.82^{+0.07}_{-0.07}\times 10^{-9}$ &  $     2.62^{+0.06}_{-0.07}\times 10^{-9}$ \\         
\noalign{\smallskip}
\hline
\noalign{\smallskip}
\multicolumn{4}{c}{ACIS-S/HETG: \textsf{powerlaw}} \\
\noalign{\smallskip}
\hline
\noalign{\smallskip}
$F_{\rm x}$($\mathrm{ergs}\,\mathrm{s}^{-1}\,\mathrm{cm}^{-2}$)\,$^{\rm \bf a}$ \dotfill  & $     92.33^{+1.97}_{-2.56}\times 10^{-12}$ &  $     52.88^{+1.94}_{-2.53}\times 10^{-12}$ &  $      25.73^{+1.38}_{-1.15}\times 10^{-12}$ \\ 
\noalign{\smallskip}      
$L_{\rm x}$($\mathrm{ergs}\,\mathrm{s}^{-1}$)\,$^{\rm \bf a}$ \dotfill  & $     29.70^{+0.63}_{-0.82}\times 10^{33}$ &  $      17.01^{+0.62}_{-0.81}\times 10^{33}$ &  $      8.28^{+0.44}_{-0.37}\times 10^{33}$ \\  
\noalign{\smallskip}      
$\dot{M}$($M_{\odot}\mathrm{yr}^{-1}$)\,$^{\rm \bf b}$ \dotfill  & $     4.03^{+0.09}_{-0.11}\times 10^{-9}$ & $     2.31^{+0.08}_{-0.11}\times 10^{-9}$ &  $     1.12^{+0.06}_{-0.05}\times 10^{-9}$ \\    
\noalign{\smallskip}
\hline
\noalign{\smallskip}
\multicolumn{4}{c}{ACIS-S/HETG: \textsf{powerlaw}+\textsf{mekal}} \\
\noalign{\smallskip}
\hline
\noalign{\smallskip}
$F_{\rm x}$($\mathrm{ergs}\,\mathrm{s}^{-1}\,\mathrm{cm}^{-2}$)\,$^{\rm \bf c}$ \dotfill  & $     120.37^{+1.89}_{-3.56}\times 10^{-12}$ &  $     62.90^{+3.07}_{-3.62}\times 10^{-12}$ &  $      55.97^{+1.92}_{-2.67}\times 10^{-12}$ \\ 
\noalign{\smallskip}      
$L_{\rm x}$($\mathrm{ergs}\,\mathrm{s}^{-1}$)\,$^{\rm \bf c}$ \dotfill  & $     38.72^{+0.60}_{-1.15}\times 10^{33}$ &  $      20.24^{+0.99}_{-1.16}\times 10^{33}$ &  $      18.00^{+0.61}_{-0.86}\times 10^{33}$ \\  
\noalign{\smallskip}      
$\dot{M}$($M_{\odot}\mathrm{yr}^{-1}$)\,$^{\rm \bf b}$ \dotfill  & $     5.25^{+0.08}_{-0.16}\times 10^{-9}$ & $     2.75^{+0.13}_{-0.16}\times 10^{-9}$ &  $     2.44^{+0.08}_{-0.12}\times 10^{-9}$ \\    
\noalign{\smallskip}
\hline
\noalign{\smallskip}
\end{tabular}
\end{center}
\begin{tablenotes}
\item[1]\textbf{Notes.} 
The unabsorbed X-ray fluxes are for the models with the best-fit parameters listed in Tables~\ref{rtcru:mekal:parameters} and \ref{rtcru:mekal:parameters2}. 
$^{\rm \bf a}$ The flux $F_{\rm x}$ derived from an unabsorbed \textsf{powerlaw} over the 0.3--8\,keV energy range, and the corresponding luminosity $L_{\rm x}$ at the distance of $D=1.64$\,kpc. $^{\rm \bf b}$ The mass-accretion rate $\dot{M}$ estimated based on the 0.3--8\,keV luminosity $L_{\rm x}$, and $M_{\rm WD} = 1.25$ M$_{\odot}$ and  $R_{\rm WD} = 7.1 \times 10^{8}$\,cm from \citet{Luna2018}. 
$^{\rm \bf c}$ The flux $F_{\rm x}$ and luminosity $L_{\rm x}$ derived from an unabsorbed \textsf{powerlaw} and \textsf{mekal} model over the 0.3--8\,keV energy range ($D=1.64$\,kpc).
\end{tablenotes}
\end{table*}

\subsection{Mass-Accretion Rate}
\label{rtcru:discussion:accretion:rate}

The accretion of gas by a WD leads to the formation of a shock front near the stellar surface that creates a hot plasma environment between the shock front and the stellar surface. 
The temperature of the hot plasma  depends only on the WD mass, while the luminosity variability is proportioned to the accretion rate. 
Accordingly, \citet{Luna2018} derived a WD mass of $M_{\rm WD} = 1.25$ M$_{\odot}$ and a WD radius of $R_{\rm WD} = 7.1 \times 10^{8}$\,cm from a maximum post-shock temperature of $53$ keV estimated from \textit{Suzaku} and \textit{NuSTAR}+\textit{Swift}.

Following \citet{Luna2007}, the mass-accretion rate can be estimated from the unabsorbed \textit{hard} X-ray luminosity $L_{\rm x}$ between 0.3 and 8\,keV:
\begin{equation}
\dot{M} = \frac{2 L_{\rm x} R_{\rm WD}} { G M_{\rm WD} }.
\label{eq:shck:4}
\end{equation}
For the $L_{\rm x}$ calculations of the HRC-S data, we did not include the \textit{soft} thermal plasma \textsf{mekal} model as it originates from either the outer layer or soft flickering variation due to its high luminosity (resulting in a nonphysical mass-accretion rate). 
As we obtained a lower limit to the hot plasma temperature in the \textit{hard} thermal plasma \textsf{mekal} model in the ACIS-S data, their mass-accretion rate $\dot{M}$ and X-ray luminosity $L_{\rm x}$ approximately correspond to a lower estimation of X-ray emission produced in an inner layer of the accretion disk. 

Table~\ref{rtcru:mekal:lumin} lists the fluxes ($F_{\rm x}$) derived from the unabsorbed \textit{hard} model over the 0.3--8\,keV energy range from the time-averaged, low/hard-state and high/soft-state spectra for both the HRC-S and ACIS-S observations. It also presents their corresponding luminosities ($L_{\rm x}$) calculated at the distance of $D=1.64$\,kpc, together with their mass-accretion rates ($\dot{M}$) calculated by Eq.~(\ref{eq:shck:4}), and the WD mass and radius from \citet{Luna2018}. 
We obtained source fluxes of $F_{\rm x} \approx 35 \times 10^{-12}$ and $92 \times 10^{-12}$\,erg\,cm$^{-2}$\,s$^{-1}$ over the 0.3--8\,keV energy range from the unabsorbed powerlaw model fitted to the time-averaged HRC-S and ACIS-S spectra, respectively. Assuming a distance of $D=1.64$\,kpc, we derived an X-ray luminosities of $L_{\rm x} \approx 11 \times 10^{33}$\,erg\,s$^{-1}$ for the HRC-S time-averaged spectrum.
For the ACIS-S data, we estimated $L_{\rm x} \approx  30 \times 10^{33}$\,erg\,s$^{-1}$ at $D=1.64$\,kpc, whereas
$44 \times 10^{33}$\,erg\,s$^{-1}$ at $D=2$\,kpc, which is slightly higher than $L_{\rm x} = 31^{+8}_{-7} \times 10^{33}$\,erg\,s$^{-1}$ derived by \citet{Luna2007} at $D=2$\,kpc. 

Our X-ray luminosities estimated at $D=1.64$\,kpc from the unabsorbed \textsf{powerlaw} model correspond to mass-accretion rates of $\dot{M} \approx 1.5\times 10^{-9} M_{\odot}$\,yr$^{-1}$ in 2015 (HRC-S) and $4\times 10^{-9} M_{\odot}$\,yr$^{-1}$ in 2005 (ACIS-S).
They are close to the mass-accretion rate $\dot{M} \approx 1.8\times 10^{-9} M_{\odot}$\,yr$^{-1}$ estimated at $D=2$\,kpc from a cooling flow model by \citet{Luna2007}. 
Keeping the hard thermal plasma emission component in the ACIS-S model (\textsf{powerlaw}+\textsf{mekal}), we also derived $\dot{M} \approx 5.3\times 10^{-9} M_{\odot}$\,yr$^{-1}$\,($D$/1.64\,kpc)$^2$ for the 2005 observation, which is within mass-accretion rates of $3.7\times 10^{-9} M_{\odot}$\,yr$^{-1}$ and $6.7\times 10^{-9} M_{\odot}$\,yr$^{-1}$ estimated at $D=2$\,kpc by \citet{Luna2018} from the two separate \textit{Suzaku} observations of low- and high-brightness epochs, respectively. They also estimated a mass-accretion rate of 
$4.3\times 10^{-9} M_{\odot}$\,yr$^{-1}$from the \textit{NuSTAR} and \textit{Swift} observations. \citet{Luna2018} pointed out that the mass-accretion rate increases during the brightening event.
However, our ACIS-S \textsf{powerlaw}+\textsf{mekal} model yielded $\dot{M} \approx 2.8\times 10^{-9} M_{\odot}$\,yr$^{-1}$ in the low/hard state and $2.4\times 10^{-9} M_{\odot}$\,yr$^{-1}$ in the high/soft state, which do not seem
to agree with higher mass-accretion rates when the source was in the high (brighter) state. Additionally, we obtained roughly similar mass-accretion rates for the low/hard- and high/soft-state HRC-S observations. 
We also noticed that the mass-accretion rate derived from the unabsorbed \textsf{powerlaw} model of the HRC-S low/hard-state spectrum was slightly higher than that in the high/soft state (see Table~\ref{rtcru:mekal:lumin}). This trend in the HRC-S mass-accretion rates is similar to those deduced from the unabsorbed \textsf{powerlaw}+\textsf{mekal} model of the ACIS-S in the low/hard and high/soft states.  

As seen in Table~\ref{rtcru:mekal:lumin}, the mass-accretion rates were roughly about $\dot{M} \approx 2.7 \times 10^{-9} M_{\odot}$\,yr$^{-1}$\,($D$/1.64\,kpc)$^2$ or  $\dot{M} \approx 4 \times 10^{-9} M_{\odot}$\,yr$^{-1}$\,($D$/2.0\,kpc)$^2$
in both the low/hard and high/soft states of the \textit{Chandra} HRC-S observations (\textsf{powerlaw}) and ACIS-S observation (\textsf{powerlaw}+\textsf{mekal}). 
This mass-accretion rate agrees with $\dot{M} \approx 4 \times 10^{-9} M_{\odot}$\,yr$^{-1}$\,($D$/2.0\,kpc)$^2$
derived from \textit{NuSTAR}+\textit{Swift} and \textit{Suzaku} low-brightness epoch \citep{Luna2018}.
The lower mass-accretion rate of $\dot{M} \approx 1.5 \times 10^{-9} M_{\odot}$\,yr$^{-1}$\,($D$/1.64\,kpc)$^2$ 
derived from the time-averaged HRC-S spectrum, and the higher mass-accretion rate 
of $\dot{M} \approx 5.3 \times 10^{-9} M_{\odot}$\,yr$^{-1}$\,($D$/1.64\,kpc)$^2$ 
from the time-averaged ACIS-S spectrum
might be related to some variations in the hard band, which were not separated in the mean spectra. 
We also notice $\dot{M} \approx 6.7 \times 10^{-9} M_{\odot}$\,yr$^{-1}$ 
estimated at $D=2$\,kpc from the \textit{Suzaku} high-brightness epoch \citep{Luna2018}, which is comparable to the value derived
for the time-average ACIS-S observation in 2005 when the source in the hard excess was apparently brighter than the HRC-S observations in 2015. 
The variation in brightness could be partially related to the absorbing materials within the line of sight that obscure the X-ray source brightness in different epochs, and does not seem to have any strong correlation with the mass-accretion rate.

\subsection{Flickering Variations}

The light curves of the HRC-S/LETG (see Figure~\ref{fig:rtcru:light}) show that the soft bands ($S$: 0.3--4\,keV) have variations with larger amplitudes than the hard bands ($H$: 4--8\,keV). However, it can be seen 
in Figure~\ref{fig:rtcru:light2} that the light curves of the ACIS-S/HETG data had variations with similar amplitudes in both the soft and hard bands. 
We should note that the hard energy band of the source in the ACIS-S observation in 2005 was much brighter than the HRC-S observation in 2015.
We also see that the source exhibits short flickering (brightening) variations on hour timescales without any periodic modulations, which could be related to either rapid outbursts or obscuring absorption materials. 
These brightening variations appear in the $S$ band of the HRC-S observations in 2015 (red triangles
in Figure~\ref{fig:rtcru:light}) and in the $S$ and $H$ bands of the ACIS-S observation in 2005 (see Figure~\ref{fig:rtcru:light2}). 
They are mostly in the high/soft state (red triangles
in Figures~\ref{fig:rtcru:hard} and \ref{fig:rtcru:light2}).
As seen in Table~\ref{rtcru:mekal:lumin}, 
the unabsorbed  X-ray luminosities $L_{\rm x}$ were roughly the same in the low/hard and high/soft states
of the HRC-S (\textsf{powerlaw}) and ACIS-S (\textsf{powerlaw}+\textsf{mekal}),
so variation in the absorbing material could partially be responsible for the brightening events.
We also see that the mass-accretion rates are slightly higher in the low sate, so
it does not seems to be a big contributor to the flickering events.

As seen in Tables~\ref{rtcru:mekal:parameters} and \ref{rtcru:mekal:parameters2}, the X-ray source was much brighter in the hard band (4--8\,keV) when it was in the low/hard states due to higher normalization factors ($K_{p}$). However, it is less brighter in the hard band when it was in the high/soft state containing more flickerings, so 
the brightening might originate from mechanical processes in the accretion disk.
A slight increase in the mass-accretion rate might make the X-ray source harder and fainter. 
Our analysis of the HRC-S and ACIS-S data revealed that the plasma temperature in the low/hard state was higher than in the high/soft state, so the plasma was hotter when the source was fainter.
As the brightening events are anticorrelated with the plasma temperatures, they should originate from the outer layers of the accretion disk, so either rapid outbursts or absorbing materials 
could generate them. 

\subsection{Origin of the Iron Lines}

The HRC-S and ACIS-S spectra show a strong fluorescent iron K$\alpha$ line at 6.4\,keV (see Figure~\ref{fig:rtcru:ironlines}) that could be an indication of either the reflection off an accretion disk or dense clumpy materiel within the line of sight. 
As seen in Tables~\ref{rtcru:mekal:parameters} and \ref{rtcru:mekal:parameters2}, the Fe K$\alpha$ line was twice stronger in the HRC-S time-averaged spectrum compared to the ACIS-S.
While the Fe K$\alpha$ line measurements have large uncertainties in the low/hard and high/soft states due to lower count rates, 
it can be seen that it is probably stronger in the high/soft state than the low/hard state in the HRC-S epoch in 2015, while
it is opposite in the ACIS-S epoch. We should note that the HRC-S epoch does not have the same brightness as the ACIS-S.
The Fe K$\alpha$ line could be emitted mostly from the pre-shock accreting matter in the HRC-S epoch. 
As the brightening was higher in the high/soft-state HRC-S events, the fluorescent
iron K$\alpha$ emission line could be more intense due to denser clumpy materiel in the outer layers. 
Nevertheless, we see a stronger Fe K$\alpha$ line in the low/hard-state ACIS-S events compared to its high/soft-state events.
A higher accretion rate (see Table~\ref{rtcru:mekal:lumin}) and slightly lower absorbing column density in the hard excess (see Table~\ref{rtcru:mekal:parameters2}) might explain this discrepancy in the Fe K$\alpha$ line.

Similarly, the photo-ionized iron lines (\fexxv\ and \fexxvi) are likely stronger in the HRC-S epoch in 2015 than 
the ACIS-S epoch in 2005. 
However, we should also consider high uncertainties in the HRC-S measurements. It is also possible that the hard excess of the source in the HRC-S epoch was harder, 
so the hard radiation produced the stronger \fexxv\ and \fexxvi\ lines. 
The unabsorbed X-ray luminosities measured over the 0.3--8 keV
energy range (see the powerlaw models presented in Table~\ref{rtcru:mekal:lumin}) suggest that the hard excess of the X-ray source was at least three times luminous 
in the ACIS-S epoch than the HRC-S epoch. However, the powerlaw photon index $\Gamma$ was approximately the same in different \textit{Chandra} observations in 2005 and 2015 (see Tables~\ref{rtcru:mekal:parameters} and \ref{rtcru:mekal:parameters2}). 
We also found extremely weak constraints (see Figures~A2--A5)  on the \fexxv\ and \fexxvi\ lines in the low/hard and high/soft states of the HRC-S and ACIS-S epochs (due to the low count rate), so we could not interpret them according to different states.

\subsection{Comparison of ACIS-S and HRC-S Epochs}

As seen in Figures~\ref{fig:rtcru:light}--\ref{fig:rtcru:hard2}, the hard excess of the source was brighter and has a higher X-ray luminosity (see Table~\ref{rtcru:mekal:lumin}) in 2005 compared to 2015. We should note that the ACIS-S/HETG effective area dramatically drops at energies below 1\,keV, so it cannot detect any thermal plasma components in the soft band (0.3--4\,keV). However, the HRC-S/LETG instrument has more effective area in the soft band than the hard band, so 
it provides fewer details about possible thermal plasma components in the hard band (4--8\,keV).
Although the HRC-S/LETG instrument has a significant background noise, it has a better sensitivity than the ACIS-S/LETG to detect any plasma emission below 2\,keV, so the HRC-S/LETG observation is expected to reveal the thermal emission in the soft band. 

As seen in Tables~\ref{rtcru:mekal:parameters} and \ref{rtcru:mekal:parameters2}, our thermal plasma \textsf{mekal} modeling of the \textit{Chandra} observations of RT\,Cru suggests two plasma temperatures of $kT \sim 1.3$ and $9.6$\,keV in the HRC-S and ACIS-S time-averaged spectra, respectively. 
Moreover, we obtained $kT \sim 1.1$ (HRC-S) and $10.3$\,keV (ACIS-S) from the \textsf{apec} model fitted to the time-averaged spectra. Our spectral analyses imply that the soft thermal plasma emission component with $kT \sim 1.3$ was obscured by some dense clumpy materials with a column density of $N_{\rm H,mk} \sim 76 \times 10^{22}$\,cm$^{-2}$ in the HRC-S/LETG epoch (2015) when the hard excess of the source was fainter than the 2005 epoch. 
However, it seems that the hard thermal plasma emission component with $kT \sim 10$ was obscured by less dense absorbing materials with $N_{\rm H,mk} \sim 12 \times 10^{22}$\,cm$^{-2}$ in the ACIS-S/HETG epoch (2005) when the source was brighter in the hard band.
We notice that the fitted \textsf{powerlaw} model has a column density of $N_{\rm H,p} \sim 5 \times 10^{22}$\,cm$^{-2}$ for the HRC-S/LETG (2015), which is significantly lower than  
$N_{\rm H,p} \sim 100 \times 10^{22}$\,cm$^{-2}$ derived for the ACIS-S/HETG time-averaged spectrum (2005).

We also fitted the thermal plasma model to the low/hard-state and high/soft-state events, which were classified based on the HR diagrams shown in Figures~\ref{fig:rtcru:hard} and \ref{fig:rtcru:hard2} under the criteria defined in \S\,\ref{rtcru:analysis:hardness}. 
It can be seen in Tables~\ref{rtcru:mekal:parameters} and \ref{rtcru:mekal:parameters2} that the plasma temperature was higher in the low/hard state of both the HRC-S (2015) and ACIS-S (2005). Fitting \textsf{mekal} and \textsf{apec} to the low/hard-state HRC-S spectrum yields plasma temperatures of
$kT=1.5$ and 1.3\,keV, respectively, which are slightly higher than those derived from the time-averaged HRC-S spectrum.
For the low/hard-state ACIS-S spectrum, we also derived $kT=46^{+13}_{-17}$\,keV (\textsf{mekal}) and 
$kT=51^{\cdots}_{-42}$\,keV (\textsf{apec}), which are comparable to the maximum post-shock temperature of $kT_{max}=53 \pm 4$\,keV derived by \citet{Luna2018} from the \textit{Suzaku} and \textit{NuStar}+\textit{Swift} observations.
In the high/soft-state spectra, we obtained lower plasma temperatures of $kT=0.6$ (HRC-S) and 4.4\,keV (ACIS-S). 

The powerlaw indexes of the hard band (4--8\,keV) were $\Gamma=1.7$ in 2015 and $\Gamma=1.8$ in 2005, so they were roughly similar in the time-averaged spectra. 
The fluorescent iron emission line at 6.4 keV, which could be emitted from either reflection off an accretion disk or dense absorbing material, was measured to be twice stronger in the HRC-S time-averaged spectrum compared to the ACIS-S. 
Stronger Fe K$\alpha$ line might be an indication 
of higher neutral dense clumps. Interestingly, the soft thermal plasma emission component ($kT \sim 1.3$) identified in the HRC-S is likely obscured by dense clumpy materials. By contrast, the hard thermal plasma emission component ($kT \sim 10$) detected in the ACIS-S is likely absorbed by less dense obscuring materials, when the Fe K$\alpha$ line was by a factor of 2 lower than that measured in the HRC-S.
Nevertheless, this is in contradiction to absorbing column densities derived from the absorbed \textsf{powerlaw} model (see Tables~\ref{rtcru:mekal:parameters} and \ref{rtcru:mekal:parameters2}).
It is possible that both the soft ($kT \sim 1.3$) and hard ($kT \sim 10$) thermal plasma emission components are emitted from outer layers of the accretion disk having dense clumps, which are not the same material blocking the hard excess in the \textsf{powerlaw} models.

\section{Summary  and Conclusion}
\label{rtcru:conclusion}

To summarize, we conducted a detail spectral analysis of the hard X-ray emitting symbiotic system RT\,Cru observed 
using the \textit{Chandra} HRC-S/LETG for a total exposure of 79\,ks in 2015, and ACIS-S/HETG for 49\,ks in 2005.
We have modeled the hard excess above 4\,keV with a powerlaw model partially covered by a local absorber, in addition to the Fe K$\alpha$, Fe\,{\sc xxv}, and Fe\,{\sc xxvi} lines using three Gaussian functions. We have extended our spectral analysis to the soft excess below 4\,keV by including a thermal plasma model fully covered by another local absorber. Additionally, we applied the same modeling procedures to the spectra generated from events separated into the low/hard and high/soft states according to the HR diagrams. 
Our key findings obtained from \textit{Chandra} HRC-S/LETG and ACIS-S/HETG are as follows:

1.~Our hardness ratio analysis of the light curves of the soft ($S$: 0.3--4\,keV) and hard ($H$: 4--8\,keV) bands revealed two distinct states (see Figures~\ref{fig:rtcru:hard} and ~\ref{fig:rtcru:hard2}), namely \textit{low/hard} and \textit{high/soft} states. The X-ray source shows aperiodic flickering variations mostly from the soft excess in the HRC-S epoch (see Figure~\ref{fig:rtcru:light}), but also from the hard excess in the ACIS-S epoch (see Figure~\ref{fig:rtcru:light2}). 
We found an anticorrelation relation between the hardness ratio ($H/S$) and total brightness ($H+S$) in both the HRC-S and ACIS-S epochs, implying that the X-ray source was fainter when it was harder, and vice versa.

2.~The time-averaged HRC-S/LETG and ACIS-S/HETG spectra likely have the same hard X-ray continuum above 4\,keV described
by a \textsf{powerlaw} model with a photon index of about $\Gamma \sim 1.7$--1.8 over the 4--8\,keV energy range. While the X-ray source was fainter in the HRC-S epoch in 2015, its K$\alpha$ fluorescent iron emission line at 6.4\,keV is twice as brighter compared to the ACIS-S epoch in 2005. A higher value of the Fe K$\alpha$ line may be an indication for a denser neutral absorber located in the outer layers. The soft excess is more obscured by heavily dense absorbing materials in the HRC-S epoch, while we have no knowledge about the physical condition of the soft band in the ACIS-S epoch. By contrast, the hard excess is less obscured in the HRC-S, while it is heavily absorbed in the ACIS-S epoch (see Tables~\ref{rtcru:mekal:parameters} and \ref{rtcru:mekal:parameters2}).

3.~In the low/hard state when the X-ray source was fainter and harder, the spectra were fitted by a powerlaw phenomenological model over 4--8\,keV having photon indexes of $\Gamma \sim 2.4$ and $2$ in the HRC-S and ACIS-S epochs, respectively.  In the high/soft state when the X-ray source was brighter and softer, the spectra were fitted by lower powerlaw indexes $\Gamma \sim 1.7$ (HRC-S) and $0.9$ (ACIS-S) over the 4--8\,keV energy range.  
This is in contrast to the photon index behavior is typically seen in the low/hard and high/soft states of X-ray binary over 0.3--8\,keV. 
However, we derived roughly the same X-ray luminosities from the unabsorbed \textsf{powerlaw} model for the low/hard and high/soft states in the HRC-S (see Table~\ref{rtcru:mekal:lumin}), and the similar X-ray luminosities from the unabsorbed \textsf{powerlaw}+\textsf{mekal} model for the low/hard and high/soft states in the ACIS-S. 
Interestingly, these unabsorbed hard X-ray luminosities are very close to $\sim 60 \times 10^{-12}$\,$\mathrm{ergs}\,\mathrm{s}^{-1}\,\mathrm{cm}^{-2}$.

4.~The thermal emissions of the time-averaged HRC-S/LETG and ACIS-S/HETG spectra were well reproduced by
thermal plasma \textsf{mekal} models with $kT \sim 1.3$ and $9.6$\,keV, respectively. For the HRC-S/LETG, 
we also derived $kT \sim 1.5$\,keV in the low/hard state (fainter and harder) and $kT \sim 0.6$\,keV in the high/soft state (brighter and softer).  For the ACIS-S/HETG, we estimated $kT \sim 46$ and $4$\,keV in the low/hard and high/soft states, respectively.

5.~From the unabsorbed \textsf{powerlaw} model fitted to the HRC-S time-averaged spectrum, we derived a mass accretion rate of $\dot{M} \sim 1.5\times 10^{-9} M_{\odot}$\,yr$^{-1}$ (at $D=1.64$\,kpc), whereas for the ACIS-S time-averaged spectrum, we obtained $\dot{M} \sim 4\times 10^{-9} M_{\odot}$\,yr$^{-1}$ from the unabsorbed \textsf{powerlaw} model and $\dot{M} \sim 5.3\times 10^{-9} M_{\odot}$\,yr$^{-1}$ from the unabsorbed \textsf{powerlaw}+\textsf{mekal} model. However, we obtained roughly $\dot{M} \sim 2.7\times 10^{-9} M_{\odot}$\,yr$^{-1}$ for low/hard and high/soft states of the HRC-S (\textsf{powerlaw}) and the ACIS-S (\textsf{powerlaw}+\textsf{mekal}), so the discrepancy between the time-averaged HRC-S and ACIS-S might be related to some variations in the hard band of the ACIS-S and strong soft flickerings in the HRC-S, which were not separated from each other in the time-averaged spectra. 

In conclusion, our Bayesian \textit{low-count} X-ray spectral analysis of the \textit{Chandra} HRC-S/LETG and ACIS-S/HETG observations
provided the first evidence for the soft thermal plasma emission component ($kT \sim 1.3$\,keV), in addition to the reconfirmation of the hard thermal plasma emission component ($kT \sim 10$\,keV) previously identified. This soft emission component is possibly largely obscured by highly dense materials ($N_{\rm H,mk} > 5 \times 10^{23}$\,cm$^{-2}$) within the line of sight. Future observations by 
\textit{Chandra} ACIS-S/LETG with a significantly longer exposure time will be able to verify this soft thermal plasma emission. In particular, the \textit{Arcus} mission \citep{Smith2016}, recently admitted to Phase A study, would offer a more effective area at energies below 1\,keV, which would allow us to reaffirm its presence in the soft excess.
Interestingly, some other hard X-ray emitting symbiotics have also been found to contain two distinct (soft and hard) thermal plasma emission components, namely CH\,Cyg \citep[$kT=0.2$, $0.7$ and 7.3\,keV;][]{Ezuka1998}, 
MWC\,560 \citep[$kT=0.18$ and $11.26$\,keV;][]{Stute2009}, and 
SS73\,17 \citep[$kT=1.12$ and $9.9$\,keV;][]{Eze2010}. 
In particular, jets have been detected in CH\,Cyg \citep{Corradi2001,Sokoloski2003,Galloway2004,Karovska2007,Karovska2010}
and MWC\,560 \citep{Tomov1990,Tomov1992,Schmid2001,Lucy2018}, while \citet{Eze2010} also
suggested a possible unseen jet in SS73\,17. Although the soft thermal plasma emission component
could be associated with a jet in the symbiotic system, \citet{Stute2009} pointed out that the $0.8$--1\,keV soft thermal component 
can also be emitted from shocks (with velocities of $\sim 1000$\,km\,s$^{-1}$) created by colliding winds
of accreting flows. However, we have not been able to find clear evidence of a jet in RT\,Cru through these data.

\section*{Acknowledgments}

Support for this work was provided by the National Aeronautics and Space Administration (NASA) through \textit{Chandra} Award Number GO5-16023X issued by the \textit{Chandra} X-ray Center, which is operated by the Smithsonian Astrophysical Observatory for and on behalf of the NASA under contract NAS8-03060. 
JJD and VLK were supported by the NASA contract NAS8-03060 to the \textit{Chandra} X-ray Center. 
The scientific results reported in this article are based on observations made by the \textit{Chandra} X-ray Observatory, associated with program number 619012, 
and data obtained from the \textit{Chandra} Data Archive. 
This work has made use of data from the European Space Agency (ESA) mission
{\it Gaia}, 
processed by the {\it Gaia}
Data Processing and Analysis Consortium (DPAC). 
We thank the anonymous referee for
helpful suggestions and corrections that greatly improved the paper.

\section*{Data Availability}

The observational raw data underlying this article are available in the \textit{Chandra} Data Archive (CDA) at 
\url{https://cda.harvard.edu/chaser/},
and the processed data will be shared on reasonable request to the corresponding author.



\section*{Supplementary Material}
\label{rtcru:supplementary}

\subsection*{\textbf{Appendix A:} Bayesian Low-Count X-ray Spectral Analysis}


The following figures are available for the electronic edition of this article:
~\\
\textbf{Figures A1--A5.} The same as Figure~\ref{fig:rtcru:mcmc1}, but for the HRC-S/LETG time-averaged spectrum, and the ACIS-S/HETG and HRC-S/LETG \textit{low/hard}-  and \textit{high/soft}-state spectra, respectively. 
~\\
\textbf{Figures A6--A9.} The same as Figure~\ref{fig:rtcru:mcmc7}, but for  the HRC-S/LETG \textit{high/soft}-state spectrum,
the ACIS-S/HETG time-averaged spectrum, and the ACIS-S/HETG  \textit{low/hard}-  and \textit{high/soft}-state spectra, respectively.

\label{lastpage}

\newpage



\textbf{\LARGE Supplementary Material}

\appendix
\section{Bayesian Low-Count X-ray Spectral Analysis}
\label{rtcru:sec:mcmc}

\newpage 

\begin{figure*}
\begin{center}
\includegraphics[width=5.00in, trim = 0 0 0 0, clip, angle=270]{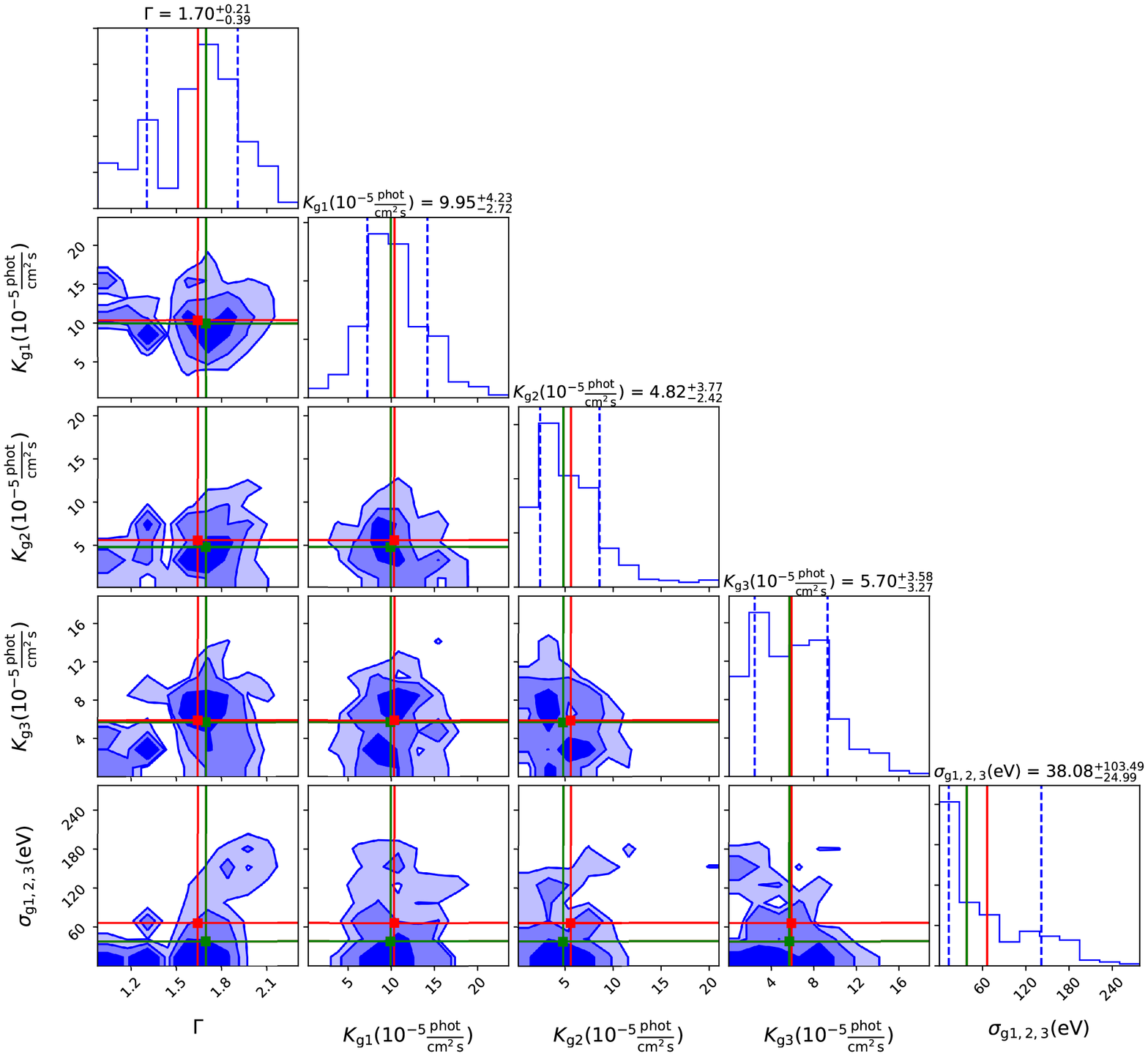}%
\caption{The same as Figure~6, but for the HRC-S/LETG time-averaged spectrum. 
\label{fig:rtcru:mcmc:a1}%
}
\end{center}
\end{figure*}

\begin{figure*}
\begin{center}
\includegraphics[width=5.00in, trim = 0 0 0 0, clip, angle=270]{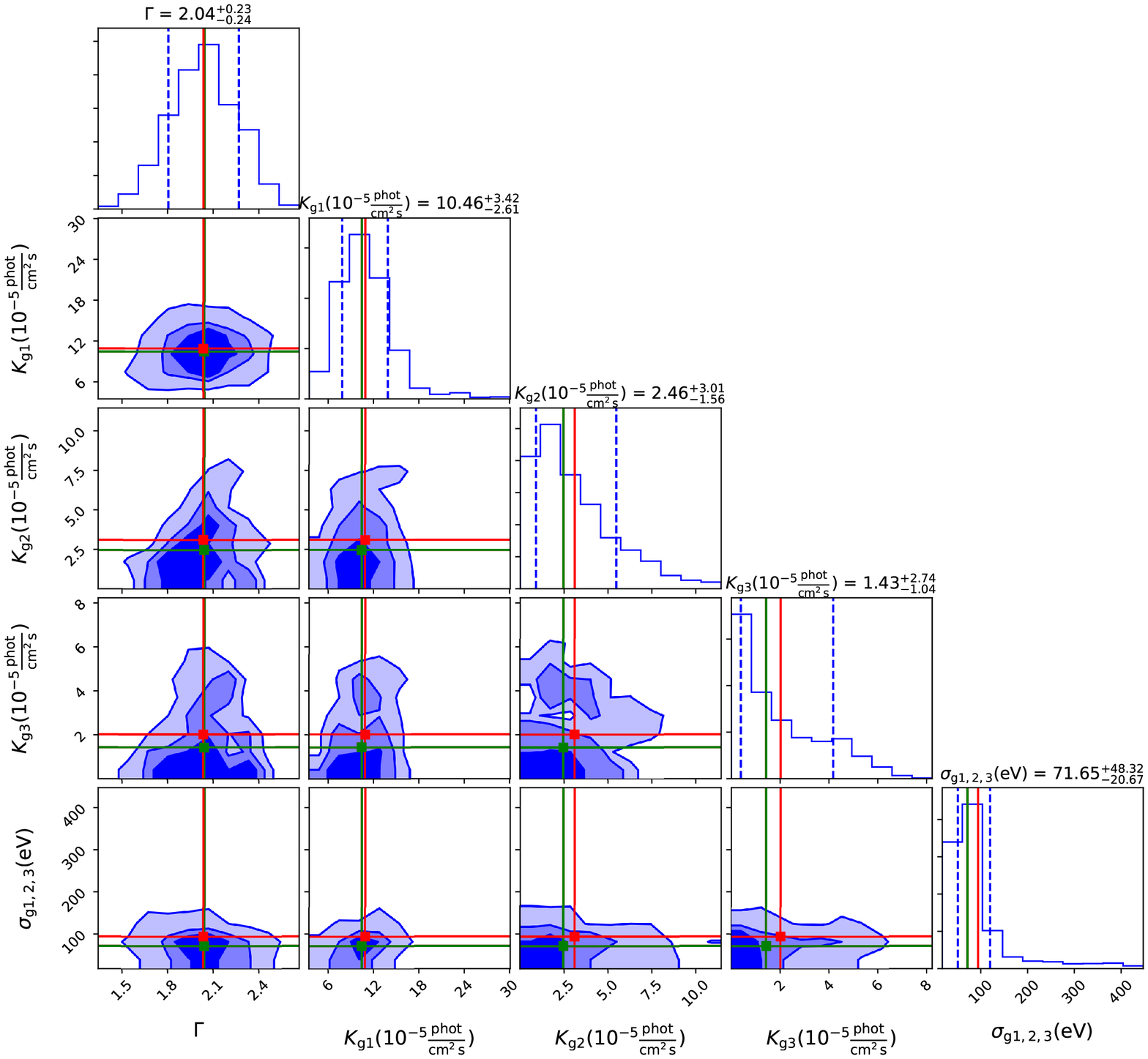}%
\caption{The same as Figure~6, but for the ACIS-S/HETG \textit{low/hard}-state spectrum. 
\label{fig:rtcru:mcmc:a2}%
}
\end{center}
\end{figure*}

\begin{figure*}
\begin{center}
\includegraphics[width=5.00in, trim = 0 0 0 0, clip, angle=270]{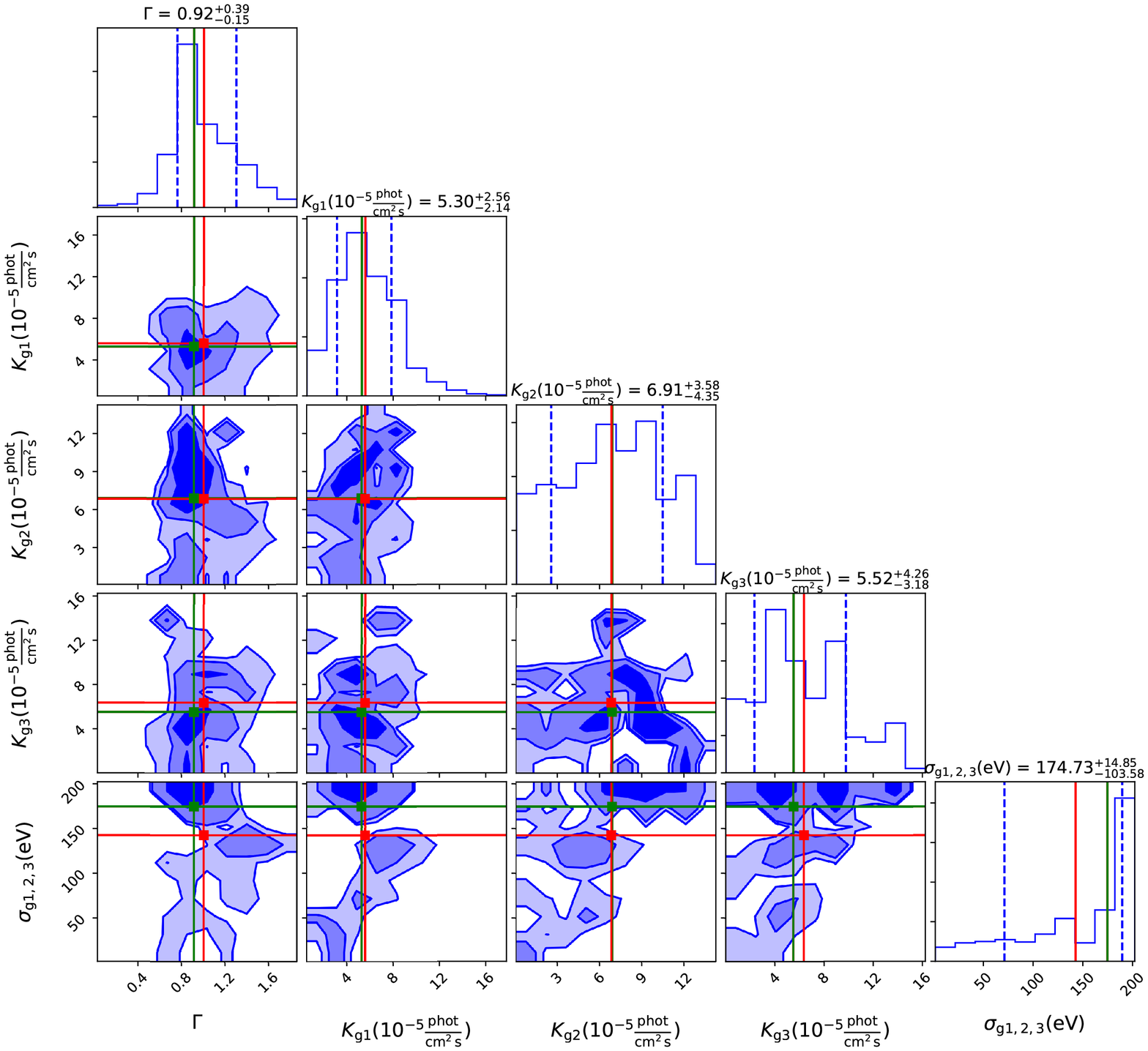}%
\caption{The same as Figure~6, but for the ACIS-S/HETG \textit{high/soft}-state spectrum. 
\label{fig:rtcru:mcmc:a3}%
}
\end{center}
\end{figure*}

\begin{figure*}
\begin{center}
\includegraphics[width=5.00in, trim = 0 0 0 0, clip, angle=270]{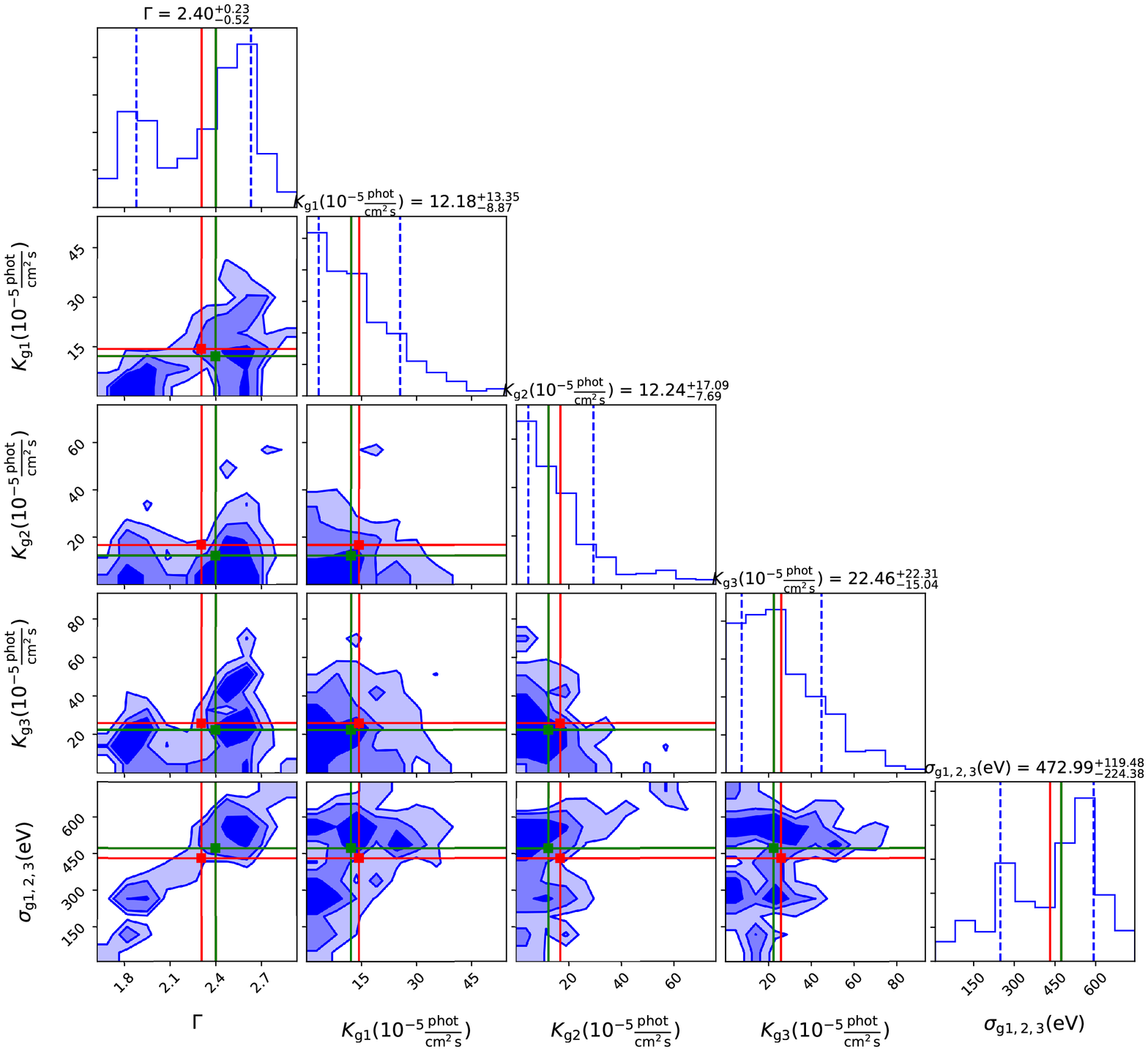}%
\caption{The same as Figure~6, but for the HRC-S/LETG \textit{low/hard}-state spectrum. 
\label{fig:rtcru:mcmc:a4}%
}
\end{center}
\end{figure*}

\begin{figure*}
\begin{center}
\includegraphics[width=5.00in, trim = 0 0 0 0, clip, angle=270]{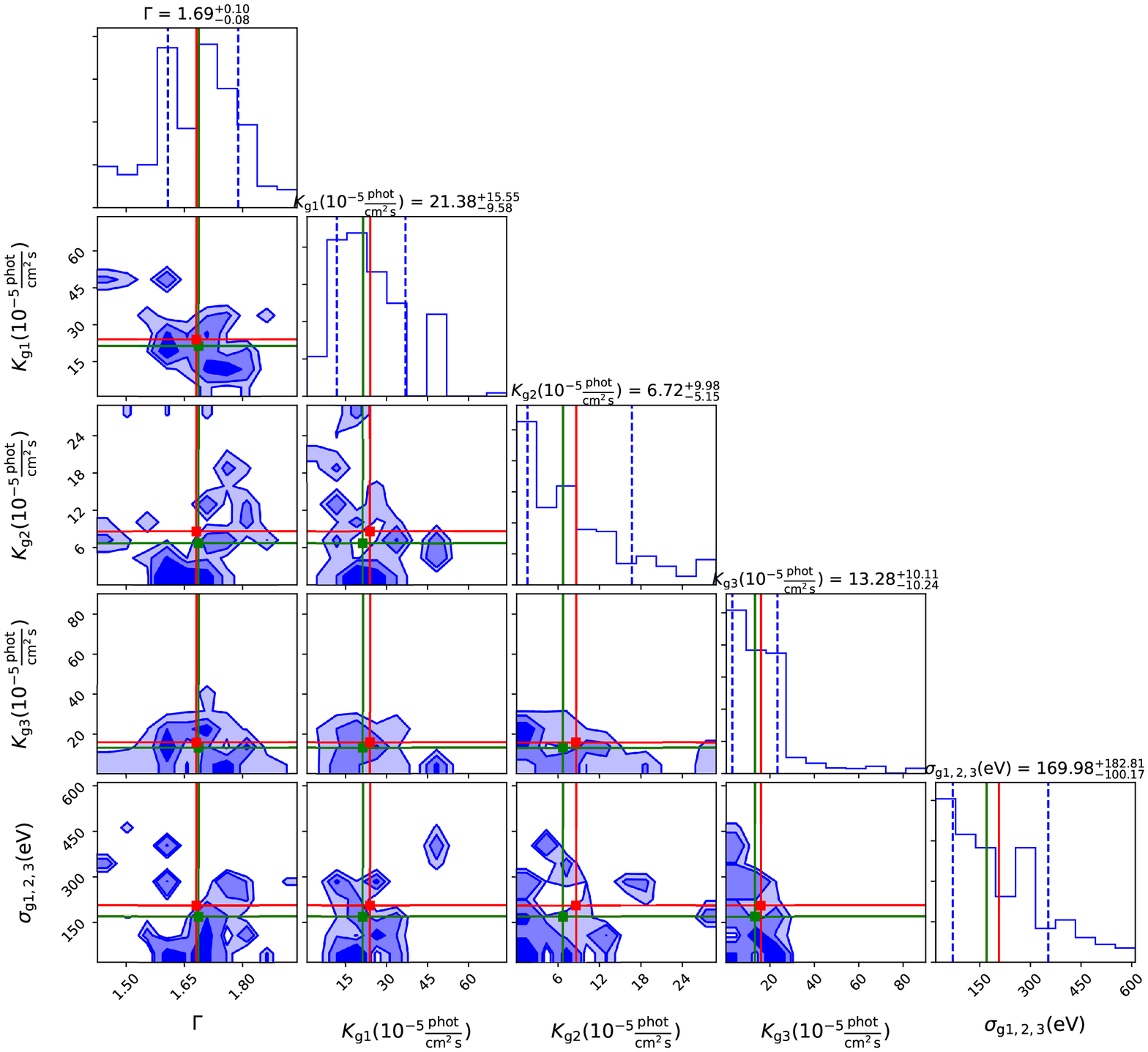}%
\caption{The same as Figure~6, but for the HRC-S/LETG \textit{high/soft}-state spectrum. 
\label{fig:rtcru:mcmc:a5}%
}
\end{center}
\end{figure*}

\newpage


\begin{figure*}
\begin{center}
\includegraphics[width=5.in, trim = 0 0 0 0, clip, angle=270]{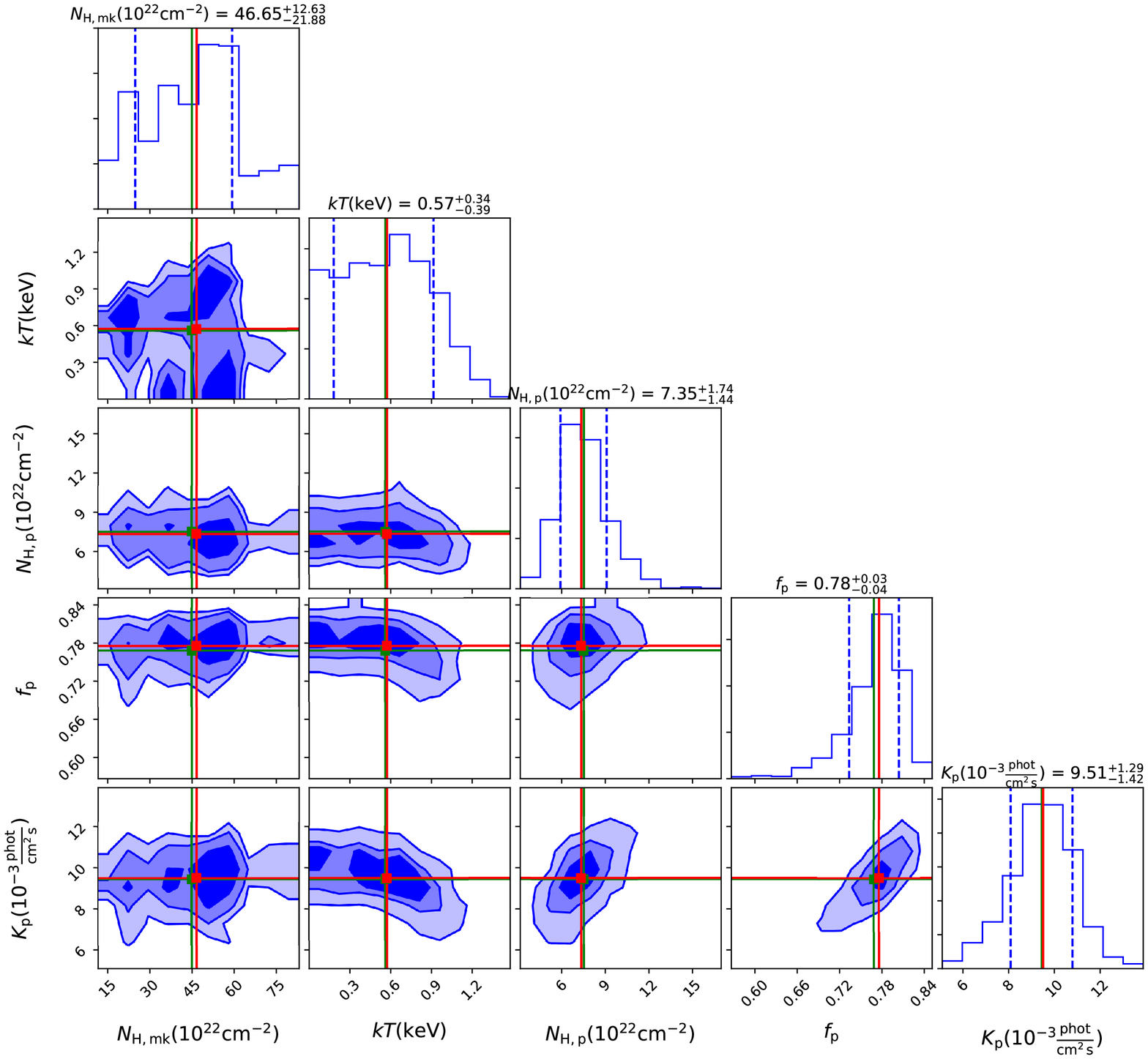}%
\caption{The same as Figure~8, but for the HRC-S/LETG \textit{high/soft}-state spectrum. 
\label{fig:rtcru:mcmc:a7}%
}
\end{center}
\end{figure*}

\begin{figure*}
\begin{center}
\includegraphics[width=5.in, trim = 0 0 0 0, clip, angle=270]{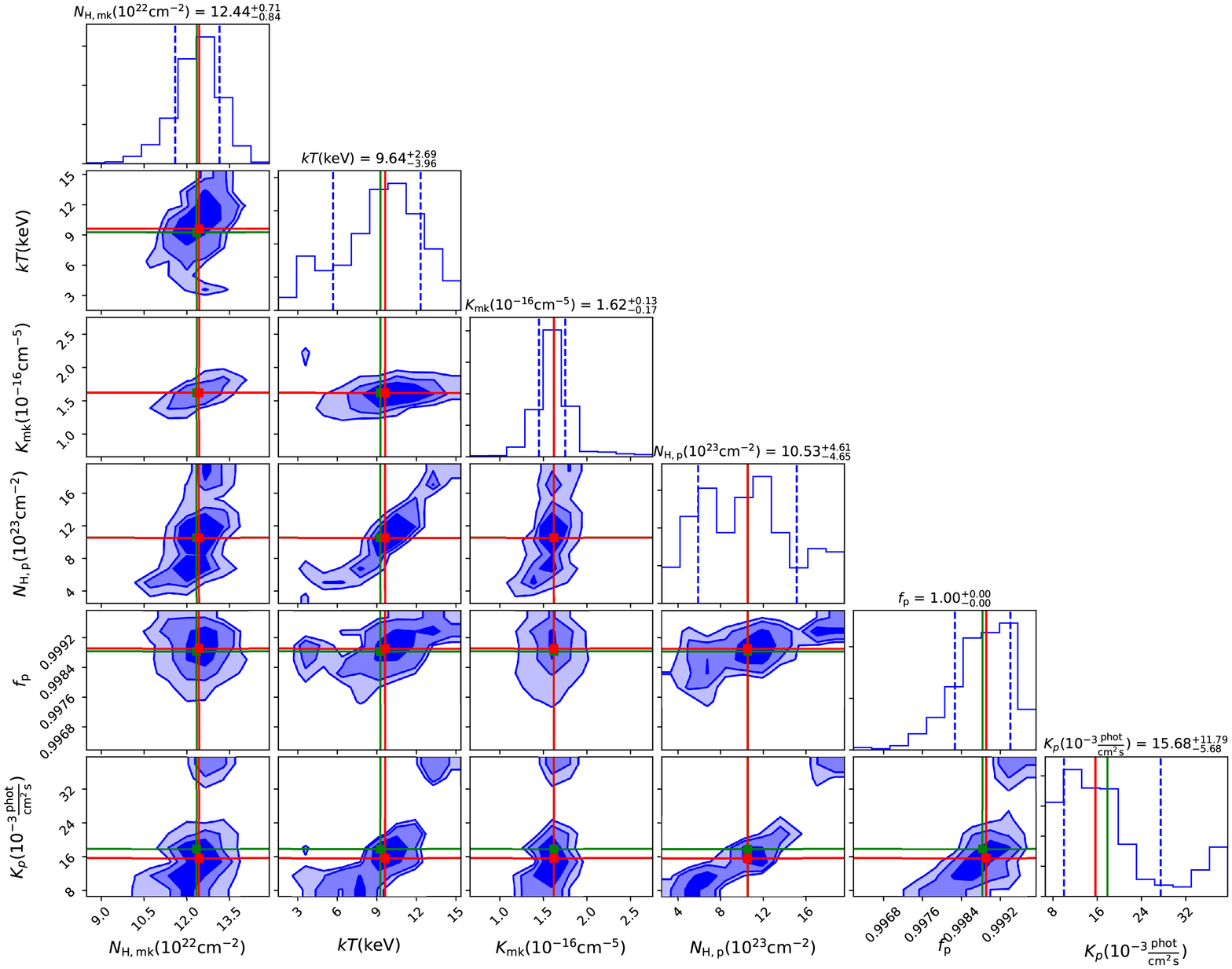}%
\caption{The same as Figure~8, but for the ACIS-S/HETG time-averaged spectrum. 
\label{fig:rtcru:mcmc:a8}%
}
\end{center}
\end{figure*}

\begin{figure*}
\begin{center}
\includegraphics[width=5.in, trim = 0 0 0 0, clip, angle=270]{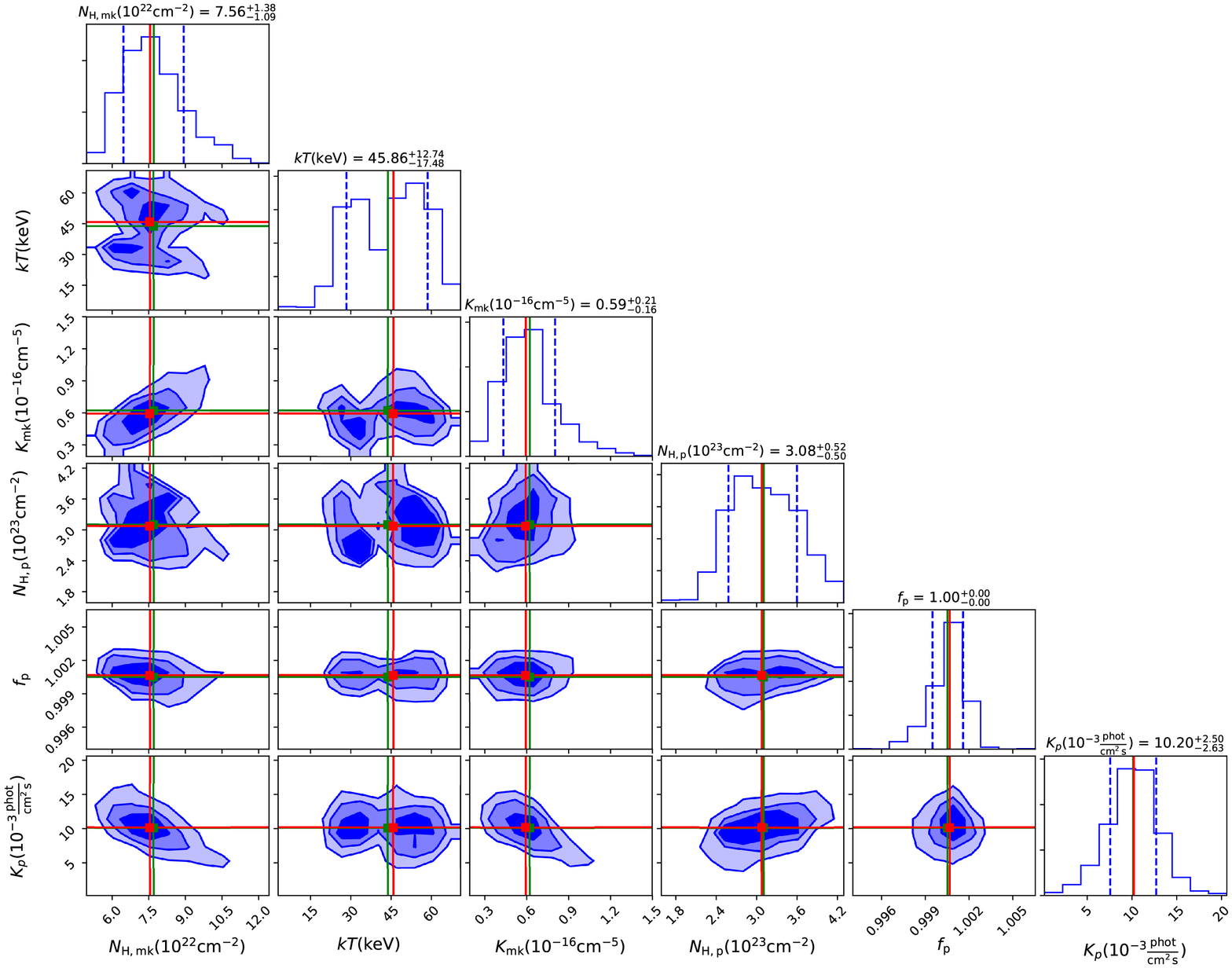}%
\caption{The same as Figure~8, but for the ACIS-S/HETG \textit{low/hard}-state spectrum. 
\label{fig:rtcru:mcmc:a9}%
}
\end{center}
\end{figure*}

\begin{figure*}
\begin{center}
\includegraphics[width=5.0in, trim = 0 0 0 0, clip, angle=270]{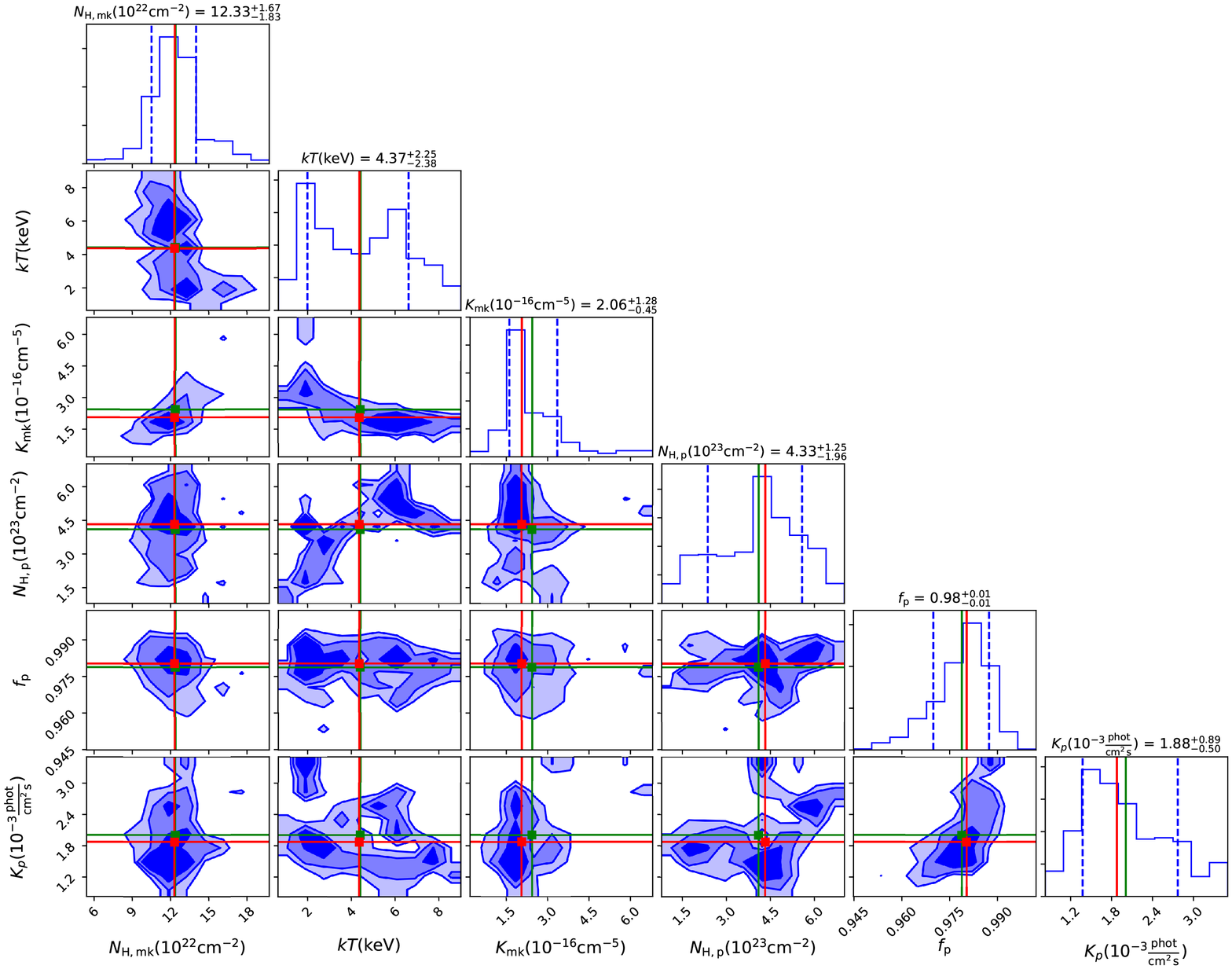}%
\caption{The same as Figure~8, but for the ACIS-S/HETG \textit{high/soft}-state spectrum. 
\label{fig:rtcru:mcmc:a10}%
}
\end{center}
\end{figure*}

\bsp	


\begin{thebibliography}{}
\makeatletter
\relax
\def\mn@urlcharsother{\let\do\@makeother \do\$\do\&\do\#\do\^\do\_\do\%\do\~}
\def\mn@doi{\begingroup\mn@urlcharsother \@ifnextchar [ {\mn@doi@}
  {\mn@doi@[]}}
\def\mn@doi@[#1]#2{\def\@tempa{#1}\ifx\@tempa\@empty \href
  {http://dx.doi.org/#2} {doi:#2}\else \href {http://dx.doi.org/#2} {#1}\fi
  \endgroup}
\def\mn@eprint#1#2{\mn@eprint@#1:#2::\@nil}
\def\mn@eprint@arXiv#1{\href {http://arxiv.org/abs/#1} {{\tt arXiv:#1}}}
\def\mn@eprint@dblp#1{\href {http://dblp.uni-trier.de/rec/bibtex/#1.xml}
  {dblp:#1}}
\def\mn@eprint@#1:#2:#3:#4\@nil{\def\@tempa {#1}\def\@tempb {#2}\def\@tempc
  {#3}\ifx \@tempc \@empty \let \@tempc \@tempb \let \@tempb \@tempa \fi \ifx
  \@tempb \@empty \def\@tempb {arXiv}\fi \@ifundefined
  {mn@eprint@\@tempb}{\@tempb:\@tempc}{\expandafter \expandafter \csname
  mn@eprint@\@tempb\endcsname \expandafter{\@tempc}}}

\bibitem[\protect\citeauthoryear{{Allen}}{{Allen}}{1984}]{Allen1984}
{Allen} D.~A.,  1984, Proc. Astron. Soc. Australia, \href
  {http://adsabs.harvard.edu/abs/1984PASAu...5..369A} {5, 369}

\bibitem[\protect\citeauthoryear{{Anders} \& {Grevesse}}{{Anders} \&
  {Grevesse}}{1989}]{Anders1989}
{Anders} E.,  {Grevesse} N.,  1989, \mn@doi [\gca]
  {10.1016/0016-7037(89)90286-X}, \href
  {https://ui.adsabs.harvard.edu/abs/1989GeCoA..53..197A} {53, 197}

\bibitem[\protect\citeauthoryear{{Arnaud}}{{Arnaud}}{1996}]{Arnaud1996}
{Arnaud} K.~A.,  1996, in {Jacoby} G.~H.,  {Barnes} J.,  eds,  ASP Conf. Ser.
  Vol. 101, Astronomical Data Analysis Software and Systems V. Astron. 
  Soc. Pac., San Francisco, p.~17

\bibitem[\protect\citeauthoryear{{Arnaud} \& {Raymond}}{{Arnaud} \&
  {Raymond}}{1992}]{Arnaud1992}
{Arnaud} M.,  {Raymond} J.,  1992, \mn@doi [\apj] {10.1086/171864}, \href
  {http://adsabs.harvard.edu/abs/1992ApJ...398..394A} {398, 394}

\bibitem[\protect\citeauthoryear{{Arnaud} \& {Rothenflug}}{{Arnaud} \&
  {Rothenflug}}{1985}]{Arnaud1985}
{Arnaud} M.,  {Rothenflug} R.,  1985, \aaps, \href
  {http://adsabs.harvard.edu/abs/1985A%26AS...60..425A} {60, 425}

\bibitem[\protect\citeauthoryear{{Bailer-Jones}, {Rybizki}, {Fouesneau},
  {Mantelet}  \& {Andrae}}{{Bailer-Jones} et~al.}{2018}]{Bailer-Jones2018}
{Bailer-Jones} C.~A.~L.,  {Rybizki} J.,  {Fouesneau} M.,  {Mantelet} G.,
  {Andrae} R.,  2018, \mn@doi [\aj] {10.3847/1538-3881/aacb21}, \href
  {https://ui.adsabs.harvard.edu/abs/2018AJ....156...58B} {156, 58}

\bibitem[\protect\citeauthoryear{{Belczy{\'n}ski}, {Miko{\l}ajewska}, {Munari},
  {Ivison}  \& {Friedjung}}{{Belczy{\'n}ski} et~al.}{2000}]{Belczynski2000}
{Belczy{\'n}ski} K.,  {Miko{\l}ajewska} J.,  {Munari} U.,  {Ivison} R.~J.,
  {Friedjung} M.,  2000, \mn@doi [\aaps] {10.1051/aas:2000280}, \href
  {http://adsabs.harvard.edu/abs/2000A%26AS..146..407B} {146, 407}

\bibitem[\protect\citeauthoryear{{Bohlin}, {Savage}  \& {Drake}}{{Bohlin}
  et~al.}{1978}]{Bohlin1978}
{Bohlin} R.~C.,  {Savage} B.~D.,   {Drake} J.~F.,  1978, \mn@doi [\apj]
  {10.1086/156357}, \href {http://adsabs.harvard.edu/abs/1978ApJ...224..132B}
  {224, 132}

\bibitem[\protect\citeauthoryear{{Brassington} et~al.,}{{Brassington}
  et~al.}{2008}]{Brassington2008}
{Brassington} N.~J.,  et~al., 2008, \mn@doi [\apjs] {10.1086/591527}, \href
  {https://ui.adsabs.harvard.edu/abs/2008ApJS..179..142B} {179, 142}

\bibitem[\protect\citeauthoryear{{Brinkman} et~al.,}{{Brinkman}
  et~al.}{2000}]{Brinkman2000}
{Brinkman} A.~C.,  et~al., 2000, \mn@doi [\apjl] {10.1086/312504}, \href
  {http://adsabs.harvard.edu/abs/2000ApJ...530L.111B} {530, L111}

\bibitem[\protect\citeauthoryear{{Canizares} et~al.,}{{Canizares}
  et~al.}{2005}]{Canizares2005}
{Canizares} C.~R.,  et~al., 2005, \mn@doi [\pasp] {10.1086/432898}, \href
  {https://ui.adsabs.harvard.edu/abs/2005PASP..117.1144C} {117, 1144}

\bibitem[\protect\citeauthoryear{{Cash}}{{Cash}}{1979}]{Cash1979}
{Cash} W.,  1979, \mn@doi [\apj] {10.1086/156922}, \href
  {https://ui.adsabs.harvard.edu/abs/1979ApJ...228..939C} {228, 939}

\bibitem[\protect\citeauthoryear{{Chernyakova}, {Courvoisier}, {Rodriguez}  \&
  {Lutovinov}}{{Chernyakova} et~al.}{2005}]{Chernyakova2005}
{Chernyakova} M.,  {Courvoisier} T.~J.-L.,  {Rodriguez} J.,   {Lutovinov} A.,
  2005, ATel, \href {http://adsabs.harvard.edu/abs/2005ATel..519....1C} {519, 1}

\bibitem[\protect\citeauthoryear{{Cieslinski}, {Elizalde}  \&
  {Steiner}}{{Cieslinski} et~al.}{1994}]{Cieslinski1994}
{Cieslinski} D.,  {Elizalde} F.,   {Steiner} J.~E.,  1994, \aaps, \href
  {http://adsabs.harvard.edu/abs/1994A%26AS..106..243C} {106, 243}

\bibitem[\protect\citeauthoryear{{Corradi}, {Munari}, {Livio}, {Mampaso},
  {Gon{\c c}alves}  \& {Schwarz}}{{Corradi} et~al.}{2001}]{Corradi2001}
{Corradi} R.~L.~M.,  {Munari} U.,  {Livio} M.,  {Mampaso} A.,  {Gon{\c c}alves}
  D.~R.,   {Schwarz} H.~E.,  2001, \mn@doi [\apj] {10.1086/323062}, \href
  {http://adsabs.harvard.edu/abs/2001ApJ...560..912C} {560, 912}

\bibitem[\protect\citeauthoryear{{Danehkar} et~al.,}{{Danehkar}
  et~al.}{2018}]{Danehkar2018}
{Danehkar} A.,  et~al., 2018, \mn@doi [\apj] {10.3847/1538-4357/aaa427}, \href
  {https://ui.adsabs.harvard.edu/abs/2018ApJ...853..165D} {853, 165}

\bibitem[\protect\citeauthoryear{{Ducci}, {Doroshenko}, {Suleimanov},
  {Niko{\l}ajuk}, {Santangelo}  \& {Ferrigno}}{{Ducci}
  et~al.}{2016}]{Ducci2016}
{Ducci} L.,  {Doroshenko} V.,  {Suleimanov} V.,  {Niko{\l}ajuk} M.,
  {Santangelo} A.,   {Ferrigno} C.,  2016, \mn@doi [\aap]
  {10.1051/0004-6361/201628242}, \href
  {http://adsabs.harvard.edu/abs/2016A%26A...592A..58D} {592, A58}

\bibitem[\protect\citeauthoryear{Dun\'{e}r, Hartwig  \& M\"{u}ller}{Dun\'{e}r
  et~al.}{1911}]{Duner1911}
Dun\'{e}r Hartwig  M\"{u}ller 1911, \mn@doi [Astron. Nachr.]
  {10.1002/asna.19111900402}, 190, 57

\bibitem[\protect\citeauthoryear{{Eze}}{{Eze}}{2014}]{Eze2014}
{Eze} R.~N.~C.,  2014, \mn@doi [\mnras] {10.1093/mnras/stt1947}, \href
  {https://ui.adsabs.harvard.edu/abs/2014MNRAS.437..857E} {437, 857}

\bibitem[\protect\citeauthoryear{{Eze}, {Luna}  \& {Smith}}{{Eze}
  et~al.}{2010}]{Eze2010}
{Eze} R.~N.~C.,  {Luna} G.~J.~M.,   {Smith} R.~K.,  2010, \mn@doi [\apj]
  {10.1088/0004-637X/709/2/816}, \href
  {https://ui.adsabs.harvard.edu/abs/2010ApJ...709..816E} {709, 816}

\bibitem[\protect\citeauthoryear{{Ezuka}, {Ishida}  \& {Makino}}{{Ezuka}
  et~al.}{1998}]{Ezuka1998}
{Ezuka} H.,  {Ishida} M.,   {Makino} F.,  1998, \mn@doi [\apj]
  {10.1086/305626}, \href {http://adsabs.harvard.edu/abs/1998ApJ...499..388E}
  {499, 388}

\bibitem[\protect\citeauthoryear{{Foreman-Mackey}}{{Foreman-Mackey}}{2016}]{Foreman-Mackey2016}
{Foreman-Mackey} D.,  2016, \mn@doi [J. Open Source Softw.]
  {10.21105/joss.00024}, \href
  {https://ui.adsabs.harvard.edu/abs/2016JOSS....1...24F} {1, 24}

\bibitem[\protect\citeauthoryear{{Foster}, {Ji}, {Smith}  \&
  {Brickhouse}}{{Foster} et~al.}{2012}]{Foster2012}
{Foster} A.~R.,  {Ji} L.,  {Smith} R.~K.,   {Brickhouse} N.~S.,  2012, \mn@doi
  [\apj] {10.1088/0004-637X/756/2/128}, \href
  {https://ui.adsabs.harvard.edu/abs/2012ApJ...756..128F} {756, 128}

\bibitem[\protect\citeauthoryear{{Freeman}, {Doe}  \&
  {Siemiginowska}}{{Freeman} et~al.}{2001}]{Freeman2001}
{Freeman} P.,  {Doe} S.,   {Siemiginowska} A.,  2001, in {Starck} J.-L.,
  {Murtagh} F.~D.,  eds, Proc. SPIE Conf. Ser. Vol. 4477, Astronomical Data Analysis. pp
  76--87 (\mn@eprint {} {astro-ph/0108426}), \mn@doi{10.1117/12.447161}

\bibitem[\protect\citeauthoryear{{Fruscione} et~al.,}{{Fruscione}
  et~al.}{2006}]{Fruscione2006}
{Fruscione} A.,  et~al., 2006, in {Silva} D.-R.,  {Doxsey} R.~E.,  eds,
  Proc. SPIE Conf. Ser. Vol. 6270, Observatory Operations: Strategies, Processes, 
  and Systems p. SPIE, Bellingham, p. 62701V, \mn@doi{10.1117/12.671760}

\bibitem[\protect\citeauthoryear{{Gaia Collaboration} et~al.,}{{Gaia
  Collaboration} et~al.}{2016}]{GaiaCollaboration2016}
{Gaia Collaboration} et~al., 2016, \mn@doi [\aap]
  {10.1051/0004-6361/201629272}, \href
  {https://ui.adsabs.harvard.edu/abs/2016A&A...595A...1G} {595, A1}

\bibitem[\protect\citeauthoryear{{Gaia Collaboration} et~al.,}{{Gaia
  Collaboration} et~al.}{2018}]{GaiaCollaboration2018}
{Gaia Collaboration} et~al., 2018, \mn@doi [\aap]
  {10.1051/0004-6361/201833051}, \href
  {https://ui.adsabs.harvard.edu/abs/2018A&A...616A...1G} {616, A1}

\bibitem[\protect\citeauthoryear{{Galloway} \& {Sokoloski}}{{Galloway} \&
  {Sokoloski}}{2004}]{Galloway2004}
{Galloway} D.~K.,  {Sokoloski} J.~L.,  2004, \mn@doi [\apjl] {10.1086/424925},
  \href {http://adsabs.harvard.edu/abs/2004ApJ...613L..61G} {613, L61}

\bibitem[\protect\citeauthoryear{{Garmire}, {Bautz}, {Ford}, {Nousek}  \&
  {Ricker}}{{Garmire} et~al.}{2003}]{Garmire2003}
{Garmire} G.~P.,  {Bautz} M.~W.,  {Ford} P.~G.,  {Nousek} J.~A.,   {Ricker} Jr.
  G.~R.,  2003, in {Truemper} J.~E.,  {Tananbaum} H.~D.,  eds,  Proc. SPIE Conf. Ser.
  Vol. 4851, X-Ray and Gamma-Ray Telescopes and Instruments for 
  Astronomy. SPIE, Bellingham. pp
  28--44, \mn@doi{10.1117/12.461599}

\bibitem[\protect\citeauthoryear{{Harris} et~al.,}{{Harris}
  et~al.}{2020}]{Harris2020}
{Harris} C.~R.,  et~al., 2020, \mn@doi [Nature] {10.1038/s41586-020-2649-2},
  \href {https://ui.adsabs.harvard.edu/abs/2020arXiv200610256H} {585, 357}

\bibitem[\protect\citeauthoryear{Hastings}{Hastings}{1970}]{Hastings1970}
Hastings W.~K.,  1970, \mn@doi [Biometrika] {10.1093/biomet/57.1.97}, 57, 97

\bibitem[\protect\citeauthoryear{{Hinkle}, {Fekel}  \& {Joyce}}{{Hinkle}
  et~al.}{2009}]{Hinkle2009}
{Hinkle} K.~H.,  {Fekel} F.~C.,   {Joyce} R.~R.,  2009, \mn@doi [\apj]
  {10.1088/0004-637X/692/2/1360}, \href
  {http://adsabs.harvard.edu/abs/2009ApJ...692.1360H} {692, 1360}

\bibitem[\protect\citeauthoryear{{Houck} \& {Denicola}}{{Houck} \&
  {Denicola}}{2000}]{Houck2000}
{Houck} J.~C.,  {Denicola} L.~A.,  2000, in {Manset} N.,  {Veillet} C.,
  {Crabtree} D.,  eds,  ASP Conf. Ser. Vol. 216, Astronomical Data Analysis
  Software and Systems IX. Astron. Soc. Pac., San Francisco, p.~591

\bibitem[\protect\citeauthoryear{{Houk}}{{Houk}}{1963}]{Houk1963}
{Houk} N.,  1963, \mn@doi [\aj] {10.1086/108948}, \href
  {http://adsabs.harvard.edu/abs/1963AJ.....68..253H} {68, 253}

\bibitem[\protect\citeauthoryear{{Huenemoerder} et~al.,}{{Huenemoerder}
  et~al.}{2011}]{Huenemoerder2011}
{Huenemoerder} D.~P.,  et~al., 2011, \mn@doi [\aj]
  {10.1088/0004-6256/141/4/129}, \href
  {https://ui.adsabs.harvard.edu/abs/2011AJ....141..129H} {141, 129}

\bibitem[\protect\citeauthoryear{{Hunter}}{{Hunter}}{2007}]{Hunter2007}
{Hunter} J.~D.,  2007, \mn@doi [Comput. Sci. Eng.] {10.1109/MCSE.2007.55},
  \href {https://ui.adsabs.harvard.edu/abs/2007CSE.....9...90H} {9, 90}

\bibitem[\protect\citeauthoryear{{Kaastra}}{{Kaastra}}{1992}]{Kaastra1992}
{Kaastra} J.~S.,  1992, An X-Ray Spectral Code for Optically Thin Plasmas.
Internal SRON-Leiden Report, updated version 2.0

\bibitem[\protect\citeauthoryear{{Karovska}, {Carilli}, {Raymond}  \&
  {Mattei}}{{Karovska} et~al.}{2007}]{Karovska2007}
{Karovska} M.,  {Carilli} C.~L.,  {Raymond} J.~C.,   {Mattei} J.~A.,  2007,
  \mn@doi [\apj] {10.1086/516772}, \href
  {http://adsabs.harvard.edu/abs/2007ApJ...661.1048K} {661, 1048}

\bibitem[\protect\citeauthoryear{{Karovska}, {Gaetz}, {Carilli}, {Hack},
  {Raymond}  \& {Lee}}{{Karovska} et~al.}{2010}]{Karovska2010}
{Karovska} M.,  {Gaetz} T.~J.,  {Carilli} C.~L.,  {Hack} W.,  {Raymond} J.~C.,
   {Lee} N.~P.,  2010, \mn@doi [\apjl] {10.1088/2041-8205/710/2/L132}, \href
  {http://adsabs.harvard.edu/abs/2010ApJ...710L.132K} {710, L132}

\bibitem[\protect\citeauthoryear{{Kashyap}, {Kennea}, {Karovska}  \& {Chandra
  Calibration}}{{Kashyap} et~al.}{2013}]{Kashyap2013}
{Kashyap} V.,  {Kennea} J.~A.,  {Karovska} M.,   {Chandra Calibration} 2013, in
  AAS/High Energy Astrophysics Division Vol. 13. p. 126.57

\bibitem[\protect\citeauthoryear{{Kennea}, {Mukai}, {Sokoloski}, {Luna},
  {Tueller}, {Markwardt}  \& {Burrows}}{{Kennea} et~al.}{2009}]{Kennea2009}
{Kennea} J.~A.,  {Mukai} K.,  {Sokoloski} J.~L.,  {Luna} G.~J.~M.,  {Tueller}
  J.,  {Markwardt} C.~B.,   {Burrows} D.~N.,  2009, \mn@doi [\apj]
  {10.1088/0004-637X/701/2/1992}, \href
  {http://adsabs.harvard.edu/abs/2009ApJ...701.1992K} {701, 1992}

\bibitem[\protect\citeauthoryear{{Kenyon}}{{Kenyon}}{1986}]{Kenyon1986}
{Kenyon} S.~J.,  1986, {The Symbiotic Stars}.
Cambridge University Press

\bibitem[\protect\citeauthoryear{{Kenyon} \& {Fernandez-Castro}}{{Kenyon} \&
  {Fernandez-Castro}}{1987}]{Kenyon1987}
{Kenyon} S.~J.,  {Fernandez-Castro} T.,  1987, \mn@doi [\aj] {10.1086/114379},
  \href {http://adsabs.harvard.edu/abs/1987AJ.....93..938K} {93, 938}

\bibitem[\protect\citeauthoryear{{Kholopov}}{{Kholopov}}{1987}]{Kholopov1987}
{Kholopov} P.~N.,  1987, {General Catalogue of Variable Stars.}.
Nauka, Moscow, USSR

\bibitem[\protect\citeauthoryear{{Kukarkin} et~al.,}{{Kukarkin}
  et~al.}{1969}]{Kukarkin1969}
{Kukarkin} B.~V.,  et~al., 1969, {General Catalogue of Variable Stars.
  Volume 1. Constellations Andromeda - Grus.}, Sternberg State Astronomy 
  Institute of the Moscow State University, Moscow, USSR

\bibitem[\protect\citeauthoryear{{Lagarias}, {Reeds}, {Wright}  \&
  E.}{{Lagarias} et~al.}{1998}]{Lagarias1998}
{Lagarias} J.~C.,  {Reeds} J.~A.,  {Wright} M.~H.,   E. W.~P.,  1998, \mn@doi
  [SIAM J. Optim.] {10.1137/S1052623496303470}, 9, 112

\bibitem[\protect\citeauthoryear{{Leavitt} \& {Pickering}}{{Leavitt} \&
  {Pickering}}{1906}]{Leavitt1906}
{Leavitt} H.~S.,  {Pickering} E.~C.,  1906, Harvard College Observatory
  Circular, \href {http://adsabs.harvard.edu/abs/1906HarCi.120....1L} {120, 1}

\bibitem[\protect\citeauthoryear{{Lee} et~al.,}{{Lee} et~al.}{2011}]{Lee2011}
{Lee} H.,  et~al., 2011, \mn@doi [\apj] {10.1088/0004-637X/731/2/126}, \href
  {https://ui.adsabs.harvard.edu/abs/2011ApJ...731..126L} {731, 126}

\bibitem[\protect\citeauthoryear{{Liedahl}, {Osterheld}  \&
  {Goldstein}}{{Liedahl} et~al.}{1995}]{Liedahl1995}
{Liedahl} D.~A.,  {Osterheld} A.~L.,   {Goldstein} W.~H.,  1995, \mn@doi
  [\apjl] {10.1086/187729}, \href
  {http://adsabs.harvard.edu/abs/1995ApJ...438L.115L} {438, L115}

\bibitem[\protect\citeauthoryear{{Lucy}, {Knigge}  \& {Sokoloski}}{{Lucy}
  et~al.}{2018}]{Lucy2018}
{Lucy} A.~B.,  {Knigge} C.,   {Sokoloski} J.~L.,  2018, \mn@doi [\mnras]
  {10.1093/mnras/sty929}, \href
  {https://ui.adsabs.harvard.edu/abs/2018MNRAS.478..568L} {478, 568}

\bibitem[\protect\citeauthoryear{{Luna} \& {Sokoloski}}{{Luna} \&
  {Sokoloski}}{2007}]{Luna2007}
{Luna} G.~J.~M.,  {Sokoloski} J.~L.,  2007, \mn@doi [\apj] {10.1086/522576},
  \href {http://adsabs.harvard.edu/abs/2007ApJ...671..741L} {671, 741}

\bibitem[\protect\citeauthoryear{{Luna}, {Sokoloski}  \& {Mukai}}{{Luna}
  et~al.}{2008}]{Luna2008}
{Luna} G.~J.~M.,  {Sokoloski} J.~L.,   {Mukai} K.,  2008, in {Evans} A.,
  {Bode} M.~F.,  {O'Brien} T.~J.,   {Darnley} M.~J.,  eds,  ASP Conf. Ser. Vol.
  401, RS Ophiuchi (2006) and the Recurrent Nova Phenomenon. Astron. Soc. Pac., San Francisco, 
  p.~342 (\mn@eprint
  {arXiv} {0711.0725})

\bibitem[\protect\citeauthoryear{{Luna} et~al.,}{{Luna}
  et~al.}{2018}]{Luna2018}
{Luna} G.~J.~M.,  et~al., 2018, \mn@doi [\aap] {10.1051/0004-6361/201832592},
  \href {https://ui.adsabs.harvard.edu/abs/2018A&A...616A..53L} {616, A53}

\bibitem[\protect\citeauthoryear{{Luri} et~al.,}{{Luri}
  et~al.}{2018}]{Luri2018}
{Luri} X.,  et~al., 2018, \mn@doi [\aap] {10.1051/0004-6361/201832964}, \href
  {https://ui.adsabs.harvard.edu/abs/2018A&A...616A...9L} {616, A9}

\bibitem[\protect\citeauthoryear{{Metropolis}, {Rosenbluth}, {Rosenbluth},
  {Teller}  \& {Teller}}{{Metropolis} et~al.}{1953}]{Metropolis1953}
{Metropolis} N.,  {Rosenbluth} A.~W.,  {Rosenbluth} M.~N.,  {Teller} A.~H.,
  {Teller} E.,  1953, \mn@doi [\jcp] {10.1063/1.1699114}, \href
  {https://ui.adsabs.harvard.edu/abs/1953JChPh..21.1087M} {21, 1087}

\bibitem[\protect\citeauthoryear{{Mewe}, {Gronenschild}  \& {van den
  Oord}}{{Mewe} et~al.}{1985}]{Mewe1985}
{Mewe} R.,  {Gronenschild} E.~H.~B.~M.,   {van den Oord} G.~H.~J.,  1985,
  \aaps, \href {http://adsabs.harvard.edu/abs/1985A%26AS...62..197M} {62, 197}

\bibitem[\protect\citeauthoryear{{Mewe}, {Lemen}  \& {van den Oord}}{{Mewe}
  et~al.}{1986}]{Mewe1986}
{Mewe} R.,  {Lemen} J.~R.,   {van den Oord} G.~H.~J.,  1986, \aaps, \href
  {http://adsabs.harvard.edu/abs/1986A%26AS...65..511M} {65, 511}

\bibitem[\protect\citeauthoryear{{Mikolajewski}, {Mikolajewska}  \&
  {Khudiakova}}{{Mikolajewski} et~al.}{1990}]{Mikolajewski1990}
{Mikolajewski} M.,  {Mikolajewska} J.,   {Khudiakova} T.~N.,  1990, \aap, \href
  {http://adsabs.harvard.edu/abs/1990A%26A...235..219M} {235, 219}

\bibitem[\protect\citeauthoryear{{Mor{\'e}}}{{Mor{\'e}}}{1978}]{More1978}
{Mor{\'e}} J.~J.,  1978, {The Levenberg-Marquardt Algorithm: Implementation and
  Theory}.
Lecture Notes in Mathematics, Berlin Springer Verlag, pp 105--116,
  \mn@doi{10.1007/BFb0067700}

\bibitem[\protect\citeauthoryear{{Muerset}, {Nussbaumer}, {Schmid}  \&
  {Vogel}}{{Muerset} et~al.}{1991}]{Muerset1991}
{Muerset} U.,  {Nussbaumer} H.,  {Schmid} H.~M.,   {Vogel} M.,  1991, \aap,
  \href {http://adsabs.harvard.edu/abs/1991A%26A...248..458M} {248, 458}

\bibitem[\protect\citeauthoryear{{Muerset}, {Wolff}  \& {Jordan}}{{Muerset}
  et~al.}{1997}]{Muerset1997}
{Muerset} U.,  {Wolff} B.,   {Jordan} S.,  1997, \aap, \href
  {https://ui.adsabs.harvard.edu/abs/1997A&A...319..201M} {319, 201}

\bibitem[\protect\citeauthoryear{{Murray} et~al.,}{{Murray}
  et~al.}{1997}]{Murray1997}
{Murray} S.~S.,  et~al., 1997, in {Siegmund} O.~H.,  {Gummin} M.~A., eds,
  Proc. SPIE Conf. Ser. Vol. 3114, EUV, X-Ray, and Gamma-Ray Instrumentation for
  Astronomy VIII. SPIE, Bellingham, pp 11--25, \mn@doi{10.1117/12.283772}

\bibitem[\protect\citeauthoryear{{Park}, {Kashyap}, {Siemiginowska}, {van Dyk},
  {Zezas}, {Heinke}  \& {Wargelin}}{{Park} et~al.}{2006}]{Park2006}
{Park} T.,  {Kashyap} V.~L.,  {Siemiginowska} A.,  {van Dyk} D.~A.,  {Zezas}
  A.,  {Heinke} C.,   {Wargelin} B.~J.,  2006, \mn@doi [\apj] {10.1086/507406},
  \href {http://adsabs.harvard.edu/abs/2006ApJ...652..610P} {652, 610}

\bibitem[\protect\citeauthoryear{{Phillips}, {Mewe}, {Harra-Murnion},
  {Kaastra}, {Beiersdorfer}, {Brown}  \& {Liedahl}}{{Phillips}
  et~al.}{1999}]{Phillips1999}
{Phillips} K.~J.~H.,  {Mewe} R.,  {Harra-Murnion} L.~K.,  {Kaastra} J.~S.,
  {Beiersdorfer} P.,  {Brown} G.~V.,   {Liedahl} D.~A.,  1999, \mn@doi [\aaps]
  {10.1051/aas:1999282}, \href
  {http://cdsads.u-strasbg.fr/abs/1999A%26AS..138..381P} {138, 381}

\bibitem[\protect\citeauthoryear{{Plucinsky} et~al.,}{{Plucinsky}
  et~al.}{2008}]{Plucinsky2008}
{Plucinsky} P.~P.,  et~al., 2008, \mn@doi [\apjs] {10.1086/522942}, \href
  {https://ui.adsabs.harvard.edu/abs/2008ApJS..174..366P} {174, 366}

\bibitem[\protect\citeauthoryear{{Protassov}, {van Dyk}, {Connors}, {Kashyap}
  \& {Siemiginowska}}{{Protassov} et~al.}{2002}]{Protassov2002}
{Protassov} R.,  {van Dyk} D.~A.,  {Connors} A.,  {Kashyap} V.~L.,
  {Siemiginowska} A.,  2002, \mn@doi [\apj] {10.1086/339856}, \href
  {https://ui.adsabs.harvard.edu/abs/2002ApJ...571..545P} {571, 545}

\bibitem[\protect\citeauthoryear{{Schmid}, {Kaufer}, {Camenzind}, {Rivinius},
  {Stahl}, {Szeifert}, {Tubbesing}  \& {Wolf}}{{Schmid}
  et~al.}{2001}]{Schmid2001}
{Schmid} H.~M.,  {Kaufer} A.,  {Camenzind} M.,  {Rivinius} T.,  {Stahl} O.,
  {Szeifert} T.,  {Tubbesing} S.,   {Wolf} B.,  2001, \mn@doi [\aap]
  {10.1051/0004-6361:20011073}, \href
  {https://ui.adsabs.harvard.edu/abs/2001A&A...377..206S} {377, 206}

\bibitem[\protect\citeauthoryear{{Sguera} et~al.,}{{Sguera}
  et~al.}{2012}]{Sguera2012}
{Sguera} V.,  et~al., 2012, ATel, \href
  {http://adsabs.harvard.edu/abs/2012ATel.3887....1S} {3887, 1}

\bibitem[\protect\citeauthoryear{{Sguera}, {Bird}  \& {Sidoli}}{{Sguera}
  et~al.}{2015}]{Sguera2015}
{Sguera} V.,  {Bird} A.~J.,   {Sidoli} L.,  2015, ATel,
  \href {https://ui.adsabs.harvard.edu/abs/2015ATel.8448....1S} {8448, 1}

\bibitem[\protect\citeauthoryear{{Smith}, {Brickhouse}, {Liedahl}  \&
  {Raymond}}{{Smith} et~al.}{2001}]{Smith2001}
{Smith} R.~K.,  {Brickhouse} N.~S.,  {Liedahl} D.~A.,   {Raymond} J.~C.,  2001,
  \mn@doi [\apjl] {10.1086/322992}, \href
  {https://ui.adsabs.harvard.edu/abs/2001ApJ...556L..91S} {556, L91}

\bibitem[\protect\citeauthoryear{{Smith}, {Mushotzky}, {Mukai}, {Kallman},
  {Markwardt}  \& {Tueller}}{{Smith} et~al.}{2008}]{Smith2008}
{Smith} R.~K.,  {Mushotzky} R.,  {Mukai} K.,  {Kallman} T.,  {Markwardt} C.~B.,
    {Tueller} J.,  2008, \mn@doi [\pasj] {10.1093/pasj/60.sp1.S43}, \href
  {https://ui.adsabs.harvard.edu/abs/2008PASJ...60S..43S} {60, S43}

\bibitem[\protect\citeauthoryear{{Smith} et~al.,}{{Smith}
  et~al.}{2016}]{Smith2016}
{Smith} R.~K.,  et~al., 2016, in {den Herder} J.~A.,  {Takahashi} T.,   {Bautz}
  M.,  eds,  Proc. SPIE Conf. Ser. Vol. 9905, Space Telescopes and Instrumentation 
  2016: Ultraviolet to Gamma Ray. SPIE, Bellingham, p. 99054M,
  \mn@doi{10.1117/12.2231778}

\bibitem[\protect\citeauthoryear{{Sokoloski}}{{Sokoloski}}{2003}]{Sokoloski2003a}
{Sokoloski} J.~L.,  2003, in {Corradi} R.~L.~M.,  {Mikolajewska} J.,
  {Mahoney} T.~J.,  eds,  ASP Conf. Ser. Vol. 303, Symbiotic Stars Probing
  Stellar Evolution. p.~202 (\mn@eprint {} {astro-ph/0209101})

\bibitem[\protect\citeauthoryear{{Sokoloski} \& {Kenyon}}{{Sokoloski} \&
  {Kenyon}}{2003}]{Sokoloski2003}
{Sokoloski} J.~L.,  {Kenyon} S.~J.,  2003, \mn@doi [\apj] {10.1086/345901},
  \href {http://adsabs.harvard.edu/abs/2003ApJ...584.1021S} {584, 1021}

\bibitem[\protect\citeauthoryear{{Sokoloski}, {Bildsten}  \& {Ho}}{{Sokoloski}
  et~al.}{2001}]{Sokoloski2001}
{Sokoloski} J.~L.,  {Bildsten} L.,   {Ho} W.~C.~G.,  2001, \mn@doi [\mnras]
  {10.1046/j.1365-8711.2001.04582.x}, \href
  {http://adsabs.harvard.edu/abs/2001MNRAS.326..553S} {326, 553}

\bibitem[\protect\citeauthoryear{{Stella} \& {Angelini}}{{Stella} \&
  {Angelini}}{1992}]{Stella1992}
{Stella} L.,  {Angelini} L.,  1992, in {Worrall} D.~M.,  {Biemesderfer} C.,
  {Barnes} J.,  eds,  ASP Conf. Ser. Vol. 25, Astronomical Data Analysis
  Software and Systems I. Astron. Soc. Pac., San Francisco, p.~103

\bibitem[\protect\citeauthoryear{{Stute} \& {Sahai}}{{Stute} \&
  {Sahai}}{2009}]{Stute2009}
{Stute} M.,  {Sahai} R.,  2009, \mn@doi [\aap] {10.1051/0004-6361/200811176},
  \href {https://ui.adsabs.harvard.edu/abs/2009A&A...498..209S} {498, 209}

\bibitem[\protect\citeauthoryear{{Tomov}, {Kolev}, {Georgiev}, {Zamanov},
  {Antov}  \& {Bellas}}{{Tomov} et~al.}{1990}]{Tomov1990}
{Tomov} T.,  {Kolev} D.,  {Georgiev} L.,  {Zamanov} R.,  {Antov} A.,   {Bellas}
  Y.,  1990, \mn@doi [\nat] {10.1038/346637a0}, \href
  {https://ui.adsabs.harvard.edu/abs/1990Natur.346..637T} {346, 637}

\bibitem[\protect\citeauthoryear{{Tomov}, {Zamanov}, {Kolev}, {Georgiev},
  {Antov}, {Mikolajewski}  \& {Esipov}}{{Tomov} et~al.}{1992}]{Tomov1992}
{Tomov} T.,  {Zamanov} R.,  {Kolev} D.,  {Georgiev} L.,  {Antov} A.,
  {Mikolajewski} M.,   {Esipov} V.,  1992, \mn@doi [\mnras]
  {10.1093/mnras/258.1.23}, \href
  {https://ui.adsabs.harvard.edu/abs/1992MNRAS.258...23T} {258, 23}

\bibitem[\protect\citeauthoryear{{Tueller} et~al.,}{{Tueller}
  et~al.}{2005}]{Tueller2005}
{Tueller} J.,  et~al., 2005, ATel, \href
  {http://adsabs.harvard.edu/abs/2005ATel..669....1T} {669, 1}

\bibitem[\protect\citeauthoryear{{Weisskopf}, {Brinkman}, {Canizares},
  {Garmire}, {Murray}  \& {Van Speybroeck}}{{Weisskopf}
  et~al.}{2002}]{Weisskopf2002}
{Weisskopf} M.~C.,  {Brinkman} B.,  {Canizares} C.,  {Garmire} G.,  {Murray}
  S.,   {Van Speybroeck} L.~P.,  2002, \mn@doi [\pasp] {10.1086/338108}, \href
  {https://ui.adsabs.harvard.edu/abs/2002PASP..114....1W} {114, 1}

\bibitem[\protect\citeauthoryear{{Willingale}, {Starling}, {Beardmore},
  {Tanvir}  \& {O'Brien}}{{Willingale} et~al.}{2013}]{Willingale2013}
{Willingale} R.,  {Starling} R.~L.~C.,  {Beardmore} A.~P.,  {Tanvir} N.~R.,
  {O'Brien} P.~T.,  2013, \mn@doi [\mnras] {10.1093/mnras/stt175}, \href
  {http://adsabs.harvard.edu/abs/2013MNRAS.431..394W} {431, 394}

\bibitem[\protect\citeauthoryear{{Wilms}, {Allen}  \& {McCray}}{{Wilms}
  et~al.}{2000}]{Wilms2000}
{Wilms} J.,  {Allen} A.,   {McCray} R.,  2000, \mn@doi [\apj] {10.1086/317016},
  \href {http://adsabs.harvard.edu/abs/2000ApJ...542..914W} {542, 914}

\bibitem[\protect\citeauthoryear{{Worrall}, {Birkinshaw}, {O'Sullivan},
  {Zezas}, {Wolter}, {Trinchieri}  \& {Fabbiano}}{{Worrall}
  et~al.}{2010}]{Worrall2010}
{Worrall} D.~M.,  {Birkinshaw} M.,  {O'Sullivan} E.,  {Zezas} A.,  {Wolter} A.,
   {Trinchieri} G.,   {Fabbiano} G.,  2010, \mn@doi [\mnras]
  {10.1111/j.1365-2966.2010.17162.x}, \href
  {https://ui.adsabs.harvard.edu/abs/2010MNRAS.408..701W} {408, 701}

\bibitem[\protect\citeauthoryear{{Wright}}{{Wright}}{1996}]{Wright1996}
{Wright} M.,  1996, in {Griffiths} D.,  {Watson} G.,  eds, {Numerical
  Analysis}. Dundee Biennial Conf. Num. Analysis Proc..
Addison-Wesley, Harlow, pp 191--208

\bibitem[\protect\citeauthoryear{{Xu} et~al.,}{{Xu} et~al.}{2014}]{Xu2014}
{Xu} J.,  et~al., 2014, \mn@doi [\apj] {10.1088/0004-637X/794/2/97}, \href
  {https://ui.adsabs.harvard.edu/abs/2014ApJ...794...97X} {794, 97}

\bibitem[\protect\citeauthoryear{{van Dyk}, {Connors}, {Kashyap}  \&
  {Siemiginowska}}{{van Dyk} et~al.}{2001}]{vanDyk2001}
{van Dyk} D.~A.,  {Connors} A.,  {Kashyap} V.~L.,   {Siemiginowska} A.,  2001,
  \mn@doi [\apj] {10.1086/318656}, \href
  {https://ui.adsabs.harvard.edu/abs/2001ApJ...548..224V} {548, 224}

\bibitem[\protect\citeauthoryear{{van der Walt}, {Colbert}  \&
  {Varoquaux}}{{van der Walt} et~al.}{2011}]{vanderWalt2011}
{van der Walt} S.,  {Colbert} S.~C.,   {Varoquaux} G.,  2011, \mn@doi [Comput.
  Sci. Eng.] {10.1109/MCSE.2011.37}, \href
  {https://ui.adsabs.harvard.edu/abs/2011CSE....13b..22V} {13, 22}

\makeatother
\end{thebibliography}

\end{document}